\documentstyle[sprocl]{article}

\def\Journal#1#2#3#4{{#1} {\bf #2} (#3), #4}

\def\NPB{{\em Nucl. Phys.} B}
\def\PLB{{\em Phys. Lett.}  B}

\def\PRD{{\em Phys. Rev.} D}

\def\be{\begin{equation}}
\def\ee{\end{equation}}
\def\bea{\begin{eqnarray}}
\def\eea{\end{eqnarray}}
\newcommand{\AmS}{{\protect\the\textfont2
  A\kern-.1667em\lower.5ex\hbox{M}\kern-.125emS}}

\def\un#1{\relax\ifmmode\@@underline#1\else
        $\@@underline{\hbox{#1}}$\relax\fi}

\def\fracm#1#2{\hbox{\large{${\frac{{#1}}{{#2}}}$}}}

\def\ad{{\kern0.5pt
                   \alpha \kern-5.05pt \raise5.8pt\hbox{$\textstyle.$}\kern
0.5pt}}

\def\Dot#1{{\kern0.5pt
     {#1} \kern-5.05pt \raise5.8pt\hbox{$\textstyle.$}\kern
0.5pt}}

\def\a{\alpha}
\def\b{\beta}

\def\d{\delta}
\def\e{\epsilon}

\def\g{\gamma}

\def\k{\kappa}
\def\l{\lambda}
\def\m{\mu}
\def\n{\nu}
\def\o{\omega}

\def\q{\theta}
\def\r{\rho}
\def\s{\sigma}

\def\D{\Delta}

\def\G{\Gamma}

\def\L{\Lambda}
\def\O{\Omega}

\def\S{\Sigma}

\def\bo{{\raise.15ex\hbox{\large$\Box$}}}               % D'Alembertian
\def\pa{\partial}                                       % curly d
\def\iff{\leftrightarrow}                               % <-->
\def\TH{{\raise.2ex\hbox{$\displaystyle \bigodot$}\mskip-4.7mu \llap H \;}}
\def\face{{\raise.2ex\hbox{$\displaystyle \bigodot$}\mskip-2.2mu \llap {$\ddot
        \smile$}}}                                      % happy face
\def\Tilde#1{\widetilde{#1}}                    % big tilde
\def\Hat#1{\widehat{#1}}                        % big hat
\def\Bar#1{\overline{#1}}                       % big bar
\def\leftrightarrowfill{$\mathsurround=0pt \mathord\leftarrow \mkern-6mu
        \cleaders\hbox{$\mkern-2mu \mathord- \mkern-2mu$}\hfill
        \mkern-6mu \mathord\rightarrow$}
\def\dvec#1{\vbox{\ialign{##\crcr
        \leftrightarrowfill\crcr\noalign{\kern-1pt\nointerlineskip}
        $\hfil\displaystyle{#1}\hfil$\crcr}}}           % <--> accent
\def\fracm#1#2{\hbox{\large{${\frac{{#1}}{{#2}}}$}}}
\def\frac#1#2{{\textstyle{#1\over\vphantom2\smash{\raise.20ex
        \hbox{$\scriptstyle{#2}$}}}}}                   % fraction
\def\sfrac#1#2{{\vphantom1\smash{\lower.5ex\hbox{\small$#1$}}\over
        \vphantom1\smash{\raise.4ex\hbox{\small$#2$}}}} % alternate fraction
\def\bfrac#1#2{{\vphantom1\smash{\lower.5ex\hbox{$#1$}}\over
        \vphantom1\smash{\raise.3ex\hbox{$#2$}}}}       % "
\def\afrac#1#2{{\vphantom1\smash{\lower.5ex\hbox{$#1$}}\over#2}}    % "
\def\partder#1#2{{\partial #1\over\partial #2}}   % partial derivative of
\def\parvar#1#2{{\d #1\over \d #2}}               % variation of
\def\on#1#2{\mathop{\null#2}\limits^{#1}}               % arbitrary accent
\def\bvec#1{\on\leftarrow{#1}}                  % backward vector accent
\newskip\humongous \humongous=0pt plus 1000pt minus 1000pt
\def\caja{\mathsurround=0pt}
\def\eqalign#1{\,\vcenter{\openup2\jot \caja
        \ialign{\strut \hfil$\displaystyle{##}$&$
        \displaystyle{{}##}$\hfil\crcr#1\crcr}}\,}
\newif\ifdtup

\begin{document}

\title{BASIC CANON IN D = 4, N = 1 SUPERFIELD THEORY: \\Five Primer 
Lectures}

\author{S. JAMES GATES, JR.}

\address{Department of Physics, University of 
Maryland,\\ College Park, MD 20742-4111, USA\\
E-mail: gatess@wam.umd.edu}


\maketitle\abstracts{
~~~~The topic of 4D, N = 1 supersymmetry is introduced 
for the reader with a prior background in relativistic 
quantum field theory. The presentation is designed to be a 
useful primer for those who plan to later engage in serious
investigation of the area or as an overview for the generally
interested.}

\newpage
\begin{center}
{\it {Prologue}}
\end{center}
$${~}$$
~~~~At MIT when I was a graduate student looking for a topic on
which I would write my doctoral thesis, one thing I did
was to make a survey of the research literature.  There were
many topics that were of current interest; ``the Bag Model,''
the dilute gas approximation of QCD, unified electroweak models,
instanton physics and numbers of other things.  However, being
a graduate student, I figured that the best thing on which to work
must be something completely new.  After all if it was as new as
possible, then I had a fair chance to compete with everyone else 
who also did not know anything about this new topic. I very rapidly
came to the conclusion that my choice ought to be the (then) newly 
discovered class of quantum field theories which possessed a 
property called ``supersymmetry.''  I could tell it was new
because in the first article I read about it, there appeared
the notion of a gradient operator which possessed a {\it {spinorial}}
index! I had never seen anything similar in any of my classes,
so this seemed a pretty good indication that this was really
{\it {new}}.  \newline ${}$

For a while it seemed as though my approach had worked too well. 
No one with whom I spoke in the department knew anything about 
this stuff that involved  ``superspace'' and ``superfields.''  Worse 
yet, it seemed as though it was slightly disreputable to work on 
this fanciful topic especially when there was `real physics' to be 
done. So there were no obvious faculty members to whom I could 
be an apprentice. I did, however, find a thesis adviser from whom 
I had learned a lot about electroweak models and he trusted me 
enough to let me follow my nose. But he made it clear that I would 
be totally responsible for teaching myself the topic, although he 
was available to discuss things whenever I got stuck. Thus, I was 
off on the adventure of mentally going to superspace and learning 
about the lay of the land. The fact that I have been invited to give 
these lectures is a hint that I succeeded in learning something about 
this place.  
\newline ${}$

In preparing these lectures, I have decided by and large to keep 
them as simple as possible, In fact, my idea for these discussions 
is an attempt to try to recall the things that I found confusing 
when I was first teaching myself. The best way to accomplish this 
seems for me to imagine that you had met me while I was working 
on my doctoral thesis and then to tell you about the kinds of 
things on which I worked. The one major exception to this is
the final section of the text. There I discuss a form of 
supergravity with some unique properties and which was first 
presented in 1989. I am particularly moved to make this
exception because this special form of supergravity, in my
opinion, will ultimately be found to be the 4D, N = 1 
supergeometrical limit of heterotic and superstring theories.
\newline ${}$

So these discussions are designed almost exclusively for the 
student who has completed courses in relativistic quantum field 
theory, group theory, linear algebra and a few other useful 
mathematical topics and who is embarking on the quest of learning 
about supersymmetry.  As Minkowski space is the natural setting 
for a discussion of relativistic theories, so is superspace the 
natural setting for a discussion of supersymmetry. Today this 
does not seem such a radical notion.  But when I was learning 
and creating some of this stuff, there were actually places 
where graduate students, even those working on supersymmetric 
topics, were extremely discouraged from learning how to use 
superspace and superfields.
\newline ${}$

These lecture are extremely limited in their scope and choice of 
topics. The student is encouraged to view these discussions solely 
as a primer. There are by now a plethora of books available on 
supersymmetry and supersymmetrical field theory. I regard most 
books on this topic as giving a ``beginning'' to ``intermediate'' 
level treatment. As such they are pretty interchangeable.  There 
are, to my knowledge, only two books that give an ``advanced'' 
treatment. I hope that those who read these lectures will easily be 
able to go on to any real textbook of their choice having benefitted 
from my 1997 TASI lectures.
\newline ${}$

I would be remiss if I did not thank the organizers, Jon Bagger 
and K.~T.~Mahanthappa, for their kind invitation to make these 
presentations. Additional thanks are extended to Ms. Betty Krusberg 
and Ms. Delores Kight for their assistance in the preparation of the 
manuscript. Of course, any errors are my own responsibility.

\newpage
\section{First Lecture: Tyro 4D, N = 1 Superfield Theory}
{\it {Introduction}}

~~~~In this first lecture, I will endeavor to get you, the
reader, to become comfortable with the notion of superspace.
The simplest and easiest way to do this is to recall an
experience which you most likely have already had if you
are reading this. Namely, you experienced a ``paradigm shift'' 
when you first encountered special relativity and the four 
vector formulation of theories consistent with its symmetries. 
This must have forced a radical change in your notions of 
space and time which became the Minkowski space-time 
manifold ${\cal M}_4$.  In a similar way, I wish for you to 
make another paradigm shift. The best way to gain insight 
into supersymmetrical theories is to jump right into superspace 
where the Minkowski space-time manifold is replaced by the 
Salam-Strathdee superspace manifold $s{\cal M}_{(4,N)}$.  
We won't actually learn very much superfield theory in this 
lecture. But we will learn a lot about the appropriate setting 
for these theories. 
$${~}$$

\noindent
{\it {Lecture}}
\subsection{Minkowski Space: The Boundary of Superspace}

~~~~Ordinary 4D relativistic field theory is defined in terms of fields 
over a four dimensional manifold ${\cal M}_4$. The objects 
that make up this manifold are points that may be labeled by four 
numbers such as $x^{\underline m} = (x^{\underline 0}, \, x^{
\underline 1}, \,  x^{\underline 2}, \, x^{\underline 3})$.  The 4D 
Minkowski manifold is actually an equivalence class.  We regard the 
manifold with coordinates given by $X^{\underline m}$ as an 
equivalent one if 
\be
X^{\underline m} ~=~ x^{\underline n} \, \Big( K_0 \Big){}_{\underline n}
{}^{\underline m} ~+~ ( k_0 )^{\underline m} ~~~,
\ee
where $\Big( K_0 \Big){}_{\underline n}{}^{\underline m}$ and $( k_0 
)^{\underline m}$ are sets of constants such that 
\be
\eqalign{
( d X^{\underline m}) \, \eta_{\underline m \, \underline n} \,
( d X^{\underline n}) ~=~ ( d x^{\underline m}) \, \eta_{\underline m 
\, \underline m} \, ( d x^{\underline n})  ~~~, \cr
\eta_{\underline m \, \underline n} ~=~ \left(\begin{array}{cccc}
~- 1 & ~~0 &  ~~0  & \,~~~0~~ \\
~0 & ~~1 &  ~~0  & ~~0 \\
~0 & ~~0 &  ~~1  & ~~0 \\
~0 & ~~0 &  ~~0  & ~~1 \\
\end{array}\right)    {~~~~~,~~~~~~~~~~}
}\ee
which implies the following restrictions on the matrix $(K_0)_{
\underline{n}}{}^{\underline{m}}$ 
\be 
\eta_{\underline m \, \underline n} ~=~ \Big( K_0 \Big){}_{\underline m}
{}^{\underline r }   \, \eta_{\underline r \, \underline s}  \, 
\Big( K_0 \Big){}_{\underline n}{}^{\underline s} ~~~.
\ee
We can also write 
\be
X^{\underline m} ~=~ exp[\, i \fracm 12 \Big( K_0 \Big){}_{\underline r}
{}^{\underline s} L_{\underline s}{}^{\underline r} \,] \, \,
exp[\, i \,( k_0 )^{\underline m} P_{\underline m} \,]
\, x^{\underline m}  ~~~,
\ee
where $P_{\underline m}$ and $L_{\underline s}{}^{\underline r}$ are 
differential operators 
\be
\eqalign{
P_{\underline m} &\equiv ~ {1 \over i}\, \partder{~~~~}{x^{\underline 
m}}~=~ - i \, \pa_{\underline m} ~~, ~~
L_{\underline s}{}^{\underline r}  ~\equiv ~ \eta^{\underline r \, 
\underline t}\,  L_{\underline s \, \underline t} ~=~ 
\eta^{\underline r \, \underline t} \, [ ~ - i \,
( ~ x_{\underline s} \, \pa_{\underline t} ~-~
x_{\underline t} \, \pa_{\underline s} ~) ~]  ~~~.
} \label{eq:fv}
\ee
It is easy to prove that
\be
\eqalign{
\Big[ \, P_{\underline r} ~,~ P_{\underline s} \, \Big] &=~ 0 ~~~, ~~~
\Big[ \,  L_{\underline r \, \underline s}~,~ P_{\underline t} \, \Big]
~=~ i \, \Big( ~ \eta_{\underline r \, \underline t}\,  P_{\underline 
s} ~-~ \eta_{\underline s \, \underline t}\,  P_{\underline 
t} ~ \Big) ~~~, ~~~ \cr
\Big[ \,  L_{\underline r \, \underline s}~,~ L_{\underline t \, 
\underline u} \, \Big] &=~ i \, \Big( ~ \eta_{\underline r \, \underline 
t}\,  L_{\underline s \, \underline u} ~-~ \eta_{\underline s \, 
\underline t}\,  L_{\underline r  \, \underline u} ~-
~ \eta_{\underline r \, \underline u}\,  L_{\underline s \, \underline 
t} ~+~ \eta_{\underline s \, \underline u}\,  L_{\underline r  \, 
\underline t} ~ \Big) ~~~. \cr
}  \label{eq:sx}
\ee
This is the Poincar\'{e} Lie algebra.  The Poincar\'{e} algebra is
exponentiated to form the Poincar\'{e} group. The operators $P_{\underline{m}}$
and $L_{\underline{r \, s}}$ respectively generate ``translations" and 
``Lorentz rotations."  When exponentiated each of these also separately 
form groups.

All 4D fields are\footnote{Following Wigner, we may say that the definition
of an elementary particle is that \newline ${~~~~}$ it is a representation
of Lie algebras.} representations of these groups. The simplest
such fields are
\be 
\begin{array}{lllll}
\phi(x) &,& ~\psi_{(\a)}(x) &,& ~A_{\underline a}(x), ... \\
{\rm {scalar}} &,& {~~}{\rm {spinor}} &,& {~~}{\rm {vector}}
\end{array}
\ee
and this list goes on forever where fields are defined with 
increasing numbers of four-component spinor $(\a)$ or 4-vector $\underline a$
indices.  These various possibilities arise because there exist constant 
numerical matrices (${\Tilde S}_{\underline a  \, \underline b}$) which satisfy the 
same commutation algebra as that of the $L$'s.
\be 
\begin{array}{lll}
{\rm {Ex. ~ 1}}: ~~~ {\Tilde S}_{\underline a  \, \underline b} &\equiv & 0 ~~~,\\
{\rm {Ex. ~ 2}}: ~~~ {\Tilde S}_{\underline a  \, \underline b} &\equiv & i 
\fracm 14 \, (\s_{\underline a  \, \underline b})_{(\a)} {}^{(\b)} ~~~, \\
{\rm {Ex. ~ 3}}: ~~~ {\Tilde S}_{\underline a  \, \underline b} &\equiv & i 
\, (~ \eta_{\underline a  \, \underline r} \d_{\underline b} {}^{\underline s}
~-~ \eta_{\underline b  \, \underline r} \d_{\underline a}
{}^{\underline s} ~ ) ~~~. 
\end{array}  
\label{eq:eig} \ee
The total angular momentum operator for any field is $J_{\underline a  
\, \underline b} = L_{\underline a  \, \underline b} \, + \, S_{\underline a  
\, \underline b}$ where the explicit matrix form for $S_{\underline{a 
\,b}}$ depends on the spin of the field upon which $S_{\underline{a 
\,b}}$ acts.  This allows us to define an abstract Lie algebra
operator ${\cal M}_{\underline a  \, \underline b} \, = \, -i 
S_{\underline a  \, \underline b}$ so that
\be 
\eqalign{
{\cal M}_{\underline a  \, \underline b} \, \phi(x) &=~ 0 ~~~, \cr
{\cal M}_{\underline a  \, \underline b} \, \psi_{(\a)}(x) &=~ 
\fracm 14 \, (\s_{\underline a  \, \underline b})_{(\a)} {}^{(\b)}
\, \psi_{(\b)}(x) ~~~, \cr
{\cal M}_{\underline a  \, \underline b} \, A_{\underline c}(x) 
&=~ (~ \eta_{\underline a  \, \underline c} \,\d_{\underline b} {}^{\underline 
d} ~-~ \eta_{\underline b  \, \underline c}\, \d_{\underline a}
{}^{\underline d} ~ ) \, A_{\underline d}(x) ~~~,\cr
& {~~}\Dot {~} \cr & {~~}\Dot {~} \cr & {~~}\Dot {~} \cr
} \label{eq:nin}
\ee

The basic ideas of Salam-Strathdee superspace \cite{1} is that the geometrical
notion of points should be replaced by ``superpoints" and the 4D 
Minkowskian spacetime manifold is replaced by ``superspace."
\be
x^{\underline m} ~~\to ~~ z^{\underline m} ~\equiv~ \Big( ~ \q^{(\m) 
\, i} \, , \, x^{\underline m} ~\Big) ~~~,~ i ~=~ 1, \dots , N ~~~.
\ee
The quantity $\theta^{(\m)i}$ corresponds to 4$N$ Grassmann numbers (or 
anticommuting classical numbers ACN's) where the ($\mu$) index indicates 
that these carry a spinorial representation of the Lorentz group.  If $i$ 
takes on a single value, we call this ``4D, $N=1$ superspace." We want to 
discuss field theories in 4D, $N=1$ superspace.  In a sense, Minkowski 
space lies on the boundary of superspace. Any physical consequences of 
superspace are only detectable for us on this boundary.

\subsection{Grassmann Numerology}

~~~~As observed above, the extra coordinates of superspace are not ordinary 
numbers.  This distinguishes superspace from similar constructions 
such as Kaluza-Klein models.  By definition a Grassmann number $\theta$ 
is a non-trivial solution to the equation 
\be
\q ^2 = 0 ~~~.
\ee
This should be thought of as an analogy to the fundamental definition of
complex numbers 
\be
x^2 ~=~ - \, 1 ~~~.
\ee
Since we possess four of these (i.e. $\q^{(\m)}, \, (\m) \,=\,(1), \, (2), \,
(3), \, (4)  $) we define also
\be
\q^{(\a)} \, \q^{(\b)} ~=~ - \, \q^{(\b)} \, \q^{(\a)}
~~~,
\ee
and furthermore may regard these as real quantities for the sake of
simplicity
\be
\q^{(\a)} ~=~ \Big[ \, \q^{(\a)} \, \Big]^* ~~~.
\ee
We will continue the discussion of complex conjugation in superspace
later. There is a certain subtlety which occurs and is the source
of numbers of (from my viewpoint) `strange' conventions in the
literature about superspace and superfields.

\subsection{A Notational Interlude}

~~~~In order to calculate most efficiently in 4D, $N=1$ superspace, it 
is convenient to introduce a notation that is especially adapted to this
purpose.  The two-component notation of the van der Waerden formalism 
is the best.  We begin by rewriting $x^{\underline {m}}$
as a 2 $\times$ 2 hermitian matrix
\be
x^{\underline m} ~=~ \left(\begin{array}{cc}
~x^0 \, +\,  x^3  & ~~ x^1 \, -\,  i x^2 \\
{}~&~\\
~ ~x^1 \, + \, i x^2 & ~~ x^0 \, - \, x^3 \\
\end{array}\right) ~\equiv ~ x^{\m \Dot \m} ~~~. 
\label{eq:ftn}
\ee
This permits us to regard the index ${\underline m}$ as either
the components of a 4-vector or as a pair of undotted and dotted
indices as in the equation above. We will liberally make use of
this convention in all subsequent discussions. So the interesting 
point is instead of thinking of $x^{\underline m}$ as
a 4-component vector, we may also regard it as a $2 \times 2$ hermitian
matrix. (Note $ det(x^{\mu \,{\Dot \mu}}) = -x^{\underline{m}}
\eta_{\underline{mn}} x^{\underline{n}}$).

Usually one first sees spinors expressed in terms of 4-component entities.
This is mostly due to our familiarity with the Dirac equation and
its use of four component entities to describe spinors. In all of
our discussion up to this point, in the back of our minds we had
equations like
\be
\psi_{(\a)}(x) ~=~ \left(\begin{array}{c}
 \psi_{(1)}(x) \\  \psi_{(2)}(x) \\  \psi_{(3)}(x) \\  \psi_{(4)}(x) 
\end{array}\right) ~~~, 
\label{eq:sxt}
\ee
which may be multiplied by 4 $\times$ 4 ``Dirac gamma" matrices 
$ (\g_{\underline a})$. These latter quantities
satisfy the usual algebraic equation
\be
 \Big(\g_{\underline a} \Big)_{(\a)}{}^{(\b)}  \Big(\g_{\underline 
b} \Big)_{(\b)}{}^{(\g)} ~+~  \Big(\g_{\underline b} \Big)_{(\a)}
{}^{(\b)}  \Big(\g_{\underline a} \Big)_{(\b)}{}^{(\g)}
~=~ 2 \, \eta_{\underline {a \, b}} \Big( {\rm I} \Big)_{(\a)}{}^{(\g)}
~~~,
\ee
where $({\rm I} )$ denotes the identity matrix.

The ``chiral projection matrices" $P_\pm$ may be defined by first noting
(suppressing $(\a)$-type indices)
\be
\g^5 ~\equiv~ i \, \fracm 14 \, \e^{\,{\underline a} \, {\underline b}
\, {\underline c} \, {\underline d}} \, \g_{\underline a} \g_{\underline 
b} \g_{\underline c} \g_{\underline d} ~~~,
\label{eq:egh}
\ee
which implies $(\g^5){}^2 = {\rm I}$, so that 
\be
P_\pm~ \equiv~ \fracm 12 \Big( ~ {\rm I} ~\pm~ \g^5 ~ \Big) ~~~,
\ee
satisfy the equations
\be
{\rm I} ~=~ P_+ ~+~ P_- ~~,~~ P_{\pm} P_{\mp} ~=~ 0 ~~,~~ 
(P_{\pm} ){}^2 ~=~ P_{\pm}  ~~~.
\ee
All of this implies that there exist a basis where
\be
P_+ ~=~ \left(\begin{array}{cccc}
~ 1 & ~~0 &  ~~0  & \,~~~0~~ \\
~0 & ~~1 &  ~~0  & ~~0 \\
~0 & ~~0 &  ~~0  & ~~0 \\
~0 & ~~0 &  ~~0  & ~~0 \\
\end{array}\right) ~~,~~ P_- ~=~  \left(\begin{array}{cccc}
~0 & ~~0 &  ~~0  & \,~~~0~~ \\
~0 & ~~0 &  ~~0  & ~~0 \\
~0 & ~~0 &  ~~1  & ~~0 \\
~0 & ~~0 &  ~~0  & ~~1 \\
\end{array}\right)  ~~.~~
\ee
When a four component spinor is multiplied by these we find
\be 
\Big(P_+ \Big)_{(\a)}{}^{(\b)} \psi_{(\b)}(x) ~=~ 
\left(\begin{array}{c}
 \psi_{(1)}(x) \\  \psi_{(2)}(x) \\  0 \\  0
\end{array}\right) ~\equiv ~ \psi_{\a}(x) ~~~,
\label{eq:twt}
\ee
\be 
\Big(P_- \Big)_{(\a)}{}^{(\b)} \psi_{(\b)}(x) ~=~ 
\left(\begin{array}{c}
 0 \\  0 \\ \psi_{(3)}(x) \\  \psi_{(4)}(x)
\end{array}\right) ~\equiv ~ \chi_{\Dot \a}(x) ~~~.
\label{eq:twh}
\ee

The spinors $\psi_{\a}(x)$ and ${\chi}_{\Dot \a}(x)$ are independent 
with each possessing only two components.  In our notation, such
spinor indices are described by Greek letters without the accompanying
parenthesis marks. Note we use ``dotted" and ``undotted" indices to
distinguish the chirality
\be 
\Big(\g^5 \, P_+ \Big)_{(\a)}{}^{(\b)} \psi_{(\b)}(x) ~=~ + \,
\psi_{(\a)}(x) ~~~,~~~ \Big(\g^5 \, P_- \Big)_{(\a)}{}^{(\b)} 
\psi_{(\b)}(x) ~=~ - \, \psi_{(\a)}(x) ~~~. 
\ee
So $\psi_{\a}(x)$ is the positive chiral (``right-handed'') part of 
the four component spinor $\psi_{(\a)}(x)$ and $\chi_{\Dot \a}(x)$ 
is the negative chiral (``left-handed'') part.

Finally we may use $P_+$ and $P_-$ in one more way. We define
2 $\times$ 2 Pauli matrices $\s_{\underline a}$ via 
\be
\Big(P_+ \g_{\underline a} P_-\Big)_{(\a)}{}^{(\b)} ~=~
\left(\begin{array}{cc}
~0 & ~~ (\s_{\underline a})_{\a} {}^{\Dot \b} \\
{}~&~\\
0 & ~~ 0\\
\end{array}\right) ~~,~~
\Big(P_- \g_{\underline a} P_+ \Big)_{(\a)}{}^{(\b)} ~=~
\left(\begin{array}{cc}
~0 & ~~0 \\
{}~&~\\
 (\s_{\underline a})^{\b} {}_{\Dot \a} & ~~ 0\\
\end{array}\right) ~~~,
\ee
and as well observe $ P_\pm \g_{\underline a} P_\pm = 0$.
The Pauli matrices permit us to write
\be
\eqalign{
 x^{\underline 0} (\s_{\underline 0})_{\a}
{}^{\Dot \a} ~+~  x^{\underline 1} (\s_{\underline 1})_{\a}{}^{\Dot 
\a} ~+~ x^{\underline 2} (\s_{\underline 2})_{\a}{}^{\Dot \a} ~+~  
x^{\underline 3} (\s_{\underline 3})_{\a}{}^{\Dot \a} &=~
x_{\a}{}^{\Dot \a} ~~~, \cr
(\s_{\underline a})_{\a} {}^{\Dot \b} (\s_{\underline b})^{\g}
{}_{\Dot \b} ~+~ (\s_{\underline b})_{\a} {}^{\Dot \b} (\s_{
\underline a})^{\g}{}_{\Dot \b} &=~ 2\, \eta_{\underline {a \, 
b}} \Big( {\rm I} \Big)_{\a}{}^{\g} ~~~, \cr
(\s_{\underline a})^{\b} {}_{\Dot \a} (\s_{\underline b})_{\b}
{}^{\Dot \g} ~+~ (\s_{\underline b})^{\b}{}_{\Dot \a} (\s_{
\underline a})_{\b}{}^{\Dot \g} &=~ 2\, \eta_{\underline {a \, 
b}} \Big( {\rm I} \Big)_{\Dot \a}{}^{\Dot \g} ~~~,
}\ee
but as we will see these last two need never be used by a clever choice 
of conventions.

In ordinary 4D spacetime, four vector indices are raised and lowered 
with $\eta_{\underline {a \, b}}$.  What can we do in Superspace?
It turns out that some unexpected results can be obtained via some
more definitions.
\be
\eta_{\underline {a \, b}} ~\equiv~ C_{\a \, \b} \,  C_{{\Dot \a} 
\, {\Dot \b}} ~~,~~ \eta^{\underline {a \, b}} ~\equiv~ C^{\a \, 
\b} \,  C^{{\Dot \a} \, {\Dot \b}}  ~~~,
\ee
\be
C_{\a \, \b} ~=~  i \left(\begin{array}{cc}
~0  & ~~1 \\
~-1 & ~~0\\
\end{array}\right) ~~~,~~~ C^{\a \, \b} ~=~  - i \left(\begin{array}{cc}
~0  & ~~1 \\
~-1 & ~~0\\
\end{array}\right) ~~~,
\ee
\be
C_{{\Dot \a} \, {\Dot \b}} ~=~  i \left(\begin{array}{cc}
~0  & ~~1 \\
~-1 & ~~0\\
\end{array}\right) ~~~,~~~ C^{{\Dot \a} \, {\Dot \b}} ~=~ - i \left(
\begin{array}{cc} ~0  & ~~1 \\
~-1 & ~~0\\
\end{array}\right)  ~~~.
\ee
All the usual properties of $\eta_{\underline {a \, b}}$ can be
seen to be true using this representation, (i.e. $\eta_{\underline {a \, b}}
\eta^{\underline {a \, b}} = 4 , \, \eta_{\underline {a \, b}} =
\eta_{\underline {b \, a}}$, etc.)
The $C$'s can also be used to raise and lower spinor indices
\be
\q_{\a} ~=~ \q^{\b} C_{\b \, \a} ~~,~~  \q^{\a} ~=~  C^{\a \, \b} 
\q_{\b} ~~,~~ {\bar \q}_{\Dot \a} ~=~ {\bar \q}^{\Dot \b} C_{\Dot \b \, 
\Dot \a} ~~,~~  {\bar \q}^{\Dot \a} ~=~  C^{\Dot \a \, \Dot \b} 
{\bar \q}_{\Dot \b} ~~.~~
\ee
Notice that since $C_{\a \, \b} = -C_{\b \, \a }$ we must exercise 
some care, i.e.
\be
\q^{\a} \, \q_{\a} ~\ne~ \q_{\a} \, \q^{\a} ~~~,~~~
{\bar \q}^{\Dot \a} \, {\bar \q}_{\Dot \a} ~\ne~ 
{\bar \q}_{\Dot \a} \, {\bar \q}^{\Dot \a}\, 
~~~.
\ee
As well these are useful to define the ``magnitude'' of the Grassmann numbers
\be
\q^2 ~\equiv~ - \fracm 12 \q^{\a}  \,
C_{\a \, \b}\,  \q^{\b} ~~,~~ {\bar \q}{}^2 ~\equiv~ - \fracm 12 
{\bar \q}{}^{\Dot \a} \, 
C_{{\Dot \a} \, {\Dot \b}} \,\, {\bar \q}{}^{\Dot \b} ~~~.
\ee

There is one additional invariant four tensor that is very useful
in ${\cal M}_4$, the Levi-Civita tensor. This too has a representation
in terms of the spinor $C$-metrics.
\be
\eqalign{
\e_{\,{\underline a} \, {\underline b} \, {\underline c} \, {\underline d}}
&\equiv~ i \, \fracm 12 \, [ ~ C_{\a \, \b} \, C_{\g \, \d} \, 
C_{{\Dot \a} \, ({\Dot \g}} \, C_{{\Dot \d}) \, {\Dot \b}}
~-~  C_{{\Dot \a} \, {\Dot \b}}\,  C_{{\Dot \g} \, {\Dot \d}}\, 
C_{\a \, (\g} \, C_{\d) \, \b} ~] ~~~, \cr
\e^{\,{\underline a} \, {\underline b} \, {\underline c} \, {\underline d}}
&\equiv~ i \, \fracm 12 \, [ ~ C^{\a \, \b} \, C^{\g \, \d} \, 
C^{{\Dot \a} \, ({\Dot \g}} \, C^{{\Dot \d}) \, {\Dot \b}}
~-~  C^{{\Dot \a} \, {\Dot \b}}\,  C^{{\Dot \g} \, {\Dot \d}}\, 
C^{\a \, (\g} \, C^{\d) \, \b} ~] ~~~.
}\ee
Once again it is easy to verify familiar properties (i.e. $\e_{\,{\underline 
b} \, {\underline a} \, {\underline c} \, {\underline d}}
= - \e_{\,{\underline a} \, {\underline b} \, {\underline c} \, 
{\underline d}}$, etc. $\eta^{\underline {a \, b}}
\e_{\,{\underline a} \, {\underline b} \, {\underline c} \, {\underline d}}
=0$).  Above the notation $( \, )$ acting on a pair of indices is given
by
\be  
\psi_{( \a} {\Bar \chi}_{\b ) }  ~\equiv~  \psi_{ \a} \chi_{\b  }
~+~ \psi_{\b} {\Bar \chi}_{\a  } ~~~.
\ee
We shall also use the notation  $[ \, ]$ acting on a pair of 
indices
\be  
\eqalign{
\psi_{[ \a} {\Bar \chi}_{\b ] }  &\equiv~  \psi_{ \a} {\Bar \chi}_{\b  }
~-~ \psi_{\b} {\Bar \chi}_{\a  } ~=~ -  \, C_{\a \, \b} \,
\psi^{\g} {\Bar \chi}_{\g } ~~~,\cr
{\Bar \psi}_{[ {\Dot \a}} {\chi}_{{\Dot \b} ] }  &\equiv~  {\Bar \psi}_{
{\Dot  \a}} \chi_{\Dot \b  }
~-~ {\Bar \psi}_{\Dot \b} {\chi}_{\Dot \a} ~=~ -  \, C_{\Dot \a \, 
\Dot \b} \, {\Bar \psi}^{\Dot \g} {\chi}_{\Dot \g } ~~~.
}\ee

We end this notational interlude by observing that as we regard the
2 $\times$ 2 matrix in (\ref{eq:ftn}) as the definition of the `bosonic' component
of a superpoint, derivatives of fields may be calculated with
respect to this 2 $\times$ 2 matrix. In particular, when we write
the symbol $P_{\underline m}$, this may be regarded as an
instruction to differentiate with respect to the variable in
(\ref{eq:ftn}). Adhering to this convention, no Pauli nor Dirac matrices
ever appear in our formulation of supersymmetrical theories. As
consequences of this we find
\be
\eqalign{ {~~~}
\pa_{\m \Dot \m} \, \pa^{\n \Dot \m} &\equiv~ \d_{\m}{}^{\n}\, {}_\bo 
~~\to ~~ \pa_{\m \Dot \m} \, \pa{}_{\n}{}^{ \Dot \m} ~=~ C_{\m \, \n}\, 
{}_\bo ~~,~~ {}_\bo ~=~ \fracm 12\,  \pa^{\underline m} \pa_{\underline m} 
~~~, \cr
\pa_{\m \Dot \m} \, \pa^{\m \Dot \n} &\equiv~ \d_{\Dot \m}{}^{\Dot \n}\, 
{}_\bo ~~\to ~~ \pa_{\m \Dot \m} \, \pa{}^{\m}{}_{ \Dot \n} ~=~ C_{\Dot 
\m \, \Dot \n}\, {}_\bo ~~~. 
}\ee
Since no Pauli or Dirac matrices ever appear, it is never necessary 
to calculate any traces thereof.

This interlude has been rather lengthy and the notation to be used
in the subsequent discussion has now been described. It may seem
as though this set of conventions is arbitrary and unwieldy. However,
I encourage the reader to actually use it. In fact for supersymmetrical
theories it is the most efficient notation I
have seen. 

\subsection{Super Translation Generators}

~~~~Just as Minkowski space is actually an equivalence class, so too
is superspace. In the following we are going to learn the form of
the generalization  of the Poincar\' e group.  We consider our 
superpoints of 4D, $N = 1$ superspace to be represented by 
\be
z^{\underline M} ~\equiv~ \Big( ~ \q^{\m } \, ,  {\bar \q}^{\Dot \m } \, ,
\, x^{\underline m} ~\Big) ~~~,
\ee
using 2-component spinor Grassmann coordinates.  It is simplest to first 
describe the super-translations as differential operators similar to those 
in (\ref{eq:fv}).
\be
\eqalign{
P_{\underline m} &= ~  - i \, \pa_{\underline m} ~=~ - i 
\, \partder{~~~~}{x^{\m \Dot \m}}  ~~~, \cr
Q_{\m}  &= ~   i \, \Big[ ~ \partder{~~~}{\q^{\m}} ~-~i \, \fracm 12 \,
{\bar \q}^{\Dot \m} \, \pa_{\underline m} ~ \Big] ~=~i \, \Big[ ~ \pa_{\m}
~-~i \, \fracm 12 \, {\bar \q}^{\Dot \m} \, \pa_{\underline m} ~ \Big] ~~~, \cr
{\Bar Q}{}_{\Dot \m}  &= ~  \Big[ ~ \partder{~~~}{{\bar \q}^{\Dot \m}} 
~-~i \,  \fracm 12 \, \q^{\m} \, \pa_{\underline m} ~ \Big] ~=~ i \, 
\Big[ ~ {\bar \pa}_{\Dot \m} ~-~i \, \fracm 12 \,\q^{\m} \, \pa_{\underline 
m} ~ \Big] ~~~.
} \label{eq:thg}
\ee
It is a simple matter to show that the follow algebra is satisfied,
\be
\eqalign{
\Big[ \, P_{\underline r} ~,~ Q_{\n} \, \Big] &=~ 0 ~~~, ~~~
\Big[ \, P_{\underline r} ~,~ {\Bar Q}_{\Dot \n} \, \Big] ~=~ 0 ~~~, ~~~
\Big[ \, P_{\underline r} ~,~ P_{\underline s} \, \Big] ~=~ 0 ~~~,\cr
\Big\{ \, Q_{\m} ~,~ Q_{\n} \, \Big\} &=~ 0 ~~~, ~~~
\Big\{ \, {\Bar Q}_{\Dot \m} ~,~ {\Bar Q}_{\Dot \n} \, \Big\} ~=~ 0 ~~~,~~~ 
\Big\{ \, Q_{\m} ~,~ {\Bar Q}{}_{\Dot \m} \, \Big\} ~=~ i\, P_{\underline m}
~~~.}\ee 
Note the anti-commutators, this is a ``super" Lie algebra.  (Also note absence
of gamma matrices, i.e. simplicity.)

The ordinary translation generators ``create'' motions
\be
i ( k_0 )^{\underline m} P_{\underline m} \, x^{\underline n} 
~~\to~~ exp[\, i ( k_0 )^{\underline m} P_{\underline m} \, ] 
\, x^{\underline n} ~=~ \, x^{\underline n} ~+~  ( k_0 )^{\underline n}
~~~. \ee 
So it is revealing to study the motions of superpoints generated by 
the operators in (\ref{eq:thg}).
\be
\eqalign{
i\, ( k_0 )^{\underline m} \, P_{\underline m} z^{\underline N} &=~
\Big( ~ 0, \,\, 0 , \,\, ( k_0 )^{\underline m} ~\Big) ~~~, \cr
i\, \e_0^{\m}  \,Q_{\m}  z^{\underline N} &=~ 
\Big( ~ - \e_0^{\m}, \,\, 0 , \,\, i \, \fracm 12 \e_0^{\m} {\bar 
\q}^{\Dot \m} ~\Big) ~~~, \cr
i\,  \bar\e_0^{\Dot \m} \, {\Bar Q}_{\Dot \m}  z^{\underline N} 
&=~ \Big( ~ 0, \,\, - \bar\e_0^{\Dot \m} , \,\, i \, \fracm 12 
\bar\e_0^{\Dot \m} {\q}^{\m} ~\Big) ~~~,
}\ee
where we have introduced Grassmann parameters $ \e_0^{\m}$ and $
\bar\e_0^{\Dot \m}$ so that we can form a `super-translation''
parameter $( k_0 )^{\underline M} = (\, \e_0^{\m}, \, \bar\e_0^{\Dot 
\m} , \, ( k_0 )^{\underline m}\, )$. Also rather obviously it
is possible to combine the translations in a similar manner,
$P_{\underline M} = (\, Q_{\m}, \, {\Bar Q}_{\Dot \m} , \, 
P_{\underline m} \, )$. Using a more concise notation, we have
$\d  z^{\underline N} = i( k_0 )^{\underline M} P_{\underline M}
z^{\underline N} $, so that
\be
\d \q^{\m} ~=~ -  \e_0^{\m} ~~,~~ \d {\bar \q}^{\Dot \m} ~=~ - 
\bar\e_0^{\Dot \m} ~~,~~ \d x^{\underline m} ~=~ i \, \fracm 12 ~( ~ 
\e_0^{\m} \,{\bar \q}^{\Dot \m}  ~+~
\bar\e_0^{\Dot \m} \, {\q}^{\m} ~) ~~~. \label{eq:frt}
\ee

\subsection{Super Lorentz Generators}

~~~~It is also possible to generalize the orbital angular momentum
generator to a superspace orbital angular momentum generator
$L_{\underline{M \, N}}$. The only non-vanishing components of
this tensor are $L_{(\m \, \n)}$, $L_{( {\Dot \m} \, \Dot \n )}$
and $L_{\underline {m \, n}}$.  Since the Grassmann coordinates 
are relativistic spinors, it is natural that they should appear 
in the appropriate generalization,
\be
\eqalign{
L_{\underline {m \, n}} &\equiv~ C_{\m \, \n} \, L_{( \Dot \m \, 
\Dot \n )} ~+~  C_{ \Dot \m \, \Dot \n }\, L_{(\m \, \n)} ~~~, \cr
L_{(\m \, \n)}  &\equiv~ - i \, \fracm 12 \, [~ x_{(\m}{}^{\Dot \n}
\pa_{ \n ) \Dot \n} ~+~ \q_{(\m} \, \pa_{\n )}  ~]  ~~~, \cr
L_{( {\Dot \m} \, \Dot \n )} &\equiv~  - i \, \fracm 12 \, 
[~ x^{\n}{}_{( \Dot \m} \pa_{\n  \Dot \n ) } ~+~ {\bar \q}_{(\Dot \m} 
\,{\bar \pa}_{\Dot \n )}  ~]  ~~~. 
}  \label{eq:fth}
\ee
This first equation is consistent since $\underline m = \m \Dot \m$ 
and $\underline n = \n \Dot \n$. The generator $L_{\underline{m \,n}}$ 
does exactly what we expect.  Namely it implies that $x^{\underline{m}}$ 
transforms as a Lorentz 4-vector and $\theta ^{\m}$, ${\bar \theta}^{\Dot 
\m}$ transform as Lorentz spinors.

The set of operators $\{ {\cal A} \,\} = \{ \, P_{\underline M}, \, 
L_{\underline m \, \underline n} \, \}$ forms a super-Lie algebra. The basic 
point is that we may define 
\be (-1)^P ~\equiv~ 1 ~~,~~  (-1)^L ~\equiv~ 1 ~~,~~ (-1)^Q ~\equiv~
 - 1 ~~,~~ (-1)^{\Bar Q} ~\equiv~ - 1 ~~~,
\ee
and a generalized or graded commutator through
\be
[~ {\cal A}_1 \,, \, {\cal A}_2 ~ \} ~\equiv ~
 {\cal A}_1 \,{\cal A}_2 ~-~ (-1)^{ {\cal A}_1 {\cal A}_2}
\, {\cal A}_2 \, {\cal A}_1 ~~~.
\ee
Under the operation of this generalized commutator $\{ \cal A \}$
is a closed set. In addition, there is a generalized Jacobi identity
defined by
\be
\eqalign{
&(-1)^{ {\cal A}_1 {\cal A}_3}\,[~ [~ {\cal A}_1 \,, \, 
{\cal A}_2 ~ \} , \, {\cal A}_3 ~ \} ~+~
(-1)^{ {\cal A}_2 {\cal A}_1}\,[~ [~ {\cal A}_2 \,, \, 
{\cal A}_3 ~ \} , \, {\cal A}_1 ~ \} ~+~ \cr
&(-1)^{ {\cal A}_3 {\cal A}_2}\,[~ [~ {\cal A}_3 \,, \, 
{\cal A}_1 ~ \} , \, {\cal A}_2 ~ \} 
~=~ 0 ~~~,
}\ee
which the generators satisfy. 

\subsection{Superspace Conjugation}

~~~~Note that if we consider (\ref{eq:frt}) solely for the temporal 
component of $x^{\underline m}$ (thus breaking Lorentz covariance) 
we have $\d x^{\underline 0} \propto i\,  (\s^{\underline 0})_{\m \Dot 
\m} ( \e_0^{\m} \,{\bar \q}^{\Dot \m}  + \bar\e_0^{\Dot \m} \, {\q}^{\m} )$.  
At first this seems extremely problematical since $x^{\underline 0}$ 
is a real quantity,  the factor in the parenthesis is real using standard 
definitions and $(\s^{\underline 0})_{\m \Dot \m}$ is real.  Yet this 
equation suggests that the variation of  $x^{\underline 0}$ is {\it {
purely}} imaginary (and nilpotent) even if we use a real spinor basis. 

But the variation of $(ct)$ should be real! We can make this so if the 
meaning of ${}^*$ is modified. One way to do this in superspace is
to assign the following definition of how complex conjugation 
is applied
\be
\eqalign{
{}^* &\equiv~ {\cal C} \otimes {\cal O}_{\rm I}
~~~, \cr
{\cal C} &\equiv~ {\rm {replace}}~ i~ {\rm {by}}~ - i ~{\rm {and~
vice-versa}}~~~, \cr
 {\cal O}_{\rm I}  &\equiv~ {\rm {invert~the~order ~of ~all ~monomials
~in ~all ~factors}} ~~~.
}\ee
To see how this definition of the ${}^*$ operation solves the
problem, consider the brief calculation below. For the purposes
of this illustration, we can consider a truly one dimensional
world in which spinors are real one-component entities. 
\be
\eqalign{
{\cal C} \otimes {\cal O}_{\rm I}: \, [ \, i \fracm 12 \e \,  \q \,] 
&=~ \q^* \, \e^* \,  \fracm 12 \, (-i)   \cr
&=~ - i \fracm 12 \q \e \cr
&=~  i \fracm 12 \e  \q \cr 
 ~~ [ \, i \fracm 12 \e \,   \q \,]^* &=~ [ \, i \fracm 12 
\e \, \q \,]  ~~~.
}\ee
So under superspace conjugation the variation of the temporal
component (and all others) of $x^{\underline m}$ is real.

The definition of the $*$ operator (which I first introduced in my 
Ph.D. thesis \cite{2}) above may seem very capricious. It 
is not. In large numbers of places in the literature on superspace, 
this is the cause of the appearance of seemingly `inexplicable'
signs. Let us continue to study more implications of this rule.
Although we had not mentioned it before, the spinor in (\ref{eq:sxt})
may be subjected to a reality condition, $\psi_{(\a)} =
[\psi_{(\a)}]^*$. However this condition is inconsistent with
the basis in which we have given the $\g^5$-matrix in 
(\ref{eq:egh}-\ref{eq:twh}).  Thus if we wish to use this basis, 
the spinors in (\ref{eq:twt},\ref{eq:twh}) must
be complex.  The same reasoning applies to the spinorial
components of superspace. If we use 2-component spinor
coordinates for superspace, they must be complex coordinates.

The fact that we use complex Grassmann coordinates is
consistent with the definitions that we have introduced
for raising and lowering their indices with the $C$-tensors
since the latter include factors of $i$ in their definitions.
Thus we have (also recall we only use superspace conjugation)
\be
\eqalign{
(\q^{\a})^* ~=~ \bar\q^{\Dot \a} ~~{\rm {but}}~~ (\q_{\a})^* ~=~
-\, \bar\q_{\Dot \a} ~~~, \cr
(\bar\q^{\Dot \a})^* ~=~ \q^{\a} ~~{\rm {but}}~~ (\bar\q_{\Dot 
\a})^* ~=~ -\, \q_{\a} ~~~.
}\ee
Another consequence follows from
\be
\eqalign{
( \, \q^{\a} \q_{\a} \,)^* &=~ (\q_{\a} )^* (\q^{\a})^*  
~~~{\rm {s-conjugation}} \cr
&=~ -\, \bar\q_{\Dot \a} \bar\q^{\Dot \a}  \cr
&=~  \bar\q^{\Dot \a} \bar\q_{\Dot \a} ~~~~~~~~~~{\rm {Grassmann~property}}
~~~. }\ee
So that the quantity $\q^{\a} \q_{\a} ~+~ \bar\q^{\Dot \a} 
\bar\q_{\Dot \a}$ is real for us. 

Also since $\pa_{\a} \q^{\b} = \d_{\a}{}^{\b}$, it follows that
$(\pa_{\a})^* = - \bar\pa_{\Dot \a}$ (and of course $\bar\pa_{\Dot 
\a} \bar\q^{\Dot \b} = \d_{\Dot \a}{}^{\Dot\b}$, implies
$(\bar\pa_{\Dot \a})^* = - \pa_{\a}$). This is seen through
the calculation below.
\be
\eqalign{
( \, \pa_{\a} \q^{\b} \,)^* &=~ (\q^{\b} )^* \, (\bvec\pa_{\a})^*  
~~~{\rm {s-conjugation}} \cr
&=~ \bar\q^{\Dot \b} \, (\, - \, {\bvec{\bar\pa}}_{\Dot \a} \, ) \cr
&=~  \bar\pa_{\Dot \a} \, \bar\q^{\Dot \b} ~~~~~~~~~~{\rm {Grassmann~property}}
~~~. }\ee
The general rule is if a superspace tensor is ``real'' it transforms like
a product of superspace monomial coordinates under the action of
superspace conjugation. We apply the superspace conjugation rule
followed by restoring the order of the indices as they appear on
the original tensor. This is simplest demonstrated by some examples.
\be
\eqalign{
x^{\underline m} ~\sim~  \q^{\m} \, \bar\q^{\Dot \m} ~~; ~~~~
 (  \q^{\m} \, \bar\q^{\Dot \m} )^* &=~ ( \bar\q^{\Dot \m})^*
( \q^{\m})^* \cr
&=~ \q^{\m}\,  \bar\q^{\Dot \m} ~~
\to~~ ( x^{\underline m})^* ~=~ x^{\underline m} ~~~,
}\ee
\be
\eqalign{
x_{\underline m} ~\sim~  \q_{\m} \, \bar\q_{\Dot \m} ~~;~~~
(  \q_{\m} \, \bar\q_{\Dot \m} )^* &=~ ( \bar\q_{\Dot \m})^*
( \q_{\m})^* \cr
&=~ (- \q_{\m} )\, (- \bar\q_{\Dot \m} )~~
\to~~ ( x_{\underline m})^* ~=~ x_{\underline m} ~~~.
}\ee
These examples may be considered too simple, so we finish with
a multi-indexed more complicated example,
\be
\eqalign{
 T_{\a \, \b \, \Dot \g \, \underline d \,}{}^{\e}
&\sim~ \q_{\a} \, \q_{\b } \, \bar\q_{\Dot \g} \,
x_{\underline d} \,  \q^{\e} \cr
(\,  \q_{\a} \, \q_{\b } \, \bar\q_{\Dot \g} \,
x_{\underline d} \,  \q^{\e}  \, )^* &=~ (  \q^{\e} )^* \, (
x_{\underline d} )^* \, (  \bar\q_{\Dot \g} )^* \,
( \q_{\b} )^* \, ( \q_{\a })^* \cr
&=~ \bar\q^{\Dot \e}  \, x_{\underline d} \, (- \q_{\g}) 
\, (- \bar\q_{\Dot \b} )\, \, (- \bar\q_{\Dot \a} ) \cr
&=~-\,  \bar\q_{\Dot \a} \, \bar\q_{\Dot \b}  \,  \q_{\g} \,
x_{\underline d} \, \bar\q^{\Dot \e} ~~~.
}\ee
Thus if $T$ is a ``real'' superspace tensor it possesses the property
\be
(\, T_{\a \, \b \, \Dot \g \, \underline d \,}{}^{\e} \, )^* ~=~
-  \, T_{\Dot \a \, \Dot \b \, \g \, \underline d \,}{}^{\Dot \e} ~~~.
\ee

\subsection{Superfields}\label{subsec:wpp}

~~~~In 4D spacetime, fields may be defined as the fundamental dynamical
entities with the scalar field $\phi(x)$ as the simplest example.  How
do such entities appear in superspace? ${}$ \newline  ${}$ \newline \indent
Definition: A superfield is a function of $z^{\underline{m}}$ that is
analytic in $\theta^{\mu}$ and $\bar{\theta}^{\dot \mu}$.
${}$ \newline ${}$ \newline \noindent
A simple example of a superfield is provided by a real scalar superfield
$F(z)$ which may be expanded as
\be
\eqalign{
F(z) &=~ C(x) ~+~ \q^{\a} \chi_{\a}(x) ~+~  \bar\q^{\Dot \a} {\Bar 
\chi}_{\Dot \a}(x) ~-~ \q^2  M(x)  ~-~ {\bar \q}{}^2 {\Bar M}(x) \cr
&{~~~~}+~  \q^{\a} \bar\q^{\Dot \a} A_{\underline a}(x)  ~-~
{\bar \q}{}^2  \q^{\a} \l_{\a}(x) ~-~ \q^2 \bar\q^{\Dot \a} {\Bar \l}
{}_{\Dot \a}(x) ~+~ \q^2  {\bar \q}{}^2{\rm d}(x) ~~~.
} \label{eq:fyx}
\ee
The expansion terminates due to the Grassmann nature of the $\theta$'s.
The ``component fields'' $C(x)$, $\chi_{\a}(x)$, $M(x)$, $A_{\underline 
a}(x)$, $\l_{\a}(x)$ and ${\rm d}(x)$ are ordinary fields.  All the bosonic
fields are real ones (see later discussion). How do the motions which 
generate an equivalence class of superspace affect these ordinary fields?  
The answer to this is found by applying the differential operators which 
represent the generators to the expansion of the superfield. Thus in terms 
of the superfield we write,
\be
\d_Q \, F(z) ~=~ i \Big( \, \e^{\m} \,{Q}_{\m} ~+~
\bar\e{}^{\Dot \m} \, {\Bar Q}{}_{\Dot \m} \, \Big) \,  F(z)  ~~~.
\ee
where we have ``dropped'' the zero subscript on the parameters.
In terms of the fields in the expansion, this leads to
\be
\eqalign{
\d_Q \, C &=~  - \, ( ~ \e^{\a} \, \chi_{\a}~+~ \bar\e{}^{\Dot \a} \, 
{\Bar \chi}{}_{\Dot \a}~) ~~~,  \cr
\d_Q \, \chi_{\a} &=~ \e_{\a} \, M ~-~ \bar\e{}^{\Dot \a} \, 
( ~i \, \fracm 12 \pa_{\underline a}  C ~+~  A_{\underline a} ~) ~~~,  \cr
\d_Q \, M &=~ \bar\e{}^{\Dot \a} \, (~ {\Bar \l}_{\Dot \a} ~- ~ i \, \fracm 
12 \pa_{\underline a} {\Bar \chi}{}^{\Dot \a} ~) ~~~, \cr
\d_Q \,  A_{\underline a} &=~  \e^{\b} \, (~ C_{\b \, \a} {\Bar \l}_{\Dot \a}~
+ ~ i \, \fracm 12 \pa_{\b  \Dot \a} \,{\chi}{}_{\a} ~) ~+~ 
\bar\e{}^{\Dot \b} \, (~ C_{\Dot \b \, \Dot \a} {\l}_{\a} ~+~ i \, 
\fracm 12 \pa_{\a  \Dot \b} \,{\Bar \chi}{}_{\Dot \a} ~) ~~~, \cr
\d_Q \, \l_{\a} &=~  \e^{\b} \, (~ C_{\b \, \a} {\rm d} ~
+ ~ i \, \fracm 12 \pa_{\b  \Dot \a} \, A_{\a}{}^{\Dot \b} ~) ~-~ 
\bar\e{}^{\Dot \a} \, \pa_{\underline a} {\Bar M} ~~~,  \cr
\d_Q \, {\rm d} &=~   i \, \fracm 12 \, \Big[ ~\pa_{\underline a} 
 ( ~ \e^{\a} \, \l^{\a}~+~  \bar\e{}^{\Dot \a} \, 
{\Bar \l}{}^{\Dot \a}~) ~\Big] ~~~.
}\ee

This completes the discussion for the translational part of the super 
Poincar\'e group, but there are still to be discussed the effects of the 
pure Lorentz generators (\ref{eq:fth}). To do this consistently, we return to our 
discussion (\ref{eq:sx}) of the Poincar\' e group.  There we saw the parameter of 
the pure Lorentz rotations were given by $\Big( K_0 \Big){}_{\underline a \, 
\underline b}$. Similar to the result of (\ref{eq:fth}) we may introduce
new parameters $K_{\a  \, \b}$ and $K_{\Dot \a \, \Dot \b}$ via
\be
\Big( K_0 \Big){}_{\underline a \, \underline b} ~=~ \fracm 12 \, [~
C_{\Dot \a \, \Dot \b} \, K_{\a  \, \b} ~+~ 
C_{\a  \, \b} \, K_{\Dot \a \, \Dot \b} ~] ~~~.
\label{eq:fyn}
\ee
This corresponds to a decomposition of the Lorentz parameters into 
``right-handed'' (or ``self-dual'') and ``left-handed'' (or ``anti-self-dual'') 
parts. In the case of the translations, we added new Grassmann 
parameters ($\e$ and $\bar\e$) so that the fermionic translations 
could be implemented.  We do something very different for the Lorentz
rotations. Even though there exist the supertensor $L_{\underline M \, 
\underline N}$, according to (\ref{eq:fyn}) not all of its components are
independent. Thus, it is unnecessary and redundant to attempt
to introduce new parameters for its fermionic components. Instead
the motions of the superspace orbital angular momentum generator
are derived from 
\be
\d_L \, F(z) ~=~ i \, \fracm 12 \,\Big( K_0 \Big){}_{\underline a} {}^{ 
\underline b} \, L_{\underline b}{}^{ \underline a} F(z) 
~=~  i \, \fracm 12 \, [~ K_{\a}{}^{\b} L_{\b}{}^{\a}
~+~  K_{\Dot \a}{}^{\Dot \b} L_{\Dot \b}{}^{\Dot \a} ~] F(z) ~~~.
\ee
It is an easy exercise to show that this induces at the component
level the interesting result that for the components
$C$, $M$ and ${\rm d}$ transform as scalars, $\l_{\a}$
and $\chi_{\a}$ transform as spinors and $A_{\underline a}$ transforms
as a vector. 

So the superfield $F(z)$ contains component fields of different 
spins even though the superfield itself carries no `external' 
spinor or vector indices.  However, this does {\it {not}} preclude
us from considering superfields that do. In fact, such superfields
do exist and are important in numbers of applications. This
leads to the notion that the abstract spin angular momentum
operator $S_{\underline a \, \underline b}$ continues to
exist in superspace (which it does). It is usually referred to
as the ``superspin operator'' in this context.

The real SF (superfield) is not the simplest, most fundamental 4D
supersymmetry representation.  This distinction belongs to the chiral SF
$\Phi$(z).  The chiral superfield is foremost a complex superfield
whose expansion takes the form
\be
\eqalign{
\Phi(z) &=~ A(x) ~+~ \q^{\a} \psi_{\a}(x)  ~-~ \q^2  F(x)  ~
+~ i \fracm 12  \q^{\a} \bar\q^{\Dot \a} \pa_{\underline a} A(x) \cr
&{~~~~}~+~ i\, \fracm 12 \q^2 \bar\q^{\Dot \a}  \pa_{\underline a}
 \psi^{\a}(x) ~+~ \fracm 14 \q^2  {\bar \q}{}^2 {}_{\bo} A(x) ~~~.
} \label{eq:sxo}
\ee
Here the bosonic fields $A(x)$ and $F(x)$ are complex. It should be
clear that $\Phi$ is a type of `restricted' superfield since some
of the $\q$-expansion terms which appear in (\ref{eq:fyx}) are missing from
the expansion in (\ref{eq:sxo}). The changes induced among the fields
by the fermionic translation are found from
\be
\eqalign{
\d_Q \, \Phi(z) &=~ i \Big( \, \e^{\m} \,{Q}_{\m} ~+~
\bar\e{}^{\Dot \m} \, {\Bar Q}{}_{\Dot \m} \, \Big) \,  \Phi(z)  ~~~, \cr
&\to ~~ \d_Q \, A ~=~  - \,  \e^{\a} \, \psi_{\a}
~~,~~ \d_Q \, \psi_{\a} ~=~ - i \, \bar\e{}^{\Dot \a} \,
\pa_{\underline a} A ~+~
\e_{\a} \, F ~~~, \cr
&{~~~~~~~}\d_Q \, F ~=~ i \, \bar\e{}^{\Dot \a} \,~\pa_{\underline a} \, 
\psi^{\a} ~~~.
}\ee

In order to relate the chiral superfield to the general complex
superfield, it is convenient to define some new spinorial 
differential operators,
\be
D_{\m} ~\equiv~ 
\Big[ ~ \pa_{\m} ~+~i \, \fracm 12 \, {\bar \q}^{\Dot \m} 
\, \pa_{\underline m} ~ \Big] 
~~~,~~~  {\Bar D}{}_{\Dot \m} ~=~ \Big[ ~ {\bar \pa}_{\Dot 
\m} ~+~i \, \fracm 12 \,\q^{\m} \, \pa_{\underline 
m} ~ \Big] ~~~,
\ee
that are called the ``supersymmetry covariant derivatives.''
It can be seen that these are {\it {not}} linearly related to
the super translation generators $ {Q}_{\m} $ and
$ {\Bar Q}{}_{\Dot \m} $. In fact, these possess a very
different graded commutator algebra,
\be
\eqalign{
\Big[ \, P_{\underline r} ~,~ D_{\n} \, \Big\} &=~ 0 ~~~, ~~~
\Big[ \, P_{\underline r} ~,~ {\Bar D}{}_{\Dot \n} \, \Big\} ~=~ 0 
~~~ \Big[ \, D_{\m} ~,~ D_{\Dot \n} \, \Big\} ~=~ 0 ~~~, ~~~\cr
\Big[ \, Q_{\m} ~,~ D_{\n} \, \Big\} &=~ 0 ~~~, ~~~
\Big[ \, Q_{\m} ~,~ {\Bar D}{}_{\Dot \m} \, \Big\} ~=~ 0 ~~~, ~~
\Big[ \, D_{\m} ~,~ {\Bar D}{}_{\Dot \m} \, \Big\} ~=~ i \, \pa_{\underline m}
~~~.}\ee 
We define a ``super vector" gradient by
\be
D_{\underline A} ~=~ \Big( \,  D_{\a}, \, {\Bar D}_{\Dot \a} , \,
\pa_{\underline a} \, \Big) ~~~,
\ee
and this is the basic tool for constructing kinetic energy operators
for superfields. The chiral superfield satisfies the differential 
equation ${\Bar D}_{\Dot \a} \Phi = 0$. It can also be noted
that the ${}^*$ operation has the following realization on a
complex scalar superfield $H(z)$
\be
\eqalign{ D_{\a} H(z) &=~ \Big[ ~  { \pa}{}_{\a} \, + \, i \fracm 12 \, 
\bar\q^{\Dot \a} {\pa}{}_{\underline a}~ \Big] \,  H ~~~\to \cr
\Big[ D_{\a} H(z) \Big]^* ~=~ H^* \, \Big[ {\bvec D}{}_{\a} \Big]^* 
&=~  H^* \, \Big[ ~ [{\bvec \pa}{}_{\a}]^* \, + \, [{\bvec \pa}{}_{\underline 
a}]^* (\bar\q^{\Dot \a})^* \fracm 12 ( - i) ~ \Big] 
~~~ \cr
&=~  H^* \, \Big[ ~ - \,{\bvec {\bar \pa}}{}_{\Dot \a} \, - \,  i \fracm 12 \,
{\bvec \pa}{}_{\underline a} (\q^{\a})~ \Big] \cr
&=~ - \, \Big[ ~  {\bar \pa}{}_{\Dot \a} \, + \, i \fracm 12 \,
 \q^{\a} {\pa}{}_{\underline a}~ \Big] \,  H^* ~~~.
} \label{eq:sxs}
\ee
Since $ D_{\a}$ has a lower spinor index, we expect $[{\bvec D}{}_{\a}]^*
= - {\bvec {\Bar D}}{}_{\Dot \a}$ so we obtain,
\be
\eqalign{
 H^* \,\Big[ - {\bvec {\Bar D}}{}_{\Dot \a}\Big] &=~ - \, \Big[ ~ {\bar \pa}
{}_{\Dot \a} \, + \, i \fracm 12 \, \q^{\a} {\pa}{}_{\underline a}~ \Big] \,  
H^* ~~~, ~~~{\rm {or}} \cr
{ {\Bar D}}{}_{\Dot \a} H^* &=~  \Big[ ~  {\bar \pa}{}_{\Dot 
\a} \, + \, i \fracm 12 \, \q^{\a} {\pa}{}_{\underline a}~ \Big] \,  H^*
~~~.
}  \label{eq:sxv}
\ee
One can verify that all of these operatorial steps make sense by
explicitly calculating the rhs of the first line in (\ref{eq:sxs}), complex 
conjugate that resultant and then comparing to the explicit component results
obtainable from the rhs of the last line in (\ref{eq:sxv}). It is often the 
case that in other conventions in order to get agreement among these 
two result, one must insert minus signs which apparently have no 
explanation. 

\subsection{Superfields; Spin-Statistics \& Bose-Fermi Equality}

~~~~In Minkowski space, bosons (B) are fields with even numbers of 
spinorial indices
\be
B(x) \sim B^{\a_1 \cdots \a_p \, {\Dot \b}_1 \cdots {\Dot \b}_q}
~~~,~~~ p ~+~ q ~=~ 2 n ~~~,
\ee
and fermions are fields with an odd number of these indices
\be
F(x) \sim F^{\a_1 \cdots \a_p \, {\Dot \b}_1 \cdots {\Dot \b}_q}
~~~,~~~ p ~+~ q ~=~ 2 n ~+~ 1 ~~~.
\ee
These two classes of fields obey the statistics rules (classically
at least)
\be
B_1 \, B_2 ~=~ B_2 \, B_1 ~~~,~~~ F_1 \, F_2 ~=~ - F_2 \, F_1
~~~,~~~ B \, F ~=~ F \, B ~~~.
\ee
If we assume that these also hold for superfields $B(x) \to B(z)$ 
and $F(x) \to F(z)$, then we can derive the fact that a superfield 
necessarily includes fields of both statistics. Applying this to our 
prototype superfield $F$ (\ref{eq:fyx}) we find the result
\be
\eqalign{
&\{ ~ C(x) , ~ M(x), ~ A_{\underline a}(x) , ~ {\rm d}(x) ~\} ~~\in ~~ 
\{B\} ~~~, \cr
&\{ ~ \l_{\a}(x) , ~ \chi_{\a}(x) ~\} ~~\in ~~ \{F\} ~~~.}
\ee
If we replace $F \to F^{\a}$ and demand that $F_1{}^{\a}$ and
$F_2{}^{\a}$ obey fermi statistics then (with an obvious change
in notation)
\be
\eqalign{
&\{ ~ C{}^{\b}(x) , ~ M{}^{\b}(x), ~ A_{\underline a}{}^{\b}(x) , 
~ {\rm d}{}^{\b}(x) ~\} ~~\in ~~ 
\{F\} ~~~, \cr
&\{ ~ \l_{\a}{}^{\b}(x) , ~ \chi_{\a}{}^{\b}(x) ~\} ~~\in ~~ \{B\} 
~~~.}
\ee
In particular, we see that that Duffin-Kummer-Petiau fields
$\l_{\a}{}^{\b}$ and $\chi_{\a}{}^{\b}$ fields appear. A similar
result occurs if we replace $F$ according to $F \to F^{\Dot \a}$.

The other obvious feature of 4D superfields is that they always
introduce multiplets of particles of different spins but where
the total number of bosonic degrees of freedom is equal to
the total number of fermionic degrees of freedom. The source of
this can be made obvious by re-writing our prototype in a 
slightly different way,
\be
\eqalign{
F(z) &=~ C(x) ~-~ \q^2  M(x)  ~-~ {\bar \q}{}^2 {\Bar M}(x)
~+~  \q^{\a} \bar\q^{\Dot \a} A_{\underline a}(x) 
~+~ \q^2  {\bar \q}{}^2{\rm d}(x) \cr
&{~~~~}+~ \q^{\a} \chi_{\a}(x) ~+~  \bar\q^{\Dot \a} {\Bar 
\chi}_{\Dot \a}(x)  ~-~ {\bar \q}{}^2  \q^{\a} \l_{\a}(x) ~-~ 
\q^2 \bar\q^{\Dot \a} {\Bar \l} {}_{\Dot \a}(x)  ~~~,
}\ee
and it is seen that the fields are in one-to-one correspondence
with the monomials that are constructed from $\q$ and $\bar\q$.
These monomials are in turn isomorphic to a Clifford algebra
so there the total number of odd monomials is equal to the 
total number even monomials. This translates into counting
fields as follows,
\be 
\eqalign{
N(C) &=~ 1 ~~~~~~~{\rm {because}} ~ C ~=~ C^* ~~~, \cr
N(M) &=~ 2 ~~~~~~~{\rm {because}} ~ M ~\ne~ M^* ~~~, \cr
N(A_{\underline a}) &=~ 4 {~~~~~~~}{\rm {because}} ~ A_{\underline a}~=~
A_{\underline a}^* ~~~ \& ~~
{\underline a} ~=~ 0, \dots ,3 ~~~,\cr
N({\rm d}) &=~ 1 ~~~~~~~{\rm {because}} ~ {\rm d} ~=~ {\rm d}^* ~~~, \cr
\to ~~ N_B &=~ N(C) ~+~ N(M) ~+~ N(A_{\underline a}) ~+~ N({\rm d}) ~=~ 8 ~~~.
}\ee
\be 
\eqalign{
N(\l^{\a}) &=~ 4 ~~~~~~~{\rm {because}} ~ \l^{\a} ~\ne~ [\l^{\a}]^* ~~~
 \& ~~{\a} ~=~ 1,2  ~~,\cr
N(\chi^{\a}) &=~ 4 ~~~~~~~{\rm {because}} ~ \chi^{\a} ~\ne~ [\chi^{\a}]^* ~~~
 \& ~~{\a} ~=~ 1,2  ~~,\cr
\to ~~ N_F &=~ N(\l^{\a}) ~+~  N(\chi^{\a})   ~=~ 8 ~~~.
}\ee
This same type of procedure can be applied to the chiral scalar superfield 
with the result that $N_B = N_F = 4$, so the chiral multiplet is literally
a smaller representation containing fewer independent degrees of freedom than
the real scalar superfield.

The equality in the numbers of bosons versus the fermions has far reaching 
implications for the quantum behavior of supersymmetrical systems. It is 
the case that the quantum corrections to a particular theory are 
determined by Feynman diagrams containing ``loops.''  The fact that fermionic 
loops have minus signs relative to bosonic loops implies the possibility of 
some types of cancellations. 

\subsection{Superspace Integration}

~~~~Berezin was the first to introduce integration over Grassmann numbers.
It is useful to review this.  Consider a ``toy" superfunction dependent on
a single coordinate t and a single Grassmann coordinate $\theta$, then
\be
f(z) ~=~ A(t) ~+~ i \q \psi(t)
\ee
(assume $f(z) = f^*(z)$ for simplicity). Berezin's definition of integration 
was 
\be
\int d \q ~ f(z) ~=~ i \, \psi(t) ~~~,
\ee
but this is the same as
\be
\partder{~}{\q} f(z) ~=~ i \, \psi(t) ~~~.
\ee
For Grassmann numbers, integration is the same as differentiation!
Using this as a hint we may define the superspace integral by
\be
\int d^4 \q ~=~ \int d^2 \q \, d^2 {\bar \q} ~=~ \fracm 14 \,
\eta^{\underline {a \, b}} \, \partder{~~}{\q^{\a}}
 \, \partder{~~}{\q^{\b}}  \, \partder{~~}{\bar\q^{\Dot \a}}
 \, \partder{~~}{\bar\q^{\Dot \b}}
~~~. \ee
However, given the forms of $D_{\a}$ and ${\Bar D}{}_{\Dot \a}$ this 
definition is equivalent to the following (up to total derivatives)
\be
\eqalign{
\int d^4 \q ~{\cal L} &\equiv~ \lim_{\q \to \, 0}  \lim_{\bar\q \to \, 0} 
\fracm 14 \, \eta^{\underline {a \, b}} \, D_{\a} \, D_{\b}
\, {\Bar D}{}_{\Dot \a} \, {\Bar D}{}_{\Dot \b} ~{\cal L} ~~~,  \cr
&\equiv~  D^2 {\Bar D}{}^2 {\cal L} {\Big |} ~~~.
}\ee
Thus we can calculate $\int d^4 \q ~ F(z)$ for the superfield in 
(\ref{eq:fyx}) to find
\be
\int d^4 \q  ~F(z) ~=~ {\rm d}(x) ~~~,
\ee
just like Berezin's result.  This is called the ``${\rm d}$-component" of a
superfield and can be applied to {\it {any}} SF to construct an
effective supersymmetry invariant,
\be
{\d}_Q \, \Big( ~ \int d^4 \q ~{\cal L} ~\Big) ~=~ {\rm {total~div}}~~~.
\label{eq:ehw} \ee
We can say that the action is effectively invariant because this
equation shows us that the action (i.e. ${\cal S} = \int d^4
x \, \int d^4 \q \, {\cal L}$) only changes by surface terms
which are usually ignored in field theories anyway).

For the chiral SF we must be a bit more careful.  We note
\be
\int d^4 \q ~ \Phi ~=~ \int d^4 \q ~{\Bar \Phi} ~=~ 0 ~~~,
\ee
(up to total divergences), but
\be
\int d^2 \q ~\Phi ~\equiv~  \lim_{\q \to \, 0} D^2 \Phi ~=~ 
D^2 \Phi {\Big |} ~=~ F(x) ~~~,
\ee
and the super translation variation of this satisfies
\be
{\d}_Q \, \Big( ~ \int d^2 \q ~{\cal L}_c ~\Big) ~=~ {\rm {total~div}}
~~~. \label{eq:ehv}
\ee
This is called the ``$F$-component" invariant and it can be used for any 
chiral superfield. The feature that is special about a chiral SF is that it
satisfies a certain differential constraint, ${\Bar D}{}_{\Dot \a}
\Phi = 0$, which makes it unnecessary to integrate over the
complete set of superspace coordinates. It is only integrated over
a ``chiral sub-space.''

So we have three superspace measures with which to build actions
\be
\eqalign{
\int d^6 z &\equiv~ \int d^4 x ~ d^2 \q \,  {\cal L}_c ~~~,~~{\rm {if}}~
{\Bar D}{}_{\Dot \a} \, {\cal L}_c~=~ 0 ~~~, \cr
\int d^6 {\bar z} &\equiv~ \int d^4 x ~ d^2 \bar \q \,  
{\Bar {\cal L}}{}_c ~~~,~~{\rm {if}}~ D_{\a} \, {\Bar {\cal L}}{}_c
~=~ 0 ~~~, \cr
\int d^8 z &\equiv~ \int d^4 x ~ d^2 \q \,  d^2 \bar\q \, {\cal L}
~~~,~~{\rm {if}} ~{\cal L} ~{\rm {is}} ~{\rm {unrestricted}}~~~.
}\ee
 
\subsection{Putting It All Together: The Dynamics of the Chiral 
SF}

~~~~We can now contemplate the existence of SF Lagrangians
\be
{\cal L} ~=~ {\cal L}( \Phi, D_{\underline A} \Phi, \dots ) ~~~,
\ee
that are the analogs of ordinary Lagrangians.  These Lagrangians can
lead to actions with a definition of superspace integration
theory. We will end this discussion by showing how all of the pieces
we have introduced in this section are used to write supersymmetrical
field theories which turn out to remarkably similar to ordinary
relativistic field theories. For the purposes of this illustration
we treat the free massive chiral superfield \cite{4}.

As is customary this theory is described by
\be
{\cal S} ~=~ \int d^8 z ~ {\Bar \Phi} \, \Phi  ~+~ \Big[ ~ \int d^6 z ~
(\, \frac 12 \, m \, \Phi^2 \, ) ~+~ {\rm {h.\, c.}} ~\Big] ~~~.
\label{eq:hth}
\ee 
This action involves two functions, the K\"{a}hler potential ($K = 
{\Bar \Phi}  \Phi$) and the superpotential ($W = \fracm 12 m \Phi^2$)
In principle we can use explicit $\theta$-expansions to derive the
``${\rm d}$'' component of the first superfield and the ``$F$''
term of the second, but that is not efficient. Instead we note
that the component fields of $\Phi$ may be thought of as Taylor
series coefficients of $\Phi$ which are defined by
\be
A(x) ~\equiv~ \Phi {\Big |} ~~,~~ \psi_{\a}  ~\equiv~ D_{\a} 
\Phi {\Big |} ~~,~~ F(x) ~\equiv~ D^2  \Phi {\Big |} ~~.
\label{eq:eyn} \ee
Under this circumstance and recalling that the superspace integrations
$ \int d^8 z$ and $ \int d^6 z$ actually correspond to fourth-order 
and second-order spinorial differential operators, respectively, opens
the following route for calculations.
\be
\eqalign{
 \int d^6 z ~ (\, \frac 12 \, m \, \Phi^2 \, ) &=~ \fracm 14 \, m \, 
\int d^4 x D^{\a} D_{\a} \Phi^2 ~  {\Big |} ~~~ \cr
&=~ \fracm 12 \,  m \, \int d^4 x ~D^{\a}[ \Phi \,  D_{\a} \Phi ]  ~ 
{\Big |} \cr
&=~ \fracm 12 \,  m \,\int d^4 x ~[~  (D^{\a}\Phi) \,  (D_{\a} \Phi)
~+~ \Phi \,  (D^{\a} \,D_{\a} \Phi) ~]  ~  {\Big |} \cr
&=~ \fracm 12 \,  m \,\int d^4 x ~[~  \psi^{\a} \, \psi_{\a} 
~+~ A \,  (2 F) ~] \cr
&=~\int d^4 x ~[~  \fracm 12 \,  m \, \psi^{\a} \, \psi_{\a} 
~+~ m \,  A \,  F ~] ~~~.} \label{eq:nty}
\ee
In a similar manner we find
\be
\eqalign{
\int d^8 z ~ {\Bar \Phi} \, \Phi &=~  \int
d^4 x ~D^2 {\Bar D}{}^2 {\Bar \Phi} \,\Phi ~  {\Big |} ~
=~  \int d^4 x ~ {\Bar D}{}^2  D^2 {\Bar \Phi} \,\Phi ~  {\Big |} \cr
&=~  \int d^4 x ~ {\Bar D}{}^2 [~ {\Bar \Phi}\,  D^2 \,\Phi ~]  ~{\Big |}\cr
&=~  \int d^4 x ~ [~ ({\Bar D}{}^2 {\Bar \Phi})\,  (D^2 \,\Phi)
\, + \, ({\Bar D}{}^{\Dot \a} {\Bar \Phi}) \,({\Bar D}{}_{\Dot \a}
D^2 \Phi ) \, + \,  {\Bar \Phi}\,  (({\Bar D}{}^2 D^2 \,\Phi)
  ~]  ~{\Big |} ~~.
} \label{eq:nyo}
\ee
Now observe
\be
\eqalign{
{\Bar D}{}_{\Dot \a} \Phi ~=~ 0  &~~\&~~ {\rm {by~definition}}~~
D_{\a} \, {\Bar D}{}_{\Dot \a} ~+~{\Bar D}{}_{\Dot \a}
D_{\a} ~=~ i \pa_{\underline a} ~~~, \cr
&\to ~~ {\Bar D}{}_{\Dot \a} D_{\a}  \Phi ~=~ i \pa_{\underline a}
 \Phi ~~~.
}\ee
This result can be used to simplify the differential operators in 
(\ref{eq:nyo}). We basically need to ``push'' the ${\Bar D}$ operators in
the second and third terms until they act on the factors of
$\Phi$. This can be obtained systematically as
\be
\eqalign{
{\Bar D}{}_{\Dot \a} D^2  \Phi &=~ \fracm 12 \, {\Bar D}{}_{\Dot \a}
D^{\a} D_{\a} \Phi ~=~ - \fracm 12 \,  {\Bar D}{}_{\Dot \a}
D_{\a} D^{\a} \Phi \cr
&=~ \fracm 12 \,  D_{\a}  {\Bar D}{}_{\Dot \a} D^{\a} \Phi
~-~ i \fracm 12 \pa_{\underline a} D^{\a} \Phi \cr
&=~ - \fracm 12 \,  D^{\a}  {\Bar D}{}_{\Dot \a} D_{\a} \Phi
~-~ i \fracm 12 \pa_{\underline a} D^{\a} \Phi \cr
&= ~-~ i\, \pa_{\underline a} D^{\a} \Phi ~~~,}
\label{eq:nyt} \ee
where in obtaining this result we have liberally used various 
points discussed earlier in this presentation.  We use the last
line of (\ref{eq:nyt}) once more to see,
\be
\eqalign{
{\Bar D}{}^2 D^2 \, \Phi &=~ \fracm 12 \, {\Bar D}{}^{\Dot \a}
{\Bar D}{}_{\Dot \a} D^2  \Phi ~=~ - \, i \, \fracm 12 \, {\Bar 
D}{}^{\Dot \a} \pa_{\underline a} D^{\a} \Phi  \cr
&=~ - i \, \fracm 12 \, \pa_{\underline a}  \, {\Bar D}{}^{\Dot \a}  
D^{\a} \Phi \cr
&=~  \fracm 12 \, \pa_{\underline a}  \, \pa^{\underline a} \Phi
~=~ {}_\bo \Phi ~~~.
} \label{eq:nyf} 
\ee
The final results of (\ref{eq:nyt},\ref{eq:nyf}) are now substituted 
into the last two term of (\ref{eq:nyo}) and yield,
\be
\eqalign{
\int d^8 z ~ {\Bar \Phi} \, \Phi &=~  \int d^4 x ~ [ ~ {\Bar F} F
\, - \, i \, {\Bar \psi}{}^{\Dot \a} \, \pa_{\underline a} {\psi}{}^{\a} \, +
\, {\Bar A} {}_\bo A ~] ~~~.}\label{eq:ntv}
\ee
 
We have now proven that the action in (\ref{eq:hth}) is equivalent to the
following ordinary looking relativistic field theory,
\be
\eqalign{
{\cal S} &=~ \int d^4 x ~ [ ~ - \fracm 12 \, ( \pa^{\underline a} 
{\Bar A} ) \, ( \pa_{\underline a} {A} ) ~-~ i \, {\Bar \psi}{}^{\Dot 
\a} \, \pa_{\underline a}  {\psi}{}^{\a} ~+~ {\Bar F} F  \cr
&{~~~~~~~~~~~~~~~}+~  \fracm 12 \,  m \,(\,  \psi^{\a} \, \psi_{\a} 
\,+ \, {\Bar \psi}{}^{\Dot \a} \, {\Bar \psi}{}_{\Dot \a} \,) ~+~
m \, (\, A \,  F \,+ \, {\Bar A}\,  {\Bar F}\,)  ~]  ~~~.
} \label{eq:nyx}
\ee
By re-writing the fermionic translation laws of the fields in terms
of four component fermions, we can see that the real part of $A$ corresponds
to a scalar ($J^P = 0^+$) and its imaginary part corresponds
to a pseudoscalar ($J^P = 0^-$). In the presence of a nonvanishing
mass term the four component spinor corresponds to a Majorana
(or real spinor). In the massless limit the spinor may be regarded
as either a Majorana or Weyl spinor.  The example above shows that 
4D superfields are particularly sensitive the the parity assignments 
of fields. For, example it seems to be impossible to construct a 4D 
theory which contains two scalar fields and one Weyl or Majorana 
field.

There is another point about the component actions that we write 
(but derive from superspace).  It is often the case that the simplest 
definition of the component fields (similar to (\ref{eq:eyn})) for a supermultiplet 
leads to unusual normalization compared to ``standard'' (i.e. 
non-supersymmetrical definitions).  This can always be `fixed' by 
choosing the definitions of the component fields appropriately.  
We shall not (as is the usual custom) concern ourselves with this 
issue except for pointing this out.

By varying this action in the usual way we can derive the following
equations of motion,
\be
\eqalign{
&~~~{}_\bo A ~+~ m \, {\Bar F} ~=~ 0 ~~~, \cr
&- i \, \pa_{\underline a}  {\psi}{}^{\a} ~+~ m \, {\Bar \psi}{}_{\Dot \a}
~=~ 0 ~~~, \cr 
&~~~~~~~F ~+~ m \, {\Bar A} ~=~ 0 ~~~.}
\label{eq:nyv} \ee
The last of these is purely algebraic, it may be used to solve for the 
field $F$ as $F ~=~ -  m{\Bar A}$. For this reason $F$ is called an 
``auxiliary field variable'' or ``auxiliary field.'' It can be used in 
the first equation in (\ref{eq:nyv}) to write the equations of motion for $A(x)$
and ${\psi}{}^{\a}(x)$
\be
\eqalign{
&~~~\,~\,(~ {}_\bo ~-~ m^2 \,) \, A ~=~ 0 ~~~, \cr
&- i \, \pa_{\underline a}  {\psi}{}^{\a} ~+~ m \, {\Bar \psi}{}_{\Dot \a}
~=~ 0  ~~~.}
\ee
These equations show another characteristic feature of supersymmetric
theories; strict supersymmetry implies a degeneracy of the mass
spectrum between some bosons and fermions.

We can also use the equation of motion for $F$ to eliminate it
from the action in (\ref{eq:nyx}). The resultant action is called the ``on-shell''
action.
\be
\eqalign{
{\cal S}_{on-shell} &=~ \int d^4 x ~ [ ~ - \fracm 12 \, ( \pa^{\underline 
a} {\Bar A} ) \, ( \pa_{\underline a} {A} ) ~+~ m^2 {\Bar A} A
\cr
&{~~~~~~~~~~~~~~~}\,-~ i \, {\Bar \psi}{}^{\Dot \a} \, \pa_{\underline a}  
{\psi}{}^{\a} ~+~  \fracm 12 \,  m \,(\,  \psi^{\a} \, \psi_{\a} 
\,+ \, {\Bar \psi}{}^{\Dot \a} \, {\Bar \psi}{}_{\Dot \a} \,)  ~]  ~~~.
}\ee

We close this lecture by once again noting that an intuitive
picture of superspace posits that it is a mathematic structure
whose boundary is to be identified with an ordinary bosonic space.
This view is very simple and has as its implication that the
physics we experience in the universe around us occurs exclusively 
on this boundary. Later we will discuss integration theory in the 
case where the bosonic manifold is ``curved'' and we will see that
this view has a simplifying consequence for the generalization
of Berezin's integration theory.

\newpage
\section{Second Lecture: Sophomore 4D, N = 1 Superfield Theory}
{\it {Introduction}}

~~~~In the second lecture, we will discuss a number of
examples of 4D, N = 1 supersymmetrical field theories. These
include; (a.) the renormalizable interacting, scalar multiplet,
(b.) the gauge-vector multiplet, (c.) K\" ahler models, (e.)
the simplex of super p-forms and (f.) the nonminimal multiplet.
The main goal is to show how field theory in 4D, N = 1 superspace
is very, very similar to field theory in Minkowski space. Along
the way we will learn how to construct Yang-Mills gauge theories
and non-linear $\s$-models.

$${~}$$

\noindent
{\it {Lecture}}
\subsection{The 4D, $N$ = 1 Interacting Chiral Multiplet}

~~~~We can generalize the superpotential $W(\Phi)$ of the previous
section by considering an action of the form,
\be
{\cal S} ~=~ \int d^8 z ~ K( \Phi, \,{\Bar \Phi})  ~+~ \Big[ ~ 
\int d^6 z ~ W( \Phi) ~+~ {\rm {c.\, c.}} ~\Big] ~~~,
\ee 
where $W(\Phi) = \fracm 12 m \Phi^2 + \fracm 1{3!} \l \Phi^3$. Using 
the same techniques as before (\ref{eq:nty}) but applied to $W$, we derive
\be
\eqalign{
{\cal S} &=~ \int d^4 x ~ \Big\{ ~ - \fracm 12 \, ( \pa^{\underline a} 
{\Bar A} ) \, ( \pa_{\underline a} {A} ) ~-~ i \, {\Bar \psi}{}^{\Dot 
\a} \, \pa_{\underline a}  {\psi}{}^{\a} ~+~ {\Bar F} F  \cr
&{~~\,~~~~~~~~~~~~~~~}+~ [ ~ (\, m \, + \, \l A\, )
\,(\,  \fracm 12 \, \psi^{\a} \, \psi_{\a} 
\,+ \,  A \,  F \, ) ~+~ {\rm {c.\,c.}}
 ~] ~ \Big\} ~~~,
}\ee
and using the new algebraic equation of motion for $F$ leads to the
action
\be
\eqalign{
~~{\cal S} &=~ \int d^4 x ~ \Big\{ ~ - \fracm 12 \, ( \pa^{\underline a} 
{\Bar A} ) \, ( \pa_{\underline a} {A} ) ~-~ i \, {\Bar \psi}{}^{\Dot 
\a} \, \pa_{\underline a}  {\psi}{}^{\a} ~+~ \fracm 12 \, m \,(\,  
\psi^{\a} \, \psi_{\a} \, + \,  {\Bar \psi}{}^{\Dot \a} \, {\Bar 
\psi}{}_{\Dot \a} \,) \cr
&{~~\,~~~~~~~~~~~~~~~~~}~+~  m^2 \, A {\Bar A} ~
-~ \fracm 12 \, m \l \, (\, A {\Bar A}{}^2 \,+ \,
{\Bar A} A^2 \, ) ~-~ \fracm 14 \l \, A^2 {\Bar A}{}^2 \cr
&{~~\,~~~~~~~~~~~~~~~~~~}+~ \fracm 12 \l \, ( \, A \psi^{\a} \, \psi_{\a} 
 \, + \, {\Bar A}  {\Bar \psi}{}^{\Dot \a} \,{\Bar 
\psi}{}_{\Dot \a} \,) ~  \Big\} ~~~.
}\ee
We see a complex spin-0 field
\be 
\begin{array}{llll}
&A ~=~ {1 \over \sqrt 2} [~ a(x) &+& i ~b(x) ~] \\
&{~~~~~~~~~~~~}{\rm {scalar}} &,& {\rm {pseudoscalar}},  
\end{array}
\ee
and a Majorana spinor,
\be
\psi_{(\a)}(x) ~=~ 
\left(\begin{array}{c}
 \psi_{\a}(x) \\   {\Bar \psi}{}_{\Dot \a} (x)
\end{array}\right)  ~~~,
\ee
both with a mass $m$.  These fields interact via a ``$\varphi^4$-term''
(coupling constant $\fracm 14 \l$), a ``Yukawa-term'' (coupling constant
$\fracm 12 \l$) and a ``$\varphi^3$-term'' (coupling constant $\fracm 12
m \l$). So in addition to imposing the condition $m_{Spin-0}
= m_{Spin-1/2}$ supersymmetry also imposes a set of relationships
among the coupling constants of the $\varphi^4$-, $\varphi^3$- and Yukawa 
interaction terms. 

\subsection{U(1) Symmetry in 4D, N = 1 Superspace}\label{subsec:wpq}

~~~~Since $A$ is complex, at first glance it might seem possible to
realize the U(1) symmetry of QED in the form
\be
\Big[ A \Big]' ~=~ e^{i g \a} A ~~~,~~~  \Big[ \psi_{\a} \Big]' 
~=~ e^{i g \a} \psi_{\a}  ~~~.
\ee
There are a couple of reasons why this is incorrect. First we
note that unless $m = \l = 0$, these transformations are 
{\it {not}} symmetries of the action above. Next even if we 
permitted $m = \l = 0$, it is still not possible to use these
transformation to realize a U(1) symmetry.  One reason for this
is because the transformation $\Big[ A \Big]' ~=~ e^{i \a} A$
``rotates'' a scalar into a pseudoscalar. Although an axial
U(1) charge (or ${\rm U}_A(1)$ charge) has this property, the 
U(1) charge of QED does not possess this property. So in order
to introduce a rigid U(1) symmetry we must modify the starting
point.

Let us consider the alternate action
\be
\eqalign{
{\cal S} &=~ \int d^8 z ~ [ ~ {\Bar \Phi}_+
\Phi_+ ~+~ {\Bar \Phi}_- \Phi_-  ~]  ~+~ 
\int d^6 z ~ [ ~ W( \Phi_+ , \, \Phi_-) ~+~ {\rm {c.\, c.}} ~] ~~~,
}\ee 
which is obtained by introducing {\it {two}} independent chiral
scalar superfields $\Phi_+$ and $\Phi_-$. The first two terms in
the action clearly possess a rigid U(1) invariance.  For the sake 
of renormalizability the polynomial superpotential $ W( \Phi_+ , 
 \Phi_-)$ must be of order three or less. By a proper choice of 
this polynomial, this action can be made to be invariant under
\be
\Big[ \Phi_+ \Big]' ~=~  e^{i g \a} \Phi_+ ~~~,~~~ 
\Big[ \Phi_- \Big]' ~=~  e^{- i g \a} \Phi_- ~~~.~~~ 
\ee 
The unique choice of such a polynomial of order three or less
is $W( \Phi_+ , \, \Phi_-) = m \,  \Phi_+   \Phi_-$.
The two scalars $a_+ (x)$ and $a_- (x)$ may be regarded as the
real and imaginary part of a complex scalar required to realize
the U(1) symmetry of QED.  Similarly, we can now define a Dirac spinor
\be
\psi_{(\a)}(x) ~=~ 
\left(\begin{array}{c}
 \psi_{\a \, + }(x) \\    \psi_{\a \, - }(x)
\end{array}\right)  ~=~  
\left(\begin{array}{c}
 D_{\a} \Phi_+ {\Big |} \\  {\Bar D}_{\Dot \a} {\Bar \Phi}_- {\Big |} 
\end{array}\right) 
~~~, \label{eq:nhh}
\ee
and under the rigid U(1) symmetry this transforms as
\be
\Big[ \psi_{(\a)}(x) \Big]' ~=~ e^{i g \a}~
\psi_{(\a)}(x) ~~~~.
\ee
We can identify this Dirac spinor as the electron. It is accompanied
by two scalar ($a_+ (x), a_- (x)$) and two pseudoscalar ($b_+ (x), 
b_- (x)$) collectively called ``selectrons.'' This is another example 
of the bose-fermi equality of supersymmetrical theories. It is well 
known that the ordinary Dirac field describes four dynamical fermionic 
degrees of freedom.   The selectrons are the corresponding bosonic 
degrees of freedom.

We now wish to ``gauge'' this symmetry by replacing the constant
$\a$ by a chiral superfield $\L$ (i.e. ${\Bar D} {}_{\Dot \a} \L = 0$)
so that the local U(1) transformation laws of the matter superfields
$\Phi_+$ and $\Phi_-$ may be cast into the form
\be
\Big[ \Phi_+ \Big]' ~=~  e^{i g \L} \Phi_+ ~~~,~~~ 
\Big[ \Phi_- \Big]' ~=~  e^{- i g \L} \Phi_- ~~~.~~~ 
\ee 
For the superpotential term this causes no problems,
\be
\Big[ \Phi_+ \,  \Phi_- \Big]' ~=~  \Big[ e^{i g \L} \Phi_+  e^{- i g \L} 
\Phi_- \Big] ~=~  \Big[ \Phi_+  e^{i g \L} \,  e^{- i g \L} 
\Phi_- \Big] ~=~  \Phi_+ \,  \Phi_- ~~~.~~~ 
\ee 
However for the K\" ahler potential we see
\be
\Big[~ {\Bar \Phi}_+ \,  \Phi_+ 
\, + \, {\Bar \Phi}_- \,  \Phi_- ~\Big]' ~=~ 
\Big[~ {\Bar \Phi}{}_+  e^{- i g {\bar \L}} \,  e^{ i g \L} \Phi_+ \, 
+ \, {\Bar \Phi}{}_-  e^{i g {\bar \L} } \,  e^{- i g \L} 
\Phi_-  ~\Big] ~~~.
\ee

This is a major problem since $\L \ne {\Bar \L}$ implies that the
K\" ahler potential terms are {\it {not}} gauge invariant.  The solution
to this problem is to introduce a ``photon superfield.'' If we recall
(\ref{eq:fyx}), it is pretty clear how this should be done. There we saw that
a real scalar superfield contains a spin-1 field. Via a number of
arguments, we can show that the spin-1 field that appears in 
(\ref{eq:fyx}) is not a vector but an axial vector. So as a first step, we need
to introduce a pseudoscalar superfield to be denoted by $V$ (its
$\q \bar\q$ term is a vector). Further by noting that $V \sim \q^{\a} 
\bar\q^{\Dot \a} A_{\underline a}$ and $\L \sim  i \fracm 12 \q^{\a} 
\bar\q^{\Dot \a} \pa_{\underline a} \a(x)$ we see that
\be
\Big[ V \Big]' ~=~ V ~-~ i (\, \L ~-~ {\Bar \L} \,) ~~\to~~
\Big[ A_{\underline a}(x) \Big]' ~=~ \Big[ A_{\underline a}(x) \Big]
\, +  \, \fracm 12 \pa_{\underline a} [\, \a(x) \, + \, {\Bar \a}(x)
\,] ~~~. \ee
The pseudoscalar superfield can be used to modify the K\" ahler
potential term according to,
\be
K ~=~ \Big[~ {\Bar \Phi}_+ \,  e^{g V} \, \Phi_+ 
\, + \, {\Bar \Phi}_- \,  e^{- g V} \, \Phi_- ~\Big] 
~~~,
\ee 
and under the U(1) gauge transformation we now find
\be
\eqalign{
\Big[~ {\Bar \Phi}_+ \,  e^{g V}\,  \Phi_+ \, + \, {\Bar 
\Phi}_- \,  e^{-g V} \,  \Phi_- ~\Big]' &=~ 
\Big[~ {\Bar \Phi}{}_+  e^{- i g {\bar \L}} \,  e^{g V'}\,  
e^{ i g \L} \Phi_+ \, \cr 
&{~~~~~~~}+ \, {\Bar \Phi}{}_-  e^{i g {\bar \L} }\,  
e^{- g V'} \,  e^{- i g \L} \Phi_-  ~\Big] \cr
&=~ 
\Big[~ {\Bar \Phi}{}_+ \,  e^{g V}\, \Phi_+ \, + \, 
{\Bar \Phi}{}_- \, e^{- g V} \,  \Phi_-  ~\Big] ~~~.
}\ee
So the matter field terms in a gauge invariant supersymmetric version 
of QED can be written as \cite{7}
\be
\eqalign{
{\cal S}_{QED-matter} &=~ \int d^8 z ~ [~ {\Bar \Phi}_+ \,  e^{g V}\,  
\Phi_+ \, + \, {\Bar \Phi}_- \,  e^{-g V} \,  \Phi_- ~] \cr
&{~~~~} ~+~ 
\int d^6 z ~ [ ~ m\,  \Phi_+ \,  \Phi_-  ~+~ {\rm {c.\, c.}} ~] 
~~~. }\label{eq:nxt}
\ee
Apparently minimal coupling of Yang-Mills symmetries in superspace 
corresponds to the insertion of the factors involving $V$. As we shall
later see this action can be shown to contain the usual Dirac action 
for the four component spinor defined in (\ref{eq:nhh}).  

\subsection{Review of Maxwell Theory in Minkowski Space}

~~~~We now require a generalization of the Maxwell action, More generally
it is useful to possess a generalization of the Yang-Mills action. 
To begin this discussion let us review ordinary 4D Maxwell theory.
As is customary, we can introduce a covariant derivative
\be
\nabla_{\underline a} ~= ~ \pa_{\underline a} ~-~ i A_{\underline a}
\, t ~~~, 
\label{eq:nsv} \ee 
where $t$ is the abstract U(1) Lie algebra generator. If a matter
field $\phi(x)$ transforms as
\be 
\eqalign{
\Big[ \phi \Big]' &=~ exp[ i Q \l(x) ] \, \phi(x)  \cr
& \approx ~ \phi(x) ~+~ i \, Q \l(x) \phi(x)  ~~ {\rm {if }} \l << 1 
\cr
\to \d \phi &\equiv ~ \Big[ \phi \Big]' ~-~ \phi ~=~ i Q \l(x) \, \phi(x) ~~, 
}\ee
then by definition $[ t \, , \, \phi ] = Q \phi$. Although not usually
introduced in such discussions, $t$ allows us to regard $\nabla_{\underline a}$
as a real operator if $t^* = - t$. This follows from
\be
\eqalign{
\Big( ~ [ \, t \, , \, \phi \,] ~ \Big)^* &=~  Q \,  \Big( \, 
\phi \, \Big)^* \cr
[ \, t^* \, , \, {\Bar \phi} \,]  &=~  Q \,  {\Bar \phi}  \cr
[ \, t \, , \, {\Bar \phi} \,]  &= ~  - \, Q \,  {\Bar \phi}  
~~~, }\ee
and this is what we want since $\phi(x)$ is associated with a
particle possessing charge $Q$ then the anti-particle should
be associated with ${\Bar \phi}(x)$ possessing charge $-Q$.
Thus,
\be
\eqalign{
\Big(\nabla_{\underline a} \, \Big)^*  &= ~ \Big[  \,
\pa_{\underline a} ~-~ i A_{\underline a} \, t \, \Big]^* \cr
&=~ \pa_{\underline a} ~+~ i A_{\underline a} \, t^* \cr 
&=~ \pa_{\underline a} ~-~ i A_{\underline a} \, t  ~~~. }\ee
The field strength $f_{\underline a \, \underline b}$ is defined
by
\be
\eqalign{
- i \, f_{\underline a \, \underline b} \, t ~\equiv~
[ \,  \nabla_{\underline a}  \,\, , \,\, \nabla_{\underline b}  
\, \} &=~ [ \, \pa_{\underline a} ~-~ i A_{\underline a} \, t \,
,  \, \pa_{\underline b} ~-~ i A_{\underline b} \, t   \,]  \cr
&= - i ( \, \pa_{\underline a} A_{\underline b} \, - \,
 \pa_{\underline b}  A_{\underline a} \, ) t \, - \,  
[ \,  A_{\underline a} \, t \, ,  A_{\underline b} \, t   \,] \cr
&=~- i ( \, \pa_{\underline a} A_{\underline b} \, - \,
 \pa_{\underline b}  A_{\underline a} \, ) t \,-\,   A_{\underline a} \,
[ \, t \, , \, A_{\underline b} \,  \,] \, t
\, \cr 
&{~\,~\,~}- \,  A_{\underline b} \, [ \,  t \, ,  A_{\underline a} \, t  
 \,]  \,-\,   A_{\underline a} \,  A_{\underline b} \,
[ \, t \, , \, t   \,] 
~~~.
}\ee 
All of the terms quadratic in the gauge field will vanish if
we define
\be
[ \, t \, , \, A_{\underline a} \,  \,] ~\equiv~ 0 ~~~,~~~ 
[ \, t \, , \, t   \,]  ~\equiv~ 0 ~~~, ~~~ \to ~
f_{\underline a \, \underline b} ~=~  ( \, \pa_{\underline a} 
A_{\underline b} \, - \, \pa_{\underline b}  A_{\underline a} 
\, ) ~~~. \label{eq:ntw}
\ee
(for nonabelian groups this changes) and the Maxwell action
is simply
\be
{\cal S}_{Maxwell} ~=~ - \fracm 18 \, \int d^4 x ~  f^{\underline a \, 
\underline b}  f_{\underline a \, \underline b} ~~~.
 \label{eq:ntww} \ee
(The strange-looking factor of 1/8 is a consequence of our definitions
in section \ref{subsec:wpp}.)   

\subsection{A U(1) Covariant Derivative in 4D, N = 1 Superspace}

~~~~In the first lecture, we saw that there exists in superspace a
``supergradient operator'' $D_{\underline A} \equiv \Big(D_{\a},\, 
{\Bar D}{}_{\Dot \a} \,,\, \pa_{\underline a} \Big)$ that generalizes
the ordinary gradient $\pa_{\underline a}$.  So it is natural to
define a 4D, N = 1 U(1) covariant operator
\be
\nabla_{\underline A} ~\equiv~ D_{\underline A} ~-~ i \G_{\underline
A} t ~~~, 
\label{eq:nwi}
\ee
and use it to define a superspace field strength $F_{\underline A
 \, \underline B}$.  There is something that needs attention prior
to deriving the explicit form of this. We note
$$
[\,   D_{\a}\, \, , \, \, D_{\b} \, \} ~=~ 0 ~~~,~~~
[\, {\Bar D}_{\Dot \a}\, \, , \, \, {\Bar D}_{\Dot \b} 
\, \} ~=~ 0 ~~~,~~~ [\,  D_{\a}\, \, , \, \, {\Bar D}{}_{\Dot 
\a} \, \} ~=~ i \, \pa_{\underline a} ~~~, 
$$
$$  
[ \, {D}_{ \a} ~,~ \pa_{\underline b} \, \} ~=~0  ~~~,~~~
[ \, {\Bar D}{}_{\Dot \a} ~,~ \pa_{\underline b} \, \} ~=~ 
0 ~~~,
$$
\be
[ \,  \pa_{\underline a}  \,\, , \,\, \pa_{\underline b}  
\, \} ~=~ 0 
~~~. \label{eq:itv}
\ee
All of these results can be collectively written in the form
\be
\Big[ ~ D_{\underline A} \, , \, D_{\underline B} ~ \Big\}
~=~ C_{\underline A \, \underline B}{}^{ \underline C}
\, D_{\underline C} ~~~,
\ee
for some appropriate coefficients $C_{\underline A \, \underline 
B}{}^{ \underline C}$ known as the ``object of anholonomity''
or more simply as the ``anholonomity.'' The values of these
coefficients are simply read off by comparing this definition
with the explicit graded commutators above (\ref{eq:itv}). The appearance
of these is important due to the definition of the field
strength in superspace. It is defined by
\be
\Big[ ~ \nabla_{\underline A} \, , \, \nabla_{\underline B} 
~ \Big\} ~=~  C_{\underline A \, \underline B}{}^{ \underline C}
\, \nabla_{\underline C} ~-~ i \,  F_{ {\underline A} \, 
{\underline B}} \, t ~~~,
\ee
where
\be
F_{{\underline A} \, {\underline B}} \, t ~=~ D_{\underline A} \, \G_{
\underline B} \, t - (-)^{{\underline A} \, {\underline B}}
D_{\underline B} \, \G_{\underline A} \, t ~-~ 
C_{\underline A \, \underline B}{}^{ \underline C}
\, \G_{\underline C} \, t ~~~.
\ee    

\subsection{Finding and Solving Covariant Constraints}

~~~~In the presence of local symmetries, the prior definition of
a chiral superfield is not obviously gauge covariant (well
it really is by a certain definition) and this suggests that
the differential constraint used to define a chiral superfield
should be changed according to 
\be
{\Bar \nabla}{}_{\Dot \a} \Phi ~=~ 0 ~~~,
\ee
and since this condition is satisfied then $ {\Bar \nabla}{}_{\Dot 
\a} {\Bar \nabla}{}_{\Dot \b} \Phi ~=~ 0$. If we symmetrize
this last result on the ${\Dot \a}$ and ${\Dot \b}$ indices then
use the definition of the superspace field strength we find
\be
0 ~=~ \Big[ ~ {\Bar \nabla}{}_{\Dot \a} \, , \, {\Bar \nabla}{}_{\Dot \b}
~ \Big\}  \Phi ~=~ F_{\Dot \a \, \Dot \b} \, t \, \Phi  ~=~
 F_{\Dot \a \, \Dot \b} \,  Q \, \Phi ~~~.
\ee
This final equation is called an ``integrability condition
\cite{5}.'' It has three solutions;

${}$\newline \indent
(A.) $Q=0$  (The matter superfield carries no charge.)

${}$\newline \indent
(B.) $ F_{\Dot \a \, \Dot \b} = 0$  (This looks crazy. In ordinary
Maxwell theory if we \newline \indent ${~~\,~~~}$ impose the condition 
$ F_{\underline a \, \underline b}= 0 $  then its solution is 
$A_{\underline a} = \pa_{\underline a } a$ \newline 
\indent ${~~\,~~~}$ and there is no photon in the theory.)

${}$\newline \indent
(C.) $ \Phi = 0$.  (There is no matter superfield at all.)

${}$\newline \noindent
Thus, at first it seems as though the notation of a superspace
U(1) covariant derivative has brought us to a bad ending.

However, a pleasant surprise is to be had by studying the second
option in more detail.  We are able to solve the restriction in option
(b.) by following the example of ordinary 4D spacetime.
\be
\eqalign{
F_{\Dot \a \, \Dot \b}\, t  &=~ 0 ~~\to ~~ {\Bar D}{}_{\Dot 
\a} \G_{\Dot \b} ~+~ {\Bar D}{}_{\Dot \b} \G_{\Dot \a} ~=~ 
0 ~~~,\cr
&\to ~ \G_{\Dot \a} t ~=~  i \,( {\Bar D}{}_{\Dot \a} \, {\Bar 
\O} ) t~=~ i \, e^{- \, {\Bar \O} t} {\Bar D}{}_{\Dot \a}  e^{ 
\, {\Bar \O} t} ~~~,\cr
&\to ~ \G_{\a}t ~=~  i( \, D_{\a} \, {\Bar \O} ) t ~=~
i \, e^{- \, \O t} {D}{}_{\a}  e^{ \O t} ~~~, 
}\ee
where we have obtained the last line by applying the superspace
conjugation operator to the penultimate line.  Similarly to 
ordinary Minkowski space, the solution yields a scalar superfield. 
However, unlike ordinary Minkowski space, the scalar 
superfield\footnote{In 1980 when W. Siegel and I first introduced
the term ``pre-potential \cite{6},'' it was in \newline ${~~~~~}$ 
the context of quantities of this type that we meant it.} $\O$ is 
complex.  This seemingly minor difference is crucially important 
as we shall shortly see.

Let us continue by examining the form of $ F_{\a \, \Dot \b}\,$,
\be
\eqalign{
 F_{\a \, \Dot \b} &=~ D_{\a} \G_{\Dot \a} ~+~ {\Bar D}{}_{\Dot 
\a} \G_{\a} ~-~ C_{\a \, \Dot \a}{}^{\underline C} \, \G_{\underline 
C} ~~~ \cr
&=~ {D}{}_{\a} \G_{\Dot \a} ~+~ {\Bar D}{}_{\Dot \a} \G_{\a} ~-~ 
i \, \G_{\underline a} \cr
&=~ i {D}{}_{\a} {\Bar D}_{\Dot \a}{\Bar \O} ~+~ i {\Bar 
D}{}_{\Dot \a} D_{\a} \O ~-~ i \, \G_{\underline a} 
~~~.
}\ee
The interesting point regarding this field strength is that the
gauge field which ``points'' in the Minkowski space direction
appears algebraically. This means if we make the definition
\be
\G_{\underline a} ~=~  {\Bar D}{}_{\Dot \a} D_{\a} \O ~-~
{D}{}_{\a} {\Bar D}_{\Dot \a}{\Bar \O} ~~~,
\ee
then $ F_{\a \, \Dot \b} = 0 $. So $\G_{\underline a} \ne 0$
even though  $ F_{\a \, \b} =  F_{\Dot \a \, \Dot \b} =  
F_{\a \, \Dot \b} = 0 $.  However, the more important matter is 
whether this choice for $\G_{\underline a}$  describes a
``pure gauge'' configuration.

Writing $ \O = \fracm 12 ( V \, + \, i W)$ where $V = V^*$
and $W = W^*$ we obtain
\be
\eqalign{
\G_{\underline a} &=~ -\,  \fracm 12 \Big( \, [ ~ D_{\a} \, , \, 
{\Bar D}{}_{\Dot \a} \,\, ] V \,\Big) ~+~ i \fracm 12 \,
\Big( \, \{ ~ D_{\a} \, , \, {\Bar D}{}_{\Dot \a} \, \} W \,\Big)
\cr
&=~ - \, \fracm 12 \Big( \, [ ~ D_{\a} \, , \, 
{\Bar D}{}_{\Dot \a} \, ] V \,\Big) ~-~ \fracm 12 \,
\pa_{\underline a} W ~~~,
}\ee
and only the second term is seen to be a pure gauge
configuration. If the superfield $V$ here is the same as the one
that was used to covariantize the kinetic energy terms for the 
chiral scalar superfields $\Phi_+$ and $\Phi_-$, we have found 
the relation between the Yang-Mills prepotential $V$ and the
``usual'' connections and field strengths that exist as a basis
for ordinary electromagnetism.  Without loss of generality, we
may set the quantity $W$ to zero.

With the superspace connections all determined as described
above, we can now calculate the remaining components of
$F_{\underline A \underline B}$. The next one is $F_{\a
\underline b}$ where we find,
\be
\eqalign{
F_{\a \, \underline b} &=~ D_{\a} \G_{\underline b} ~-~
\pa_{\underline b} \G_{\a}  \cr
&=~  \fracm 12 \,\Big( \,-\,  D_{\a} \, [ ~ D_{\b} \, {\Bar D}{}_{
\Dot \b} ~-~ {\Bar D}{}_{\Dot \b} \, D_{\b} ]\, V ~-~ \, i\, 
 \pa_{\underline b} D_{\a} V \Big) \cr
&=~  C_{\a \, \b} D^2 {\Bar D}{}_{\Dot \b} V \cr
&\equiv ~ i \, C_{\a \, \b} {\Bar W}_{\Dot \b} ~~~.
} \label{eq:nrv}
\ee
The superfield ${\Bar W}_{\Dot \b}$ must be anti-chiral due to the
identity $D_{\a} D{}^2 = 0$. By conjugation $W_{\a}$ is a chiral
superfield.  After a similar set of calculations we find
\be
F_{\underline a \, \underline b} ~=~ \fracm 12 
\Big[ ~ C_{\a \, \b} \, ( \, {\Bar D}{}_{( \Dot \a} {\Bar W}_{\Dot \b )})
~+~ C_{\Dot \a \, \Dot \b} \, ( \, D_{(\a} W_{\b )} )~ \Big] ~~~.
\ee

At this point, it is useful to employ a little dimensional
analysis.
\be
\eqalign{
&[ x^{\underline m} ] ~=~ - 1 ~~,~~ [ d^4 x ] ~=~ - \, 4 ~~,~~
[\pa_{\underline m}] ~=~ 1 ~~,~~
 [ D_{\a} ] ~=~ [ {\Bar D}_{\Dot \a} ] ~=~ \fracm 12 ~~~, \cr
&[ D^2 ] ~=~ [ {\Bar D}{}^2 ] ~=~ 1 ~~,~~ \to ~ [ \int d^2 \q ]
~=~ [\int d^2 {\bar \q}] ~ = ~ 1 ~~~,}\ee
\be
\eqalign{
&[\G_{\underline a}] ~=~ 1 ~~,~~ [\G_{\a}] ~=~ [\G_{\Dot \a}]
~=~ \fracm 12 ~~\to ~ [F_{\underline a \, \underline b}] ~=~ 2 ~~,~~
[F_{\a \, \underline b}] ~=~ \fracm 32 ~~~, \cr
& \to [ \int d^4 x ~ F^{\underline a \, \underline b} F_{\underline 
a \, \underline b}] ~=~ - \, 4 ~+~ 2 ~+~ 2 ~=~ 0 ~~.
}\ee
This last equation implies that in these units, the ordinary
Maxwell action is dimensionless (since the units of $f$ equal
those of $F$).  From the non-vanishing components of 
$F_{\underline A \, \underline B}$ there are only two quadratic 
action that can be formed
\be
[ \int d^4 x \, d^2 \q ~ W^{\a} W_{\a} ] ~=~ -\, 4 ~+~ 1 ~+~ \fracm 32
~+~ \fracm 32 ~=~ 0 ~~~,
\ee
\be
[ \int d^4 x \, d^2 \q \, d^2 {\bar \q} ~ 
F^{\underline a \, \underline b} F_{\underline a \, \underline b} ] ~=~ 
-\, 4 ~+~ 1 ~+~ 1 ~+~ 2 ~+~ 2 ~=~2 ~~~.
\ee
Only the former of these has the same dimensions as the ordinary
Maxwell action and we are led to suggest
\be
{\cal S}^{N = 1}_{Maxwell} ~=~ \fracm 14 \, \Big[ ~  
\int d^4 x \, d^2 \q ~ W^{\a} W_{\a} ~+~ {\rm {h.\,c.}} ~\Big]  ~~~.
\label{eq:nfn}
\ee

We can check whether this is correct by deriving the component
fields terms described by this action. In order to do this we
must define the component fields through the superfield $W_{\a}$.
In turn this requires knowledge of the form of
all of its spinorial derivatives. There is a systematic way in 
which such results are derived.  This technique is called ``solving 
the superspace Bianchi identities.'' It would take us out of the 
introductory nature of these lectures to discuss this technique.
So let's just use the results that can be derived,
\be
\eqalign{
D_{\a} W_{\b} &=~ \fracm 12 \, C^{\Dot \g \, \Dot \d}
\, F_{\a \Dot \g \,\, \b \Dot \d} ~-~ i\, C_{\a \,\b} ~{\rm d}
~~~, \cr
D^2 W_{\a} &=~ - i \, \pa_{\underline a} {\Bar W}{}^{\Dot \a} ~~~.
}\ee
Thus, the components of the 4D, N = 1 Yang-Mills supermultiplet are 
defined by
\be
\eqalign{
W_{\a} {\Big |} &\equiv~ \l_{\a}(x) ~~~, \cr
D_{\a} W_{\b} {\Big |} &\equiv~  \fracm 12 \, C^{\Dot \g \, \Dot \d}
\, ( ~ f_{\underline a  \, \underline b}(x) ~+~ {\Tilde 
f}_{\underline a  \, \underline b}(x)  ~ )  ~-~ i C_{\a \,\b} 
~{\rm d} {\Big |} ~~~,\cr
~~~ {\Tilde f}_{\underline a  \, \underline b}(x) 
&\equiv~  i \fracm 12  \e_{\underline a  \, \underline b}{}^{\underline c  \, 
\underline d} f_{\underline c  \, \underline d}(x) 
 ~~~~,\cr
{\rm d} {\Big |}  &\equiv~ {\rm d}(x) ~~~, \cr
D^2 W_{\a}  {\Big |}  &\equiv~ - i \, \pa_{\underline a} 
{\Bar \l}{}^{\Dot \a} (x) ~~~.} \label{eq:nfh}
\ee
In these equations, the quantity denoted by $f_{\underline a  \, 
\underline b}$ is exactly the same as that which appears in (\ref{eq:ntw}). 
It is the Maxwell field strength.  The fermionic field $\l_{\a}$ is known 
as the ``gaugino.'' After we evaluate (\ref{eq:nfn}), it will turn out that
${\rm d}$ is an auxiliary field. In principle at this stage we could 
follow a path analogous to that which led from (\ref{eq:hth}) to 
(\ref{eq:nty}). It is more prudent, however, for us to first discuss 
the case of Yang-Mills theory, evaluate the component action in this 
case and then retrieve the Maxwell theory as a special case.  

\subsection{Local Lie Algebraic Theories: Yang-Mills}

~~~~As is well known, all of the results in (\ref{eq:nsv}-\ref{eq:ntww}) can 
be generalized to non-abelian groups.  For the field theorist, the most 
fundamental quantities are fields. The simplest is the scalar field $\phi(x)$.  
If this field carries some representation of a Yang-Mills gauge group, we 
must add an index to it $\phi^{\rm I}(x)$. The abstract Lie algebraic generator 
$t_{\cal A}$ acts on the scalar field according to
\be
t_{\cal A} \, \phi^{\rm I}(x) ~=~ \big( {\tilde t}_{\cal A} \big)^{\rm I} 
{}_{\rm K} \, \phi^{\rm K}(x) ~\equiv~ [ \, t_{\cal A} ~,~ \phi^{\rm I}(x) \, ] ~~~.
\ee
In writing this equation, we adhere to the convention that $t_{\cal A}$ is
an operator which obeys a Leibnizian rule when acting on a product of
field variables, for example
\be
t_{\cal A} ( \, \phi^{\rm I}(x)  \, \phi^{\rm K}(x) \,) ~=~ (t_{\cal A} \, 
\phi^{\rm I}(x) ) \, \phi^{\rm K}(x)~+~ \phi^{\rm I}(x) \, 
(t_{\cal A} \, \phi^{\rm K}(x) ) ~~~.
\ee
On the other hand, the quantities $\big( {\tilde t}_{\cal A} \big)^{\rm I} 
{}_{\rm K}$ are a set of constants, i.e. a matrix.  The above assumptions have 
the following consequences,
\be 
t_{\cal A} t_{\cal B} \, \phi^{\rm I}(x) ~=~ \big( {\tilde t}_{\cal B} 
\big)^{\rm I} {}_{\rm K} \, t_{\cal A}\phi^{\rm K}(x)  ~=~ \big( {\tilde 
t}_{\cal B} \big)^{\rm I} {}_{\rm K} \big( {\tilde t}_{\cal A} \big)^{\rm 
K} {}_{\rm L}  \, \phi^{\rm L}(x) ~~~.
\ee
\be
\to ~ [\, t_{\cal A} ~,~  t_{\cal B}\, ]  \, \phi^{\rm I}(x) ~=~ - \, 
 \big( [\, {\tilde t}_{\cal A} ~,~  {\tilde t}_{\cal B}\, ] \big)^{\rm 
I} {}_{\rm K} \, \phi^{\rm K}(x) ~~~.
\ee
This last equation is consistent if 
\be
[\, t_{\cal A} ~,~  t_{\cal B}\, ]  ~=~ - \, i f_{\cal A \, \cal B}{}^{\cal C}
{t}_{\cal C}
~~~,~~~ \big( [\, {\tilde t}_{\cal A} ~,~  {\tilde t}_{\cal B}\, ] \big)^{\rm 
I} {}_{\rm K}  ~=~ i f_{\cal A\, \cal B}{}^{\cal C} \big( {\tilde t}_{\cal C}
\big)^{\rm I} {}_{\rm K} ~~~.
\ee

In many discussions of Yang-Mills theory, the definition of a covariant 
derivative takes the form (introducing the gauge field $A_{\underline a}
{}^{\cal I}$)
\be
\nabla_{\underline a} ~ \equiv~ {\pa}_{\underline a}  ~-~ i \, A_{\underline a}
{}^{\cal I} \, {\tilde t}_{\cal I} ~=~ {\pa}_{\underline a}  ~-~ i 
\, A_{\underline a}
~~~, 
\ee
where ${\tilde t}_{\cal I}$ is any matrix representation. The parameter of gauge
transformations has the form $ \lambda = \lambda^{\cal A} (x) {\tilde t}_{\cal A}$
so that the gauge transformation takes the form
\be
\Big( \nabla_{\underline a} \Big)' ~=~ \exp [ i \lambda ] \nabla_{\underline a}
\exp [ - i \lambda ] ~~ \to ~~  \, {A_{\underline a}}' ~=~ A_{\underline a} 
~-~i \, e^{i \lambda} ({\pa}_{\underline a} e^{- i \lambda} ) ~~~.
\label{eq:ett}
\ee
The field strength is defined by
\be
[ \,  \nabla_{\underline a}  ~,~  \nabla_{\underline b} \,] ~=~ - \, i f_{
\underline a \, \underline b} {}^{\cal I} {\tilde t}_{\cal I} ~\to ~
f_{\underline a \, \underline b} {}^{\cal I} ~=~ {\pa}_{\underline a}  
\, A_{\underline b} {}^{\cal I} ~-~ {\pa}_{\underline b}  
\, A_{\underline a} {}^{\cal I} ~-~ i \,  f_{\cal J \, \cal K}{}^{\cal I}
A_{\underline a} {}^{\cal J} \, A_{\underline b} {}^{\cal K} ~~~.
\ee
It is interesting to re-calculate $f_{\underline a \, \underline b} {}^{\cal I}$
using the abstract Lie algebra generator $t_{\cal A}$ in place of ${\tilde 
t}_{\cal A}$.
\be
\eqalign{i \, f_{\underline a \, \underline b}{}^{\cal I} {t}_{\cal I}  
&=~ [ \, {\pa}_{\underline a} ~-~ i \, A_{\underline a} {}^{\cal J} \,
{t}_{\cal J} \, , \, {\pa}_{\underline b} ~-~ i \, A_{\underline b}
{}^{\cal K} \, {t}_{\cal K} \, ] 
\cr
&=~ i \, ( {\pa}_{\underline a} \, A_{\underline b} {}^{\cal I} ~-~ 
{\pa}_{\underline b}  \, A_{\underline a} {}^{\cal I} ) t_{\cal I} ~-~ 
[ \,  A_{\underline a}{}^{\cal J} \, {t}_{\cal J} \, , \,  \, A_{\underline 
b} {}^{\cal K} \, {t}_{\cal K} \, ] \cr
&=~ i \, ( {\pa}_{\underline a} \, A_{\underline b} {}^{\cal I} ~-~ 
{\pa}_{\underline b} \, A_{\underline a} {}^{\cal I} ) t_{\cal I} 
- ~ A_{\underline a} {}^{\cal J} \, ({t}_{\cal J} \, A_{\underline
 b} {}^{\cal I} ) \, {t}_{\cal I} 
\cr
&{~~~~~}+ ~ A_{\underline b} {}^{\cal J} \, ({t}_{\cal J} \, A_{\underline 
a} {}^{\cal I} ) \, {t}_{\cal I} ~-~ A_{\underline a} {}^{\cal J} \, A_{
\underline b} {}^{\cal K} [ \, t_{\cal J} ~,~  t_{\cal K} \, ]
~~~. }\ee
This definition of the field strength will agree with the first one
if and only if the gauge field possesses a nonvanishing result under
the action of the abstract Lie algebra generator $t_{\cal J}$,
\be 
({t}_{\cal J} \, A_{\underline b} {}^{\cal I} ) ~=~ i \, 
f_{\cal J \, \cal K}{}^{\cal I}  A_{\underline b} {}^{\cal K} 
~~~. 
\ee
The fact that this condition is satisfied is equivalent to the statement
that a nonabelian gauge field is ``charged'' with regard to its own
gauge group.

\subsection{4D, N = 1 Supersymmetric Yang-Mills Theory}

~~~~We can now bring all of the experiences of the last few sections to
bare on solving the problem of finding the form of 4D, N = 1 
supersymmetric Yang-Mills theory. As an ansatz, we directly
commandeer the abelian solution,
\be
\eqalign{
&\G_{\Dot \a}{}^{\cal I} t_{\cal I} ~\equiv~  i \, e^{- \, {\Bar \O}^{
\cal I} t_{\cal I}} {\Bar D}{}_{\Dot \a}  e^{ \, {\Bar \O}^{\cal I} 
t_{\cal I}} ~~~,\cr
&\to ~ \G_{\a}{}^{\cal I} t_{\cal I} ~=~  i \, e^{- \, \O^{\cal I}
 t_{\cal I}} {D}{}_{\a}  e^{ \O^{\cal I}t_{\cal I}} ~~~,\cr
&\G_{\underline a} ~\equiv~ - i \, \Big[ \, {D}{}_{\a} \G_{\Dot \a} 
~+~ {\Bar D}{}_{\Dot \a} \G_{\a}
~-~ i \, [ \,  \G_{\a} \, , \,  \G{}_{\Dot \a} \, ]~ \Big]
~~~. }\ee
These are to be regarded as definitions of the Yang-Mills connection
superfield $\G_{\underline A}$. Due to these definitions, we find
$F_{\a \, \b}{}^{\cal I}  = F_{\a \, \Dot \b}{}^{\cal I}
= F_{\Dot \a \, \Dot \b}{}^{\cal I} = 0$. Using these, 
the 4D, N = 1 superspace Yang-Mills covariant derivatives
satisfy
$$
[\,   \nabla_{\a}\, \, , \, \, {\nabla }_{\b} \, \} ~=~ 0 ~~~,~~~
[\, {\Bar \nabla}{}_{\Dot \a}\, \, , \, \, {\Bar \nabla }_{\Dot \b} 
\, \} ~=~ 0 ~~~,~~~ [\,  \nabla_{\a}\, \, , \, \, {\Bar \nabla}{}_{\Dot 
\a} \, \} ~=~ i \nabla_{\underline a} ~~~, 
$$
$$  
[ \, {\nabla}_{ \a} ~,~ \nabla_{\underline b} \, ] ~=~
C_{\a \, \b} \, {\Bar W}{}_{\Dot \b}{}^{\cal I} \, t_{\cal I} ~~~,~~~
[ \, {\Bar \nabla}{}_{\Dot  \a} ~,~ \nabla_{\underline b} \, \} ~=~ 
C_{\Dot \a \, \Dot \b} \, {W}{}_{\b}{}^{\cal I} \, t_{\cal I} ~~~.
$$
\be
[ \,  \nabla_{\underline a}  \,\, , \,\, \nabla_{\underline b}  
\, \} ~=~ - i \, f_{\underline a \, \underline b} {}^{\cal I}\, t_{\cal I}
~~~.
\label{eq:nyvv}
\ee
(The reader is warned that the nonabelian version of $W_{\a}{}^{\cal I}$
is {\it {is}} defined by this equation {\it {not}} the abelian expression 
in (\ref{eq:nrv}).) 

With all of these results, we can now calculate the form of the
nonabelian version of (\ref{eq:nfn}). We essentially follow the same
procedure as in (\ref{eq:nty}) (here we suppress the ${\cal I}$-index for 
simplicity)
\be
\eqalign{
 \int d^6 z ~ (\, \frac 14 \, W^{\a} W_{\a} \, ) &=~ \fracm 18 \,  
\int d^4 x \nabla^{\b} \nabla_{\b} \,  W^{\a} W_{\a} ~  {\Big |} ~~~ \cr
&=~ - \, \fracm 14 \,  \int d^4 x ~\nabla^{\b}[ W^{\a} \,  \nabla_{\b} 
W_{\a} ]  ~ 
{\Big |} \cr
&=~ - \, \fracm 14 \, \int d^4 x ~[\, (\nabla^{\b}W^{\a} ) \, (\nabla_{\b} 
W_{\a}) \,-\, W^{\a} \,  (\nabla^{\b} \,\nabla_{\b} W_{\a} ) ~] {\Big |} 
~~~.}\label{eq:nvx}
\ee
A feature that may be a bit unexpected is that the superspace
measure above was replaced by superspace Yang-Mills covariant
derivatives $ d^6 z \to d^4 x \nabla^2$. This is valid because
the integrand must be a Yang-Mills invariant and thus any
apparent dependence on the spinorial connection is illusory.

The definitions of component fields for the nonabelian case are
obtained as for the U(1) theory with the appropriate generalizations 
to equation (\ref{eq:nfh}) $D_{\a}  \to  \nabla_{\a} ,~ W_{\a}  \to  
W_{\a}{}^{\cal I} ,~ \l_{\a}  \to  \l_{\a}{}^{\cal I},~ f_{
\underline a \, \underline b }  \to f_{\underline a \, 
\underline b }{}^{\cal I} ,~ {\rm d}  \to  {\rm d}{}^{\cal I} ~~~,
$ so that
\be
\eqalign{
\nabla_{\a} W_{\b}{}^{\cal I} {\Big |} &\equiv~  \fracm 12 \,
 C^{\Dot \k \, \Dot \l}
\,[~ f_{\a \Dot \k  \, \b \Dot \l }{}^{\cal I} ~+~
{\Tilde f}_{\a \Dot \k  \, \b \Dot \l}{}^{\cal I} ~]
 ~-~ i C_{\a \,\b} ~{\rm d}{}^{\cal I}  \cr
\nabla^2 W_{\a}{}^{\cal I}   {\Big |}  &\equiv~ - i \, \nabla_{\underline a} 
{\Bar \l}{}^{\Dot \a} {}^{\cal I} ~~~.}
\ee
Therefore, the action quadratic in $W_{\a}$ becomes
\be
\eqalign{
 \int d^6 z ~ (\, \frac 14 \, W^{\a} W_{\a} \, ) &=~ 
 - \, \fracm 14  \,\int d^4 x ~[~ \fracm 14 \,( f^{\underline a 
\, \underline b}{}^{\, {\cal I}} f_{\underline a \, \underline 
b}{}^{\, {\cal I}} ~+~ f^{\underline a \, \underline 
b}{}^{\, {\cal I}} {\Tilde f}_{\underline a \, \underline b}{}^{\, {\cal I}})
\cr
&{~~~~~~~~~~~~~~~~~~~~}~+~ i 2 \,  \l^{\a \, \cal I} \nabla_{\underline 
a} {\Bar \l}{}^{\Dot \a \, \cal I} ~-~ 2 \, {\rm d}^{\cal I} {\rm 
d}{}^{\cal I} ~] ~~~.}\ee
So that finally the component form of (\ref{eq:nfn}) is
takes the form
\be
{\cal S}_{SYM} ~=~   \int d^4 x ~ [~
- \fracm 18 \, f^{\underline a \, \underline b}{}^{\, {\cal I}}
\, f_{\underline a \, \underline b}{}^{\, {\cal I}} \, - \, i \l^{\a}
{}^{\, {\cal I}} \nabla_{\underline a} {\Bar \l}{}^{\Dot \a \, {\cal I}} 
\, + \, {\rm d}^{{\cal I}} {\rm d}^{{\cal I}} ~]  ~~~.
\label{eq:ofn} \ee 

\subsection{Superfield Gauge Transformations and Maurer-Cartan 
Forms}

~~~~The main difference in the gauge transformation law of ordinary
Yang-Mills theory and the superspace theory is that the latter
takes a highly non-linear form in terms of the gauge superfield $V
= V^{\cal I}t_{\cal I}$
\be
( e^V )' ~=~ e^{i {\Bar \L}} \, e^V \, e^{- \L} ~~~,
\ee
since in the non-abelian case this equation cannot be simplified we
must treat it with some care.  The reason for this is that once we write
a gauge fixing term, its gauge variation determines the ghost-antighost
propagator.  For infinitesimal $\L $ we have
\be
\d_G \, e^V  ~=~ i\, [ ~  {\Bar \L} \, e^V ~-~  e^V \, \L ~] 
 ~~~. \label{eq:exw}
\ee
Now we have the obvious identity $[ V \, , \, e^V ] = 0$, and applying 
$\d_G $ to both sides of this equation yields
\be
\eqalign{
&[ \d_G \, V \, , \, e^V ] ~=~ - \, [ V \, , \, \d_G e^V ] ~~~\to \cr
&  e^{-\fracm 12 V} \, ( \d_G \, V  ) \, e^{\fracm 12 V}  ~-~ 
\, e^{\fracm 12 V} \,(  \d_G \, V ) \, e^{- \fracm 12 V} ~=~ - \, [ V \, , \, 
e^{-\fracm 12 V} \,( \d_G e^V ) \,  e^{- \fracm 12 V} \,] ~~~,
} \label{eq:nxw}
\ee
after multiplying front and back  by $exp[- \fracm 12 V]$. We next observe
that (\ref{eq:exw}) and (\ref{eq:nxw}) can be used to show
\be \eqalign{
e^{-\fracm 12 V} \,( \d_G e^V ) \,  e^{- \fracm 12 V}  &-~ 
\, e^{\fracm 12 V} \,(  \d_G \, V ) \, e^{- \fracm 12 V} \cr
&{~~~~}=~  i\, [\, V ~,~  ~ e^{-\fracm 12 V} {\Bar \L} \, e^{\fracm 12 V} ~-~  
e^{\fracm 12 V} \, \L  \, e^{-\fracm 12 V} ~] 
~~~.  }\label{eq:nxh}
\ee

At this stage it is convenient to introduce a notational device. We can
use the symbol $L_V X \equiv [ V \, , \, X]$ defined to act on any two
matrices $V$ and $X$. Utilizing this we find
\be
\eqalign{ {~~~~} 
 e^{-\fracm 12 V} \, ( \d_G \, V ) \, e^{\fracm 12 V} &=~ exp [ - \fracm 
12 L_V] \, ( \d_G \, V) ~~~,~~~ \cr
 e^{\fracm 12 V} \, ( \d_G \, V ) \, e^{-\fracm 12 V} &=~ exp [\fracm 12 
L_V] \, ( \d_G \, V) ~~~, \cr
 e^{-\fracm 12 V} \, {\Bar \L} \, e^{\fracm 12 V} &=~ exp [ - \fracm 12 
L_V] \,  {\Bar \L}  ~~~, ~~~ \cr
 e^{\fracm 12 V} \,\L \, e^{-\fracm 12 V} &=~ exp [\fracm 12 L_V]
\,\L ~~~,
}\ee 
where these equations can be shown to be valid by use of Taylor's theorem
with respect to $V$. Using these in (\ref{eq:nxh}) then yields,
\be
\eqalign{
sinh(\fracm 12 L_V)  \, ( \d_G \, V )  &=~ - \, i\fracm 12  L_V \, [ \, 
exp(\fracm 12 L_V) \, \L ~-~ exp(- \fracm 12 L_V) \, {\Bar \L } ~] ~~ 
 \cr
&=~ - i \, \fracm 12 L_V \,  sinh(\fracm 12 L_V) \,( \,  \L \, + 
\,  {\Bar \L }\, ) \cr
&{~~~~~}- i \, \fracm 12 L_V  \, cosh(\fracm 12 L_V) \,( \,  
\L \, - \,  {\Bar \L } \,)  
~~~.}
\ee
As a final step, we formally divide this by $sinh(\fracm 12 L_V) $
\be
\eqalign{
\d_G \, V  &=~   - i \, \fracm 12 L_V \, [~ ( \,  \L \, + 
\,  {\Bar \L }\, ) \, + \, coth(\fracm 12 L_V) \,( \,  \L \, - \,  
{\Bar \L } \,)  ~] ~~~.
}\ee

Yang-Mills theories in superspace depend on the pseudoscalar pre-potential 
$V$ superfields which appear ``hidden inside'' the spinorial superconnections
as
\be
\G_{\a} ~=~ i\, e^{-  \O} (D_{\a} e^{ \O} \, ) ~~\to~~
\nabla_{\a} ~=~ D_{\a} ~-~ i \, \G_{\a} ~=~ e^{-\O} D_{\a} 
e^{ \O} ~~~.
\ee
(Recall also $\O = \fracm 12 ( V + i W).$)  This final equation is meant to 
indicate that the factor of $D_{\a}$ acts on everything that follows it.  
The operator ${\Bar \nabla}{}_{\Dot \a}$ is just the hermitian conjugate 
and $\nabla_{\underline a}$ is given by $\nabla_{\underline a} ~=~ - i 
[ \nabla_{\a} \, , \, {\Bar \nabla}{}_{\Dot \a} \}$.  From its definition, 
it should be clear that $\G_{\a}$ can be expressed as
\be
\G_{\a} ~=~ \, (D_{\a} {\O}^{\cal A} ) \, {\rm L}_{\cal A} {}^{\cal I} 
\, {\tilde t}_{\cal I} ~~~, \label{eq:nxg}
\ee
for some ``matrix'' of functions ${\rm L}_{\cal A} {}^{\cal I}$. Let us
explore the significance of these objects. The rigid gauge transformation
may be written in the form
\be
U(\a) ~=~ exp[ i \, \a^{\cal I} \, {\tilde t}_{\cal I} \,] ~~~,
\ee
whether in Minkowski space or superspace.  If ${\tilde t}_{\cal I}$ are
the generators of some compact Lie group, then all the elements in the
group may be represented by allowing the parameters $ \a^{\cal I}$ to
take on a finite range of values. Since the parameters are allowed to
vary in this range, it makes sense to consider the quantities
\be
{\rm L}^{\cal J} ~\equiv~ n_0 tr \Big\{ \, {\tilde t}^{\cal J}  \,  
exp[ - i \, \a^{\cal I} \, {\tilde t}_{\cal I} \,]\,  d \, exp[ i \, 
\a^{\cal I} \, {\tilde t}_{\cal I} \, ] \Big\} ~=~  n_0 tr \Big\{ \, 
{\tilde t}^{\cal J} U^{-1} d \, U \Big\}  ~~~.
\ee
From this definition it is clear that under ${\rm L}^{\cal J}$ is
invariant under the re-definition
\be
U(\a) ~~\to~~ {\Tilde U}(\b)  \, U(\a) ~~~, 
\label{eq:nvn}
\ee
as long as $\b$ is a constant. However, as a consequence of the
chain rule we have
\be
d U ~=~ d {\a}^{\cal K} \, \partder{U}{{\a}^{\cal K}}  ~~\to
~~ {\rm L}^{\cal J} ~=~  d {\a}^{\cal K}  n_0  tr \Big\{ \, {\tilde 
t}^{\cal J}  \, U^{-1} \, \partder{U}{{\a}^{\cal K}}  \Big\}
~=~  d {\a}^{\cal K}  {\rm L}_{\cal K}{}^{\, \cal J} ~~~.
\label{eq:nvw}
\ee
The object denoted by ${\rm L}_{\cal K}{}^{\, \cal J}$ here is
exactly the same as that in (\ref{eq:nxg}). However, the discussion 
immediately above makes it clear that this object is intrinsic to the
structure of a Lie group. In fact, it is known as the left-invariant
Maurer-Cartan form. Upon taking the determinant of this matrix,
multiplying by the volume element constructed from and integrating 
over the allowed range of $\a^{\cal I}$, we can use it to calculate
the intrinsic hyper-volume of the Lie group.

Starting from the identity $[ \a^{\cal I} {\tilde t}_{\cal I}
\, , \, U] = [ \a \, , \, U] =0$, we can repeat part of our derivation 
of the explicit form of the $\d_G V$ variation,
\be
\eqalign{
[ d \a \, , \, U(\a) ] ~=& - \, [ \a \, , \, d \, U(\a) ] ~~~\to \cr
&  U(-\fracm 12 \a) \, ( d \a  ) \, U(\fracm 12 \a)  \,-\, 
\, U(\fracm 12 \a) \,(  d \a ) \, U(- \fracm 12 \a) \cr
&{~~~~~~~~~~~~~~}=\, - \, [ \a \, , \, 
U(-\fracm 12 \a) \,( d U ) \,  U(- \fracm 12 \a) \,] ~~\to \cr
&{~~~~} sinh(\fracm 12 L_{\a}) \, d \a ~=~ \fracm 12 L_{\a} \,
U(-\fracm 12 \a) \,( d U ) \,  U(- \fracm 12 \a) ~~~,
}\ee
after multiplying front and back by $U(-\fracm 12 \a)$. We may
perform the formal `division' by $L_{\a}$ to write
\be
U^{-1} d U ~=~  U(- \fracm 12 \a) [ \,  g_1(\fracm 12 L_{\a})\,
d \a \, ] U(\fracm 12 \a) ~~~,
\ee
where the function $g_1(x) =  sinh(x)/ x$. Substituting this into
(\ref{eq:nvw}) leads to the identification,
\be
\eqalign{
{\rm L}_{\cal K}{}^{\, \cal I}(\a) &=~ i n_0 tr \Big\{ \, [ exp[\fracm 
12 L_\a] \,{\tilde t}^{\cal I} ] \, [ \,  g_1(\fracm 12 L_{\a})\,
 {\tilde t}_{\cal K} \, ] ~ \Big\} \cr
&=~ i n_0 tr \Big\{ \, {\tilde t}^{\cal I} [ g_2(\fracm 12 L_{\a})\, 
{\tilde t}_{\cal K} \, ] ~ \Big\}  ~~~,
}\ee
where $g_2(x) =  e^{-x} sinh(x)/ x$ and the constant $n_0$ is chosen 
so that ${\rm L}_{\cal K}{}^{\, \cal I}(0) = \d_{\cal K}{}^{\cal I}$. 

The invariance of ${\rm L}_{\cal K}{}^{\, \cal I}$ under (\ref{eq:nvn}) has a
very interesting extension to the supersymmetric case in (\ref{eq:nxg}). The
spinorial connection there is seen to be invariant under
\be
e^{\O} ~~\to ~~ e^{i{\Bar \L}} \, e^{\O} ~~~,~~~ D_{\a} {\Bar \L} ~=~0
~~~.
\ee
On the other hand, the superspace Yang-Mills covariant derivative
in (\ref{eq:nyvv}) is seen to transform in the usual covariant way under
\be
e^{\O} ~~\to ~~ e^{\O} \,  e^{i K} ~~~,~~~ K ~\equiv~ K^{\cal A} 
{\tilde t}_{\cal A} ~~~, ~~~  K^{\cal A} ~=~ ( K^{\cal A} )^*
~~~.
\ee
Thus, we learn that the full set of transformations under which
supersymmetric Yang-Mills theory is invariant consists of the
transformations
\be
(e^{\O})'  ~~= ~ e^{i{\Bar \L}} \, e^{\O} \, e^{i K}
~~~.
\ee
Superspace Yang-Mills theories (and every other gauge theory in 
superspace) possesses two {\it {distinct}} local symmetry groups. 
One is parametrized by chiral superfields (i.e. $\L^{\cal I}$
sometimes called the ``$\L$-gauge group'') and one by real superfields 
(i.e. $K^{\cal I}$ sometimes called the ``$K$-gauge group''). The 
transformations with respect to $K^{\cal I}$ can be used to 
``algebraically gauge away'' the purely imaginary part of $\O^{\cal 
I}$. Alternately, we may regard the superfields $V^{\cal I}$ as 
being defined by
\be
e^V ~\equiv~ e^{\O} \,  (e^{\O})^{\dag} ~~~.
\ee

\subsection{The Wess-Zumino Gauge}

~~~~Let us consider another aspect of these theories and for simplicity 
only the abelian case\footnote{The results for the non-abelian case 
are the same although the proof is more difficult.}.  Here the gauge 
transformation is $\d_G V = - i (\L \, - \,{\Bar \L})$ and components 
are defined by
\be
\eqalign{
c(x) &\equiv~ V {\Big |} ~~~,~~~ \chi_{\a}(x)  ~\equiv~ D_{\a} V {\Big |} 
~~~,
~~~ {\Bar \chi}_{\Dot \a}(x)  ~\equiv~ {\Bar D}{}_{\Dot \a} V {\Big |} 
~~~, \cr
M(x) &\equiv~ D^2 V {\Big |} ~~~,~~~ {\Bar M} ~\equiv~{\Bar D}{}^2  V 
{\Big |} ~~~, ~~~A_{\underline a} (x) ~\equiv~ - \, \fracm 12 [ ~ D_{\a} 
\, , \, {\Bar D}{}_{\Dot \a}~ ] \, V {\Big |} ~~~, \cr
\l_{\a}(x)  &\equiv~ i {\Bar D}{}^2  D_{\a} V {\Big |} ~~~, ~~~ {\Bar 
\l}_{\Dot \a}(x)  ~\equiv~ - i D^2 {\Bar D}{}_{\Dot \a} V {\Big |} ~~~,
~~~ {\rm d}(x) ~\equiv~ \fracm 12 D^{\a} {\Bar D}{}^2 D_{\a} V {\Big |} 
~~~.}\ee
The components of the gauge parameter can similarly be defined
\be
\L_1(x) ~\equiv~ \L {\Big |} ~~~, ~~~ \L_{\a} ~\equiv~ D_{\a} 
\L {\Big |} ~~~, ~~~ \L_2 ~\equiv~ D^2 \L {\Big |} ~~~.
\ee
So that the components of V change as
\be
\eqalign{
\d_G  c & =~ - i (\L_1 \, - \, {\Bar \L}_1)~~~,~~~ \d_G \chi_{\a}  ~=~ 
\L_{\a} ~~~, \cr 
\d_G M & =~ - i \L_2  ~~~,~~~ \d_G A_{\underline a} ~ =~ \fracm 12 
\pa_{\underline a} (\L_1 \, +\,  {\Bar \L}_1 )
~~~, \cr
 \d_G  \l_{\a}  & =~ 0 ~~~, ~~~ \d_G 
{\rm d} ~ =~ 0 ~~~.
}\ee
These equation show why at the component level one often finds the statement 
that the component fields $( A_{\underline a}, \, \l_{\a} , \, {\rm d}  )$ 
form the 4D, $N=1$ gauge vector multiplet. This statement is only true in a 
special gauge where $- i (\L_1  -  {\Bar \L}_1)$ is used to set $c=0$, 
$\L_{\a}$ is used to set $\chi_{\a} = 0$ and $\L_2$ is used to set $M=0$.  This
uses all of the     (the parameter superfield) except $\fracm 12 (\L_1  + {\Bar 
\L}_1 )$ which is the ordinary gauge parameter.  This special gauge is called 
``the Wess-Zumino gauge \cite{7}"  or ``WZ gauge.''

Now let us start in this gauge i.e. ($c=0$, $\chi_{\a} = 0$, $M=0$) and
note that the WZ-gauge breaks manifest supersymmetry! To see this we observe 
the supersymmetry variation of $\chi_{\a}$ is
\be
\d_Q \chi_{\a} ~=~ i \e_{\a} M ~-~ {\Bar \e}{}^{\Dot \a} \,(\, \fracm 12 
\pa_{\underline a} c ~+~ A_{\underline a} \, ) ~~~.
\ee
So if we start in the gauge where $(c,\, \chi_{\a},\, M)$ vanish, after
applying $\d_Q$ we find a ``$\chi_{\a}$'' ($\equiv - {\Bar \e}{}^{\Dot \a} 
A_{\underline a}$).  To prevent this from happening we can perform a gauge 
transformation with $\L_{\a} \equiv i {\Bar \e}{}^{\Dot \a} A_{\underline a}$ 
but this gauge transformation cause ``C" and ``M" to be non-zero after
applying $\d_G$.  These can be removed by picking $\L_2 \equiv  - 
{\Bar \e}{}^{\Dot \a} {\Bar \l}_{\Dot \a} $.  So the net result is if 
we start with only $( A_{\underline a}, \, \l_{\a} , \, {\rm d}  )$ 
we must apply
\be
\d_Q^\prime ~=~ \d_Q ~+~ \d_G ({\Tilde \L}) ~~~,
\ee
where the only components of ${\Tilde \L}$ are the ones described above.  
Clearly this will lead to a result of the form
\be
[ \,\,  \d_Q^\prime (\e_1 ) \, , \,  \d_Q^\prime (\e_2 ) ~ ] ~=~ 
\d_P ~+~ \d_G ~~~,
\ee
i.e. it must contain some dependence on the fields of ${\Tilde \L}$. 
This is an example of another statement that one often finds in the
literature on supersymmetrical gauge theories. Namely it is often
said that the closure of a supersymmetry algebra is dependent on
the fields or possibly their equations of motion. This is true
only if we work in the WZ gauge. On the other hand, if we always 
choose to work solely with the component fields $( A_{\underline a}, 
\, \l_{\a} , \, {\rm d}  )$ this is equivalent to picking the WZ gauge.
This is also why when using components, the closure of supersymmetry 
algebras almost always includes gauge transformations.  The operator 
$\d_G ({\Tilde \L})$ is called ``the compensating gauge transformation" 
and it must be introduced to maintain the WZ-gauge. 

\subsection{4D, N = 1 Supersymmetric Yang-Mills-Matter Coupling} 

~~~~One of the most studied 4D theories is quantum electrodynamics
(QED). It is of interest to construct the 4D, $N$ = 1 analog
of this. Prior to the demonstration of this theory, it is convenient
to explicitly introduce ``covariantly chiral'' superfields. The ordinary
definition of a chiral superfield was via the restriction
${\Bar D}{}_{\Dot \a} \Phi = 0$. On the other hand, a covariant 
definition was introduced in (\ref{eq:nwi}) ${\Bar \nabla}{}_{\Dot \a} 
\Phi_C = 0$ we use the label $C$ to denote covariant chirality.
In writing actions of chiral superfield matter coupled to the Yang-Mills
superfields, it is always most convenient to use covariantly chiral
superfields. Using the solution to the constraints on the
superspace covariant derivative, we see (working in the
partial choice of gauge where $W = 0$)
\be
\eqalign{
\nabla_{\a} &=~ D_{\a} ~-~ i \G_{\a}{}^{\cal I} t_{\cal I} ~=~
e^{ \fracm 12 V^{\cal I} t_{\cal I} } D_{\a} e^{- \fracm 12 
V^{\cal I} t_{\cal I} } 
~~~, \cr
{\Bar \nabla}{}_{\Dot \a} &=~ {\Bar D}{}_{\Dot \a} ~-~ i \G_{
\Dot \a}{}^{\cal I} t_{\cal I} ~=~
e^{ \fracm 12 V^{\cal I} t_{\cal I} } {\Bar D}{}_{\Dot \a} e^{- 
\fracm 12 V^{\cal I} t_{\cal I} } 
~~~.} \label{eq:nhx}
\ee
In turn this has the implication
\be
{\Bar \nabla}{}_{\Dot \a} \Phi_C ~=~ 0 ~~\to ~~ 
{\Bar D}{}_{\Dot \a} \, \Big( \, e^{- 
\fracm 12 V^{\cal I} t_{\cal I} } \Phi_C \, \Big) ~=~ 0 ~~~,
\ee
which clearly shows the quantity in the parenthesis is an
ordinary chiral superfield. Thus we may write
\be
\Phi ~=~  e^{- \fracm 12 V^{\cal I} t_{\cal I} } \Phi_C ~~~,
\ee
and if we use this definition, the explicit factors involving
the $V$ superfield in (\ref{eq:nxt}) ``disappear.'' In deriving results
this makes for a major simplification.

As one more step leading up the the derivation of the form of
chiral superfields coupled to Yang-Mills superfields, we observe
it is an exercise to show that the equations in (\ref{eq:nyvv}) imply
\be
\eqalign{
{\nabla}_{\a} {\Bar \nabla}{}^2 &=~ {\Bar \nabla}{}^2 {\nabla}_{\a} 
~-~ i \nabla_{\underline a} {\Bar \nabla}{}^{\Dot \a} ~+~ i \, W_{\a} 
{}^{\cal C} t_{\cal C} ~~~, \cr
{\Bar \nabla}_{\Dot \a} {\nabla}{}^2 &=~ {\nabla}{}^2 {\Bar \nabla}_{
\Dot \a}  ~-~ i \nabla_{\underline a} {\nabla}^{\a} ~+~ i \, {\Bar 
W}_{\Dot \a} {}^{\cal C} t_{\cal C} ~~~,  \cr
{\nabla}{}^2 {\Bar \nabla}{}^2 &=~ {\Bar \nabla}{}^2 {\nabla}{}^2 
~-~ i \fracm 12 \, \nabla_{\underline a} ( \, {\nabla}^{\a}
{\Bar \nabla}{}^{\Dot \a} \, - \, {\Bar \nabla}{}^{\Dot \a} {\nabla}^{\a}
\, ) \cr 
&{~~~~~~}+~  i \,  ( \, W^{\a \, {\cal C}} \,  {\nabla}_{\a}
 \, - \, {\Bar W}{}^{\Dot \a \, {\cal C}}{\Bar \nabla}{}_{\Dot \a}
\, ) \, t_{\cal C} ~+~ i \, \fracm 12 \, (  {\nabla}^{\a}
W_{\a}{}^{\, {\cal C}} ) \, t_{\cal C} ~~~. 
}\ee

If we assume that the matter action ${\cal S}_M$ describes Dirac
fermions then it has been almost universally accepted that it must
be of the form
$$
{\cal S}_{M} ~=~ \int d^4 x \, d^2 \q \, d^2 {\bar \q} \, 
\Big[ \, {{\Bar \Phi}_+} \, {\Phi_+} \,+\, {{\Bar \Phi}_-} \, {\Phi_-} 
\, \Big] {~~~~~~~~~~~~~~~~~~~~~~~~~}
~~$$
\be
+\, \, \Big[ \, \int d^4 x \, d^2 \q \, \, W( {{\Phi}_+}, \, 
{\Phi_-} )  \,\,+\,\, {\rm {h.\, c.}} ~\,  \Big] ~~~. 
\ee
In writing the action as above, we have made use of covariantly chiral
superfields defined by ${\Bar \nabla}_{\Dot \a} \Phi_+ = {\Bar 
\nabla}_{\Dot \a} \Phi_- = 0$. In terms of components fields this 
becomes
\be
\eqalign{
{{\cal S}}_{M} ~=~ \int d^4 x ~ \Big[  &- \frac 12 (\nabla^{
\underline a} {\Bar A_+} \, )  (\nabla_{\underline a} A_+ \, ) 
 ~-~ \frac 12 (\nabla^{\underline a} {\Bar A_-} \, ) (\nabla_{\underline 
a} {A_-} \, ) ~-~ i {\Bar \psi}_+{}^{\ad} \nabla_{\underline a} 
{\psi}^{\a}_+ \cr
&-~ i {\Bar \psi}{}^{\ad}_- \nabla_{\underline a} {\psi}_-{}^{\a} ~
+~ {\Bar F_+} F_+ ~+~ {\Bar F_-} F_-  \cr
&-~ [ \, i \,  \l^{\a \, {\cal C}}\, [  ( t_{\cal C} \, 
{\psi}_{+ \, \a} )\, A_+ \, +\,  ( t_{\cal C} \, {\psi}_{- \, \a} 
)\, {A_-} \, ] ~+~ {\rm {h. \, c.}} ~]  \cr 
&-~ [ \, \fracm 12 {W}_{,+ ,+} (A_+,\, A_-)  {\psi}_+{}^{\a}
 {\psi}_{+ \, \a} ~+~ {W}_{,+}(A_+ , \, A_-) F_+ ~+~ {\rm {h. 
\, c.}} \,] ~ \cr
&-~ [ \, \fracm 12 {W}_{,- ,-} (A_+,\, A_-)  {\psi}_-{}^{\a}
 {\psi}_{- \, \a} ~+~ {W}_{,-}(A_+ , \, A_-) F_- ~+~ {\rm {h. 
\, c.}} \,] ~ \cr
&-~ [ \,  {W}_{,+, -} (A_+, \, A_-)  {\psi}_+{}^{\a}
 {\psi}_{- \, \a} ~+~  {\rm {h. \, c.}} \, \,] ~ \cr
&+~ {\rm d}{}^{\cal C} \, [ \,  ( t_{\cal C} {\Bar A_+} ) A_+ \, + \, 
 ( t_{\cal C} {\Bar A_-} ) {A_-} \, ] ~ \Big] ~~~.
}\ee
In this expression, ${W}_{,+}$ denotes differentiation of $W$
with respect to $A_+$, ${W}_{,-}$ denotes differentiation of $W$
with respect to $A_-$, etc. The derivation of this result begins
with the expected replacement $\int d^8 z \to \int d^4 x
\nabla^2 {\Bar \nabla}{}^2$ which is followed by acting on
the K\" ahler potential term as in (\ref{eq:nyo}-\ref{eq:ntv}). 
However, care must be taken with regard to `moving' the derivatives 
past one another. This is where the commutators in (\ref{eq:nyvv})
are useful.

Once again there are seen to be auxiliary fields present whose
equations of motion are purely algebraic. The algebraic equations
can be solved to express the auxiliary fields in terms of the
propagating fields. This yields the following on-shell action
\be
\eqalign{
{\cal S}_{on-shell} ~=~  \int d^4 x ~ \Big[  &- \frac 12 (\nabla^{
\underline a} {\Bar A_+} \, )  (\nabla_{\underline a} A_+ \, ) 
 ~-~ \frac 12 (\nabla^{\underline a} {\Bar A_-} \, ) (\nabla_{\underline 
a} {A_-} \, )  \cr
&-~ i {\Bar \psi}_+{}^{\ad} \nabla_{\underline a} 
{\psi}^{\a}_+
~-~ i {\Bar \psi}{}^{\ad}_- \nabla_{\underline a} {\psi}_-{}^{\a} ~
-~ U(A_+,\, A_-)   \cr
&-~ [ \, i \,  \l^{\a \, {\cal C}}\, [  ( t_{\cal C} \, 
{\psi}_{+ \, \a} )\, A_+ \, +\,  ( t_{\cal C} \, {\psi}_{- \, \a} 
)\, {A_-} \, ] ~+~ {\rm {h. \, c.}} ~]  \cr 
&-~ [ \, \fracm 12 {W}_{,+ ,+} (A_+,\, A_-)  {\psi}_+{}^{\a}
 {\psi}_{+ \, \a} ~+~ {\rm {h. \, c.}} \,] ~ \cr
&-~ [ \, \fracm 12 {W}_{,- ,-} (A_+,\, A_-)  {\psi}_-{}^{\a}
 {\psi}_{- \, \a} ~+~ {\rm {h. \, c.}} \,] ~ \cr
&-~ [ \,  {W}_{,+, -} (A_+, \, A_-)  {\psi}_+{}^{\a}
 {\psi}_{- \, \a} ~+~  {\rm {h. \, c.}} \, \,]  ~ \Big] ~~~,
}\ee
where the potential for the spin zero fields takes the form
\be
\eqalign{
U(A_+,\, A_-) &=~ | W_{,+}(A_+ , \, A_-) \, 
|^2 ~+~ | W_{,-}(A_+ , \, A_-) \, |^2  \cr
&{~~~}~+~  \fracm 34 | \,  ( t_{\cal C} {\Bar A}_+ ) A_+ \,\, + \,\, 
 ( t_{\cal C} {\Bar A}_- ) A_-   \, |^2 ~
~~~. ~~
}\ee
For the U(1) theory, after application of the Lie algebra generators 
this becomes,
\be
\eqalign{
U(A_+,\, A_-) &=~ | W_{,+}(A_+ , \, A_-) \, 
|^2 ~+~ | W_{,-}(A_+ , \, A_-) \, |^2  \cr
&{~~~}~+~  \fracm 34 g^2 \Big[\,\, | \, {\Bar A}_+ |^2 \,\, - \,\, 
| A_- |^2   \, \Big] ~~~, ~~
}\ee
where we have introduced the gauge coupling constant $g$ and used 
the fact that $\Phi_-$ carries the opposite charge from that of 
$\Phi_+$.   The usual mass of the electron is introduced via the 
choice $W = m_e \Phi_+ \Phi_-$. The two first terms in this potential 
then describe mass terms for the four spin-0 fields.  Following the 
discussion in section \ref{subsec:wpq}, it can be seen that this 
action contains a Dirac field which transforms as 
\be
\Big[ \psi_{(\a)}(x) \Big]' ~=~ 
\left(\begin{array}{cc}
 exp[i g \L | ] & 0 \\ 0 & exp[i g {\Bar \L} |] 
\end{array}\right) 
\left(\begin{array}{c}
 D_{\a} \Phi_+ {\Big |} \\  {\Bar D}_{\Dot \a} {\Bar \Phi}_- {\Big |} 
\end{array}\right) 
~~~,
\label{eq:nni}
\ee 
under the local U(1) symmetry of the model.   For the gauge group 
U(1), the supersymmetric Yang-Mills action in (\ref{eq:ofn}) has a leading 
term which is exactly the Maxwell action.  The next term describes 
a massless (free for U(1) since $t A_{\underline a} = 0 \to t V = 0 
\to t \l_{\a} = 0$) spinor called the ``photino.''

\subsection{Other 4D, N = 1 Supersymmetric Yang-Mills Terms}

~~~~In the case of abelian groups or subgroups, the following superfield
action is gauge invariant,
\be
{\cal S}_{F-I} ~=~ \int d^4 x \, d^2 \q d^2 {\bar \q} ~ V ~=~ 
 \int d^4 x \, {\rm d} ~~~.
\ee
This is known as a Fayet-Iliopoulos term and it can become
important in considerations of symmetry breaking.

In our effort to describe supersymmetric QED, we have rushed past 
something interesting.  Notice that in order to write the supersymmetric
extension of the Yang-Mills (and Maxwell) action, we had to make use of 
the real part of (\ref{eq:nvx}). It is an obvious question to ask about the
imaginary part of this quantity?  The imaginary part of the action 
quadratic in $W_{\a}$ becomes
\be
\eqalign{
\int d^6 z \,[ \, i  (\, \frac 14 \, W^{\a} W_{\a} \, ) \, + \, 
{\rm {h.\, c.}}  \, ] &=~  \int d^4 x ~[~ \fracm 18 \,
\e^{\underline a  \underline b  \underline c  \underline d}
f_{\underline a  \underline b}{}^{\, {\cal A}} { 
f}_{\underline a  \underline b}{}^{\, {\cal A}}\, + \, \nabla_{\underline 
a} ( \l^{\a \, \cal A} {\Bar \l}{}^{\Dot \a \, \cal A} )
 ~] \cr
&\equiv~  \int d^4 x ~ \, \pa_{\underline a} \, [~ \fracm 
1{48} \, \e^{\underline a \, \underline b \, \underline c \, 
\underline d} X_{\underline a \, \underline b \, 
\underline c}^{CS} ~+~  ( \l^{\a \, \cal A} {\Bar \l}{}^{\Dot 
\a \, \cal A} ) ~] 
~~~.}\ee
On the last line above, we have introduced a quantity
$X_{\underline a  \underline b \underline c}^{CS}$ known
as the ``Chern-Simons'' three form. This quantity has
topological significance. Its integrated form is known as
the ``instanton number.''  From the way that it appears
above, it is clearly a pure surface term.  However, it is
also clear that in a supersymmetrical theory, as above,
even surface terms are often accompanied by fermionic
extensions. 

One interesting feature of the Chern-Simons form above is
that it is a single component of a superspace Chern-Simons
form $X_{\underline A \, \underline B \, \underline C}^{CS}$
all of whose components are given by
\be
\eqalign{
X_{\a \b \g}^{CS} =\ & \fracm 12 [ ~\G_{(\a} {}^{\cal I} F_{\b\g)} {}^{\cal I} - 
\fracm 13
         {f}_{{\cal I} {\cal J} {\cal K}} \G_{(\a|} {}^{\cal I} \G_{|\b|} {}^{\cal J}
         \G_{|\g)} {}^{\cal K} ~ ]  ~~, \cr  
X_{\a \b \Dot \g}^{CS} =\ & [ ~\G_{(\a} {}^{\cal I} F_{\b) \Dot 
         \g } {}^{\cal I} + {A}_{\Dot\g} {}^{\cal I} F_{\a \b } {}^{\cal I} - { 
         f}_{{\cal I} {\cal J} {\cal K}} \G_{\a} {}^{\cal I}
         \G_{\b} {}^{\cal J} {A}_{\Dot\g} {}^{\cal K} ~ ]  ~~, ~~ \cr  
X_{\a \b \underline c}^{CS} =\ & [ ~\G_{(\a} {}^{\cal I} F_{\b) \underline c } 
         {}^{\cal I} + \G_{\underline c} {}^{\cal I} F_{\a \b } {}^{\cal I} - { 
         f}_{{\cal I} {\cal J} {\cal K}} \G_{\a}{}^{\cal I} \G_{\b} {}^{\cal J} \G_{\underline 
         c} {}^{\cal K} ~ ]  ~~, ~~ \cr  
X_{\a \, {\Dot \b} \, \underline c}^{CS} =\ & [ ~\G_{\a} {}^{\cal I} F_{{\Dot \b} \underline 
         c } {}^{\cal I} + {A}_{\Dot \b} {}^{\cal I} F_{ \a \underline c } {}^{\cal I} 
         + \G_{\underline c} {}^{\cal I} F_{\a {\Dot \b}} {}^{\cal I} - { f}_{{\cal 
         I} {\cal J} {\cal K}} \G_{\a} {}^{\cal I} {A}_{\Dot \b}
         {}^{\cal J} \G_{\underline c} {}^{\cal K} ~ ]  ~~, ~~ \cr  
X_{\a \underline b \underline c}^{CS} =\ & [ ~\G_{\a} {}^{\cal I} F_{\underline b \underline 
         c } {}^{\cal I} + \G_{[ \underline b} {}^{\cal I} F_{\underline c] \a } {}^{\cal 
         I} - {f}_{{\cal I} {\cal J} {\cal K}} \G_{\a} {}^{\cal I} \G_{\underline 
         b} {}^{\cal J} \G_{\underline c} {}^{\cal K} ~ ]  ~~, ~~ \cr  
X_{\underline a \underline b \underline c}^{CS} =\ & \fracm 12 [ ~\G_{[\underline a} 
         {}^{\cal I} F_{\underline b \underline c ]}{}^{\cal I} - \fracm 13 { 
         f}_{{\cal I} {\cal J} {\cal K}} \G_{[ \underline a|} {}^{\cal I} \G_{|\underline 
         b|} {}^{\cal J} \G_{|\underline c]} {}^{\cal K} ~ ]  ~~. ~~ }  
\ee
Interestingly enough, the action of (\ref{eq:nfn}) can also be written as
\be
{\cal S}_{SYM} ~=~ \int d^4 x \, d^2 \q \, d^2 {\bar \q} ~ 
\eta^{\underline a \, \underline b} 
 ~ X_{\a \, {\Dot \a} \, \, \underline b}^{CS} ~~~.
\ee

The action that we rejected as a description of the 4D, N = 1
Maxwell theory
\be
{\cal S}(H.D.) ~=~ \int d^4 x \, d^2 \q \, d^2 {\bar \q} ~ 
F^{\underline a \, \underline b} \, F_{\underline a \, \underline b} 
~~~, \ee
contains a useful lesson.  Although we will not explicitly evaluate
this in terms of component fields, the projection techniques may
be used by the reader to find its components. Up until this
point, we have always encountered some auxiliary fields within
superfield actions. This may have led to the false impression that
all superfield actions contain auxiliary fields. This is not
the case. In the example of this action it is simple to
show that it contains a term of the form ${\rm d}{}_\bo \,{\rm d}$, 
so that the ${\rm d}$ field propagates in this action.

This illustrates the point that in general an expression of the
form
\be
{\cal S} ~=~  \int d^4 x \, \Big[ \, d^2 \q ~ f_1( W^{\a}) \,+
\, {\rm {h.\,c.}} ~ \Big] ~+~
\int d^4 x \, d^2 \q \, d^2 {\bar \q} ~ f_2 ( W^{\a},\,
 {\Bar W}{}^{\Dot \a}, \nabla_{\underline A}) ~~~,
\label{eq:wze}
\ee
will usually describe the propagation of the gauge field and
gauginos but also the propagation of the ``auxiliary'' ${\rm d}$-fields.
In complete generality such an action contains these are dynamical fields.
It is known, however, that there does exist a very special class 
of actions \cite{15} of the form above for which the ${\rm d}$-fields
do not propagate. We have referred to these as ``auxiliary-free''
actions.

\subsection{The Minimal 4D, N = 1 Supersymmetric Nonlinear $\s$-model}

~~~~In writing the kinetic energy term for the chiral scalar multiplet, we 
have taken advantage of an insight first obtained by Zumino \cite{8} in 
calling it the K\" ahler potential term.  An important class of theories 
arises by considering nonlinear $\s$-models with actions of the form
\be
{\cal S}_{\s} ~=~  \int d^4 x \, d^2 \q \, d^2 {\bar \q} ~ K(\Phi^{\cal A},
\, {\Bar \Phi}{}^{\cal A}) ~~~,
\ee
where the K\" ahler potential $K \ne \sum \Phi^{\cal A} {\Bar \Phi}{}^{\cal A}$
but instead is a general function of $\Phi^{\cal A}$ and ${\Bar \Phi}{}^{
\cal A}$.  At the level of component by use of same technique as in 
(\ref{eq:nyo}-\ref{eq:ntv}) we derive
\be
\eqalign{
{{\cal S}}_{\s} ~=~ \int d^4 x ~ \Big[  &- \frac 12 g_{{\cal A}\, {\Bar {\cal 
B}}}\, ({\cal D}^{\underline a} {\Bar A}{}^{\cal B} \, )  ({\cal D}_{\underline a} 
A{}^{\cal A} \, ) ~-~ i\, g_{{\cal A}\, {\Bar {\cal B}}} \, {\Bar \psi}{}^{\ad}{}^{\cal 
B} {\cal D}_{\underline a} {\psi}^{\a}{}^{\cal A} \cr
&+~ \fracm 14 \, {\Bar \psi}{}^{\ad}{}^{\cal A}{\Bar \psi}{}_{\ad}{}^{\cal 
B}{\psi}^{\a}{}^{\cal C}{\psi}_{\a}{}^{\cal D } {\cal R}_{{\cal C} \, {\Bar {\cal A}} \,
{\cal D} \, {\Bar {\cal B}} } ~ \Big] ~~~,
}\ee
after elimination of the auxiliary $F$-fields by their equations of motion.  In this 
expression, we have used the notations
\be
\eqalign{
g_{{\cal A}\, {\Bar {\cal B}}} &\equiv~ \partder{{}^2 K}{\Phi^{\cal A} \pa {\Bar 
\Phi}{}^{\cal B}} ~=~ K_{,{\cal A},{\Bar {\cal B}}} ~~~,\cr
{\cal D}_{\underline a} {\psi}^{\a}{}^{\cal A} &\equiv~ \pa_{\underline a} 
{\psi}^{\a}{}^{\cal A} ~-~ ({\pa}^{\underline a} {\Bar A}{}^{\cal B} \, )
\G_{{\cal B} \, {\cal C}} {}^{\cal A} \, {\psi}^{\a}{}^{\cal C}  ~~~, \cr
\G_{{\cal B} \, {\cal C}} {}^{\cal A}  &\equiv~
g^{{\cal A}\, {\Bar {\cal D}}} \, K_{,{\cal B},{\cal C},{\Bar {\cal D}}} ~~~, \cr
{\cal R}_{{\cal A} \, {\Bar {\cal C}} \, {\cal B} \, {\Bar {\cal D}} } &\equiv~
 K_{,{\cal A},{\cal B},{\Bar {\cal C}},{\Bar {\cal D}} } ~-~
g^{{\cal K}\, {\Bar {\cal K}}} \, K_{,{\cal A},{\cal B},{\Bar {\cal K}}} 
 \, K_{,{\Bar {\cal C}},{\Bar {\cal D}},{\cal K}} ~~~,
}\ee
where the quantity $g^{{\cal K}\, {\Bar {\cal K}}}$ is the matrix inverse to
$g_{{\cal K}\, {\Bar {\cal K}}}$. From the form of this component action we
see that $g_{{\cal K}\, {\Bar {\cal K}}}$ plays the role of a metric in
the space in which $\Phi^{\cal A}$ and ${\Bar \Phi}{}^{\cal A}$ are the
coordinates. If ${\cal A} = 1, \dots , p$ then this complex $p$-dimensional
space is a ``K\" ahler'' manifold. 

\subsection{4D, N = 2 K\" ahlerian Vector Multiplet Models}

~~~~As we have seen, particular collections of component fields make up N = 1
supersymmetrical theories.  Similarly some particular collections of N = 1
superfields make up N = 2 (and higher) supersymmetrical theories. It has
been known for a long time that a pair $(\Phi, V)$ can be used to provide
a realization of an N = 2 model. The action
\be
\eqalign{
{\cal S}_{Maxwell}^{N = 2} &=~ \int \, d^4 x \, d^2 \q \, d^2 {\bar \q} ~ 
{\Bar {\Phi}} {\Phi} %\cr
%&{~~~~~~~}
~+~ \Big[ ~ \int \, d^4 x \, d^2 \q \, \fracm 14 W^{\a} W_{\a} ~+~ 
{\rm {h.\, c.}} ~ \Big] ~~~,}
\label{eq:wzv}\ee
consists of the terms we have previously seen involving the propagating
fields $(A,\, \psi_{\a}, \l_{\a} , \, A_{\underline a})$ and auxiliary
fields $(F, \, {\rm d})$ and has a ``hidden'' second supersymmetry invariance. 
As the 4D, N = 1 nonlinear $\s$-model may be regarded as a generalization 
of the ``flat'' K\" ahler potential $K = {\Bar \Phi} \Phi$, it should be 
expected that the first term in the action above can be generalized in 
some way that will maintain the second supersymmetry if the Maxwell action 
is also modified. There is a special class of K\" ahler potentials for
which this possible. If the N = 1 K\" ahler potential is of the form
\be
K(\Phi, \, {\Bar \Phi}) ~=~ \Phi \Big( \partder {\Bar H}{\Bar 
\Phi} \Big) ~+~ {\Bar \Phi} \Big( \partder {H}{\Phi} \Big)  ~~~,
\ee
where $H(\Phi)$ \cite{9} is an arbitrary function, and if the Maxwell 
superfield action is modified according to
\be
W^{\a} W_{\a} ~~\to ~~ \partder{{}^2 H}{\Phi \, \pa \Phi} W^{\a} W_{\a}
~~~,
\ee
then the resulting theory\footnote{I first gave these modifications to 
the N = 1 superfield action above in 1984 \cite{9}.} retains the extra 
hidden supersymmetry.  Complex manifolds with geometries that satisfy 
this restricted form of the K\" ahler potential are now called, ``special 
K\" ahler geometries.''

Using instantons in 1988, Seiberg \cite{10} argued that the quantum effective 
action for the fundamental 4D, N = 2 supersymmetric action of (\ref{eq:wzv}) 
has the function $H(\Phi)$ of the form
\be
H(\Phi) ~=~ c_0 \, \Phi^2 \Big[ ~ 1 ~+~ c_1 ln ( {{\Phi}\over \L}) ~+~
... ~ \Big] ~~~.
\ee
In 1994, Seiberg and Witten \cite{11} presented arguments (for the SU(2) 
theory) about how duality completely determines all constants $(c_0,\, c_1, \,
\dots )$ and all terms in the ellipsis by use of a parametrical representation 
of elliptical curves. With the change of notation $H \to i {\cal F}$, the 
KVM models are now often referred to as the ``Seiberg-Witten'' effective
action. It bares the same relation to the N = 2 vector multiplet
action in (\ref{eq:wzv}) as the leading term of the chiral perturbation theory 
action for pion physics bears to the underlying QCD action.

\subsection{4D, N = 1 Supersymmetric P-form Multiplets}

~~~~The ordinary spin-1 U(1) photon $A_{\underline a}$ is a part of
a ``simplex'' of field theories. To see how the simplex arises,
let us note the gauge transformation of the photon takes the
form
\be
\d_G A_{\underline a} ~=~ \pa_{\underline a} \l(x) ~\equiv ~ 
f_{\underline a}
 ~~~, 
\ee
and involves a spin-0 field $\l(x)$ which in principle can be used
to describe a massless state (${\cal S} = - \fracm 12 \int d^4 x \, 
f^{\underline a} f_{\underline a}$). The action for the Maxwell
field is ${\cal S} = - \fracm 18 \int d^4 x \, f^{\underline a \, 
\underline b} f_{\underline a  \, \underline b}$. Comparing these 
two suggests that there might exist another field $
b_{\underline a  \, \underline b}$ whose gauge transformation
has the form of the Maxwell field strength,
\be
\d_G b_{\underline a \, \underline b} ~=~  \pa_{\underline a} 
\k_{\underline b} (x) ~-~ \pa_{\underline b} \k_{\underline a} (x)
~\equiv~ f_{\underline a \, \underline b} (\k) ~~~.
\ee
It is easy to show that the field strength for the new gauge field
(called by various names such as the ``axion,'' ``Kalb-Ramond field,''
``second rank antisymmetric tensor field,'' or ``skew symmetric
tensor field'') $b_{\underline a \, \underline b}$ is given by
\be
h_{\underline a  \, \underline b  \, \underline c} ~=~
\pa_{\underline a} b_{\underline b \, \underline c} ~+~
\pa_{\underline b} b_{\underline c \, \underline a} ~+~
\pa_{\underline c} b_{\underline a \, \underline b} ~\equiv~ 
h_{\underline a  \, \underline b  \, \underline c}(b) ~~~.
\ee
This propagation of this field can be described by an action of
the form,
\be
{\cal S}_{Axion} ~=~ \int d^4 x ~ \Big[ ~ - \, \fracm1{2 \cdot 3!} \,
h^{\underline a  \, \underline b  \, \underline c} \,
h_{\underline a  \, \underline b  \, \underline c} \, \Big] ~~~.
\ee

However, we can now start this process all over again with a new
field ${\cal G}_{\underline a  \, \underline b  \, \underline c}$ and
that will work also. Finally after all of that is done, we can repeat 
the process yet again starting with a field ${\cal K}_{\underline 
a  \, \underline b  \, \underline c \, \underline d}$, 
\be
\d_G {\cal K}_{\underline a  \, \underline b  \, \underline c \, 
\underline d} ~=~ \pa_{\underline a} \ell_{\underline b \, \underline c\, 
\underline d} ~-~ \pa_{\underline b} \ell_{\underline c \, 
\underline d \, \underline a} ~+~ \pa_{\underline c} \ell_{\underline d 
\, \underline a \, \underline b} ~-~ \pa_{\underline d} \ell_{\underline a \, 
\underline b \, \underline c} ~\equiv~ L_{\underline a  \, \underline b  
\, \underline c \, \underline d}(\ell) ~~~.
\ee

After this point the process stops. The reason is that we have ``run 
out of indices to antisymmetrize.'' We call the collection of theories
that begins with $\phi(x)$ and ends with ${\cal K}_{\underline 
a  \, \underline b  \, \underline c \, \underline d}$ ``a simplex.''
It is illustrated below \newpage
\begin{table}[h]
\caption{Simplex of Component x-space p-forms.\label{tab:I}}
\vspace{0.4cm}
\begin{center}
\footnotesize
\begin{tabular}{|c|c|c|c|}\hline
${\rm p}$ & ${\rm p}-{\rm {form}}$ &${\rm {Field}}~ {\rm {Strength}}$ & 
${\rm {Gauge}}~{\rm {Transformation}} $
\\ \hline
$ ~~0 ~~$ &  $ ~~ \phi~~$ &  $ ~~f_{\underline a} ~~$ & $ \d_G \phi ~=~ 
c_0 $  \\ \hline
$ ~~1 ~~$ &  $ ~~ A_{\underline a} ~~$ &  $ ~f_{\underline a \,
\underline b} ~~$ 
 & $ \d_G A_{\underline a} ~=~ \pa_{\underline a} \l  $  \\ \hline
$ ~~2 ~~$ &  $ ~~ b_{\underline a \, \underline b}~~$ &  $ ~~
h_{\underline a \, \underline b \, \underline c} ~~$  & $ \d_G 
b_{\underline a \, \underline b} ~= ~ f_{\underline a \,
\underline b}(\k)  $  \\ \hline
$ ~~3 ~~$ &  $ ~{\cal G}_{\underline a \, \underline b \, \underline c}~$ &  
$ ~~L_{\underline a \, \underline b \, \underline c \, \underline d} ~~$ 
 & $ \d_G {\cal G}_{\underline a \, \underline b \, \underline c}
~=~ h_{\underline a \, \underline b \, \underline c}(\m)  $  \\ \hline
$ ~~4 ~~$ &  $ ~{\cal K}_{\underline a \, \underline b \, \underline c
\, \underline d }~$ &  $ ~~ 0~~$  & $ \d_G {\cal K}_{\underline a  \, 
\underline b  \, \underline c \, \underline d} ~=~ L_{\underline a  
\, \underline b  \, \underline c \, \underline d}(\ell)  $  \\ \hline
\end{tabular}
\end{center}
\end{table}

The action for each field in the simplex takes the form
\be
{\cal S}_p ~\propto~ \int d^4 x ~ ({\rm {Field~
Strength}})^2 ~~~.
\ee
We can make this table more symmetrical in appearance if we define
some additional field strengths via
\be
{\cal G}_{\underline a} ~\equiv~ \fracm 1{3!} \e_{\underline a}{}^{\, 
\underline b  \, \underline c \, \underline d} \, {\cal G}_{\underline b 
\, \underline c \, \underline d} ~~~,~~~ {\cal K} ~\equiv~  \fracm 1{4!}
 \e^{\underline a \, \underline b  \, \underline c \, \underline d} 
{\cal K}_{\underline a \, \underline b  \, \underline c \, \underline d}
~~~.  \ee
Thus the first two columns may be written as
\begin{table}[h]
\caption{Irreducible x-space p-form simplex.\label{tab:II}}
\vspace{0.4cm}
\begin{center}
\footnotesize
\begin{tabular}{|c|c|}\hline
${\rm p}$ & ${\rm p}-{\rm {form}}$ 
\\ \hline 
$ ~~0 ~~$ &  $ ~~ \phi~~$  \\ \hline
$ ~~1 ~~$ &  $ ~~ A_{\underline a} ~~$  \\ \hline
$ ~~2 ~~$ &  $ ~~ b_{\underline a \, \underline b}~~$   \\ \hline
$ ~~3 ~~$ &  $ ~{\cal G}_{\underline a}~$   \\ \hline
$ ~~4 ~~$ &  $ ~{\cal K}~$  \\ \hline
\end{tabular}
\end{center}
\end{table}

The symmetry as we reflect this table vertically around the p = 2
entry is a realization of the ``Hodge star map.''  In the mathematical 
language, each of the gauge fields is a differential p-form. The scalar 
is a ``0-form,'' the vector is a ``1-form,'' etc. Each field strength 
is ``the exterior derivative'' of the p-form and is itself a (p+1)-form. 
Each gauge transformation is the exterior derivative of a (p-1)-form.  
Both the gauge invariance of the field strength and the Bianchi 
identities that the field strengths satisfy are a consequence of the 
``Poincar\' e  lemma'' for differential forms.

Having observed this structure in Minkowski space, it is natural
to inquire whether it is possible to reproduce this in 4D, N = 1 
superspace (and higher ones too).  The answer turns out to be 
affirmative \cite{12} at least for the 4D, N = 1 superspace.

\begin{table}[t]
\caption{Irreducible 4D, N = 1 superspace p-form simplex \label{tab:III}}
\vspace{0.4cm}
\begin{center}
\footnotesize
\begin{tabular}{|c|c|c|c| }\hline
${\rm p}$ & ${\Hat A}_p$ & ${\Hat {d A_p}}$ & $\d_G \, {\Hat A}_p$
\\ \hline
$ ~~0 ~~$ &  $ ~~ \Phi~~$ &  $ ~~i ( {\Bar \Phi}
- \Phi ) ~~$ & $ \d_G \Phi ~=~ c_0$ \\ \hline
$ ~~1 ~~$ &  $ ~~ V ~~$ &  $ ~W_{\a} \, = \, i {\Bar D} {}^2 D_{\a} V ~~$ 
& $\d_G V ~=~ - i ( \, \L \, - \, {\Bar \L} \,)$  \\ \hline
$ ~~2 ~~$ &  $ ~~ \varphi_{\a}~~$ &  $ ~ G \, = \, 
- \fracm 12 ( D^{\a} \varphi_{\a} + {\Bar D} {}^{\ad} {\Bar \varphi}_{\ad})
 ~~$ & $\d_G \varphi_{\a} ~=~ i {\Bar D}{}^2 D_{\a} {\cal X} $
\\ \hline
$ ~~3 ~~$ &  $ ~~{\Hat V} ~~$ &  $ ~~\Pi \, = \, {\Bar D}{}^2 {\Hat V} ~~$ 
&  $ \d_G {\Hat V} ~=~ - \fracm 12 ( \, D^{\a} L_{\a}
\, + \, {\Bar D}^{\Dot \a} {\Bar L}_{\Dot \a} \, )
 $ \\ \hline
$ ~~4 ~~$ &  $ ~~ {\Hat \Phi}~~$ &  $ ~~ 0~~$ &  $ \d_G {\Hat \Phi} ~=~
{\Bar D}{}^2 {\Hat M}  $\\ \hline
\end{tabular}
\end{center}
\end{table}

Of the superfields that propagate physical degrees of freedom,
$\Phi$, $\varphi_{\a}$ and ${\Hat \Phi}$ are chiral while
$V$ and ${\Hat V}$ are real unconstrained superfields. Among the
superfield parameters $\L$ and $L_{\a}$ are chiral while ${\cal X}$ and
${\Hat M}$ are real unconstrained superfields. It can be seen
that the field strength superfields $W_{\a}$, $G$ and $\Pi$
satisfy,
\be
{\Bar D}_{\Dot \a} W_{\a} ~=~ 0 ~~,~~ D^2 G ~=~ 0 ~~, ~~ G ~=~ {\Bar G}
~~,~~ {\Bar D}_{\Dot \a} \Pi ~=~ 0 ~~~.
\ee
The field strength $G$ is of particular interest. As can be seen by
comparing the two tables(\ref{tab:I}, \ref{tab:III}), this field strength 
superfield (in analogy to $W_{\a}$) contains the component level axion 
field strength. Due the algebraic and differential constraints satisfied 
by $G$, this supermultiplet is called the real linear multiplet. It was
first introduced by Siegel \cite{13} in terms of $G$ as well as its
component fields and plays a particularly important role in the theory 
of supergravity that arises as a limit of superstrings and heterotic 
strings.

The superfields ${\Hat A}_p$ in the table immediately above are additional
examples of ``pre-potentials.'' As such we have not seen the analogs
of $\G_{\underline A}$ that we know exist for the super 1-form case.
It turns out that these analogous quantities do exist.  In the case of
the Maxwell theory, we have seen how to explicitly construct the
quantities $\G_{\underline A}$ starting from $V$. We should therefore
expect that
\be
\eqalign{
b_{\underline a \, \underline b}(x) & \to ~ B_{\underline A \, \underline B}(z)
~~~, \cr
{\cal G}_{\underline a \, \underline b  \, \underline c}(x) & \to ~ 
{\cal G}_{\underline A \, \underline B  \, \underline C}(z)
~~~, \cr
{\cal K}_{\underline a \, \underline b \, \underline c \, \underline d}(x) 
& \to ~ {\cal K}_{\underline A \, \underline B \, \underline C \, 
\underline D}(z)
~~~, }\ee
where we call the superfields that appear here, ``the geometrical
super p-forms.'' It is left as an exercise to derive these from
the corresponding pre-potentials and thus derive the form of the
constraints that are satisfied by the field strength superfields
that arise as
\be
\eqalign{
h_{\underline a \, \underline b \, \underline c}(x) & \to ~ H_{\underline A 
\, \underline B \, \underline C}(z)
~~~, \cr
{ L}_{\underline a \, \underline b  \, \underline c  \, \underline d}(x) 
& \to ~ L_{\underline A \, \underline B  \, \underline C \, \underline D}(z)
~~~.}\ee
The solutions for the super 3-form field strengths are
\be
\eqalign{
H_{\a \Dot \b \, \underline c} &=~ i C_{\a \, \g } C_{\Dot \a \Dot \g}\,
G ~~~, \cr
H_{\a \underline b \, \underline c} &=~ i C_{\Dot \b \, \Dot \g } 
C_{\a  ( \b } D_{\g )} G ~~,~~ H_{\Dot \a \underline b \underline c}
~=~ - \, (H_{\a \underline b \underline c})^*  ~~~, \cr
H_{\underline a \, \underline b \, \underline c} &=~ \e_{\underline a \,
\underline b \, \underline c \, \underline d}\, [~ D^{\d} \,, \, {\Bar 
D}{}^{\Dot \d}
~ ] G ~~~, }\ee
with all remaining components vanishing. Similarly for the super 
4-form field strengths
\be
\eqalign{
L_{\a \b \underline c \, \underline d} &=~  C_{\Dot \g \Dot \d} C_{\a 
\, (\g } C_{ \d ) \, \b} \, {\Bar \Pi} ~~~, ~~~
L_{\Dot \a \, \Dot \b \, \underline c \, \underline d} ~=~ - (
L_{\a \b \underline c \, \underline d})^* ~~~, \cr
L_{\a \underline b \, \underline c \, \underline d} &=~ - \, 
\e_{\underline a \, \underline b \, \underline c \, \underline d}\, 
{\Bar D}{}^{\Dot \a} {\Bar \Pi} ~~,~~ L_{\Dot \a \, \underline b \underline c  
\underline d} ~=~ - \, (L_{\a \underline b \, \underline c \, \underline d})^*  
~~~, \cr
L_{\underline a \, \underline b \, \underline c  \, \underline d} &=~ 
i \, \e_{\underline a \, \underline b \, \underline c \, \underline 
d}\, [\, D^2 {\Pi} ~-~ {\Bar D}{}^2 {\Bar \Pi} \, ] ~~~, }\ee
again with all remaining field strength components vanishing.

\subsection{4D, N = 1 Supersymmetric Duality and the Complex
Linear Multiplet}

~~~~An example of the concept of duality (there are actually
several different such concepts) is provided by our re-examining
some of the results in the last section, especially the theory 
of the massless scalar and that of the axion. In these cases,
\be
\eqalign{ {~~~} 
&p ~=~ 0 ~~~,~~~ {\cal S}_0 ~=~ - \fracm 12 \, \int d^4 x ~ f^{\underline a} 
f_{\underline a} ~~~,~~~ f_{\underline a } ~=~ \pa_{\underline a} \phi ~~~,
\cr
&p ~=~ 2 ~~~,~~~ {\cal S}_2 ~=~  \fracm 12 \, \int d^4 x ~ h^{\underline a} 
h_{\underline a} ~~~,~~~ h_{\underline a } ~=~ \fracm 1{3!} \e_{\underline a} 
{}^{\underline b \, \underline c \, \underline d \, } h_{\underline b 
\, \underline c \, \underline d \, }~~~.
} \label{eq:wwn}
\ee
As algebraic consequences we have
\be
\eqalign{
0 & =~ \pa_{\underline a} f_{\underline b}  ~-~ \pa_{\underline b} 
f_{\underline a} ~~~, \cr
0 & =~ \pa^{\underline a} h_{\underline a} ~~~,
}\ee
and we may think of these as the respective ``Bianchi identities'' for 
the two field strengths, $ f_{\underline a}$ and $h_{\underline a}$.

On the other hand, the respective equations of motion can be derived
by simple variation of each action
\be
\eqalign{
& \parvar{S_0}{\phi} ~=~ 0 ~~ \to ~~  \pa^{\underline a} f_{\underline a}
~=~ 0 ~~~, \cr
& \parvar{S_2}{b_{\underline a \, \underline b}} ~=~ 0 ~~ \to ~~ 
\pa_{\underline a} h_{\underline b}  ~-~ \pa_{\underline b} 
h_{\underline a} ~=~ 0 ~~~.}\ee
So it is obvious that the roles of the equation of motion and Bianchi
identity become ``switched'' as we go from one field to the other.
We have called this type of duality ``Poincar\' e duality''\footnote{This
is often called ``electromagnetic duality.''} to distinguish it from
other types of ``dualities.''  The fact that in the presence of the
equations of motion (i.e. on-shell) these two systems are exactly the 
same implies that they both describe a spin-0 degree of freedom.

This example is but one of many that occur within field theories. Given
that these dual theories describe the same physical states, it is tempting
to regard them as equivalent representations. This however, is false.
In this example, it turns out that with regard to properties with respect
to conformal symmetries, these theories are very different. So in general
given two theories related by Poincar\' e duality, we must not a priori
assume that they are equivalent.

We will now show that Poincar\' e duality possess an unexpected extension 
to supersymmetrical theories. First let us observe that there should
be an expected extension. This follows from our discussion above taken
together with the results of the previous section. There we learned
that the supersymmetrical extension of the scalar multiplet is the
chiral multiplet with regard to super p-forms.  Similarly we saw
that there is a supersymmetrical extension of the axion multiplet.
Together these two facts imply that there must be a duality that
relates the chiral multiplet to the axion multiplet and this is made
obvious by the following table

\begin{table}[h]
\caption{Duality of Chiral and Tensor Supermultiplets\label{tab:IV}}
\vspace{0.4cm}
\begin{center}
\footnotesize
\begin{tabular}{|c|c|c|}\hline
{\rm {Superfield}} & {\rm {Bianchi~identity}} & {\rm {Equation~of~motion}} \\
\hline
{\rm {chiral~multiplet}} & ${\Bar D}{}^2 {D}_\a 
(\Phi + {\Bar \Phi}) = 0$   &
${\Bar D}{}^2 (\Phi + {\Bar \Phi}) = 0$ \\ \hline
{\rm {axion~multiplet}} & ${\Bar D}{}^2 G = 0$ &
${\Bar D}{}^2{D}_\a G = 0$ \\ \hline
\end{tabular}
\end{center}
\end{table}

However, there is yet another type of duality. We can observe this one
by noting the following,
\be
{\Bar D}{}_{\Dot \a} \Phi ~=~ 0 ~~ \to ~~ \Phi ~=~ {\Bar D}{}^2 {\Bar U}
~~~, \label{eq:wwu}
\ee
which follows from the algebra of the $D$'s. We may regard the complex
superfield $U$ as a ``pre-potential'' for the chiral scalar multiplet
(we will make use of this is the later discussion of quantization). This
complex superfield, since it is free of constraints, may be used to
calculate the variation of expressions that contain $\Phi$.  It follows
that 
\be
{\cal S}_0 ~=~ \int d^4 x ~ d^2 \q \, d^2 {\bar \q} ~{\Bar \Phi} \Phi ~~,
~~\& ~~ \parvar{{\cal S}_0}{\Phi} ~=~ 0  ~~\to ~~D^2 \Phi ~=~ 0 ~~~,
\label{eq:wwv} \ee
as the equation of motion.  So if we regard (\ref{eq:wwu}) as the ``Bianchi
identity'' and (\ref{eq:wwv}) as yielding the equation of motion, a dual 
theory should interchange these two and this is certainly not what is
contained in the table above.  

This other putative dual superfield \cite{14} (which will shall denote by
${\Bar \S}$) should satisfy $D^2 {\Bar \S} = 0$ as a constraint
(this is called the `linear constraint' as opposed to the chirality
constraint). The solution to this constraint is
\be
\S ~=~ {\Bar D}{}_{\Dot \a} U^{\Dot \a} ~~~.
\ee
So that we see that there is a simple solution to the constraint.
This solution also makes it clear that the $\S$ is a complex superfield.
The next problem to face is what should be the action which describes
the dynamics of this multiplet in the simplest possible manner.

The solution to this is provided by looking back at the discussion
that surrounded the ordinary axion and scalar. There it can be seen
that their actions (\ref{eq:wwn}) look almost identical with the exception
of the minus sign. This suggests that a similar stratagem may apply
to the supersymmetrical theory. 
\be
{\cal S}_{NM} ~=~ -~ \int d^4 x \, d^2 \q \, d^2 {\Bar \q} ~ {\Bar \S} 
\, \S ~~~.
\ee
Let us also propose the definitions of the component fields for this multiplet 
as
\be
\eqalign{ {~~~}
&\,B ~\equiv~  \S \, | ~~,~~ {\Bar \zeta}_{\dot \a} ~\equiv~ {\Bar D}_{
\ad} \S \, | ~~~~, ~~~~
{\r}_{\a} ~\equiv~ {D}_{\a}  \S \, | ~~,~~ H ~\equiv~ 
{D}^2 \S \, | ~~~~,\cr 
&p_{\underline a} ~\equiv~ {\Bar D}_{\dot \a} D_{\a} \S \, | ~~,~~ 
{\Bar \b}_{\ad} ~\equiv~ \frac 12  D^{\a} {\Bar D}_{\ad} D_{\a} \S 
\, | ~~~~. }
\ee
With this action, we evaluate the component fields as in previous cases
to find
\be
\eqalign{ {~~~~}
{\cal S}_{NM} &=~ \int d^4 x ~ \Big[ \, - \frac 12 (\pa^{\underline a} 
{\Bar B} \, ) (\pa_{\underline a} B \, ) ~-~ i \, {\Bar \zeta}{}^{\a} 
\pa_{\underline a} {\zeta}{}^{\ad} ~-~ {\Bar H} H  \, {~~~~~~~~~~~} \cr
&{~~~~~~~~~~~~~~~}\, ~+~ {\Bar p}^{\underline a} p_{\underline a} ~+~ 
{\b}^{\a} {\r}_{ \a} ~+~ {\Bar \b}{}^{\ad} {\Bar \r}_{\ad} ~ \, \, 
\Big]  ~~~~. }
\ee
On-shell we find the superfield equation of motion ${\Bar D}_{\Dot \a}
{\Bar \S} = 0$ or at the component level
\be
\pa^{\underline d} \pa_{\underline d} B ~=~ 0 ~~,~~ - i \, \pa_{\underline a}  
{\Bar \zeta}{}^{\a}  ~=~ 0 ~,~~ H~=~ 0 ~,~~ p_{\underline a} ~=~ 0 ~,~~
{\Bar \r}_{\Dot \a} ~=~ 0 ~,~~ {\b}_{\a} ~=~ 0 ~~,
 \ee
and off-shell since $p_{\underline a} \ne {\bar p}_{\underline a}$, the action 
contains 12 bosons and 12 fermions. It is now obvious that $\S$ also describes 
a ``scalar multiplet'' (i.e. one where the propagating component fields have 
spins no greater than 1/2) we have named this the ``nonminimal scalar 
multiplet.'' As indicated by the following table we have indeed accomplished
our goal of a second unexpected duality

\begin{table}[h]
\caption{Duality of Chiral and Nonminimal Scalar Multiplets\label{tab:V}}
\vspace{0.4cm}
\begin{center}
\footnotesize
\begin{tabular}{|c|c|c| }\hline
${~}$ & ${\rm Constraint}$ & ${\rm Equation~of~Motion}$  \\ \hline
${\rm {Chiral~SF}}$ & $  {\Bar D}_{\Dot \a} \Phi = 0
$ & $ D^2 \Phi = 0$  \\ \hline
${\rm {Nonminimal~SF}}$ & $D^2 {\Bar \S} = 0$ & $ {\Bar D}_{\Dot \a} 
{\Bar \S} = 0$  
\\ \hline
\end{tabular}
\end{center}
\end{table}

Presently, there are known two formal distinction that occur using
the nonminimal multiplet.

The nonminimal multiplet can be used in conjunction with chiral multiplets 
to describe a second version of supersymmetric QED. One striking difference 
in such a construction is that the Dirac spinor contained in such a model 
is defined by 
\be
\psi_{(\a)}(x) ~\equiv~  
\left(\begin{array}{c}
 D_{\a} \Phi {\Big |} \\  {\Bar D}_{\Dot \a} {\S} {\Big |} 
\end{array}\right) 
~~~, 
\ee
which transforms under the U(1) symmetry as
\be
\Big[ \psi_{(\a)}(x) \Big]' ~=~ 
 exp[i g \L | ] ~ \psi_{(\a)}(x) 
~~~.
\ee 
The difference between this and (\ref{eq:nni}) is clear.

In general, if one considers an action 
of the form
\be
{\cal S}_{C-H.D.} ~=~ \int d^4 x \, d^2 \q \, d^2 \bar \q ~ f
\Big( \, \Phi, \,{\Bar \Phi} , \, D_{\underline A}  \Big) ~~~,
\label{eq:fff}
\ee
it is usually the case that such an expression has the consequence
\be
{ {\d {\cal S}_{C-H.D.}} \over {\d F}} ~=~ 0 ~~~, ~~\to {\rm 
{dynamical~equation~for~}} F  ~~~. \ee
On the other hand, it has been shown that it is very simple to construct
a large class of auxiliary-free actions \cite{16} of the form
\be
{\cal S}_{CNM-H.D.} ~=~ \int d^4 x \, d^2 \q \, d^2 \bar \q ~ f
\Big( \, \Phi, \,{\Bar \Phi} ,  \, \S, \,{\Bar \S} ,\, 
D_{\underline A}  \Big) ~~~,
\label{eq:ggg}
\ee
such that {\it {none}} of the auxiliary fields propagate even
though the physical fields can possess arbitrarily high order
derivatives. We thus have a second example of a 4D, N = 1
supersymmetric system (c.f. comment below (\ref{eq:wze})). These
two sets of observations when taken together show that there
are, at least at the level of classical actions, theories
with fields of spin less than or equal to one with manifest 
supersymmetry, higher derivative interaction for propagating
fields and {\it {no}} propagating auxiliary fields.
The significance of this as well as the observation
of the preceding paragraph are still to be investigated.

We end this lecture with a slight warning. Although we have met
many, many 4D, N = 1 supersymmetrical multiplets. The ones we have
seen here are by no means exhaustive. There are many more such
multiplets that have appeared only in the research literature.
The reasons for the existence of the plethora of 4D, N = 1
superfield representations remains a mystery. Perhaps a more
complete theory will rule out the use of many of these other
representations.

\newpage
\section{Third Lecture: A Non-Supersymmetric Interlude on the Lie 
\newline ${~~~~~~~~~~~\,~~}$ Algebraic Origin of Gravitational 
Theories}

{\it {Introduction}}

~~~~~~Also while I was a graduate student, I acquired a book entitled
``Gravitation and Cosmology: Principles and Applications of The 
General Theory of Relativity'' by Steven Weinberg.  Among the
comments in the preface I read, \newline ${~}$\newline
\indent \indent{\it {I found that in most textbooks geometrical 
ideas were given a \newline \indent starring role, so that a 
student...would come away with the impress- \newline \indent ion 
that this had something to do with the fact that space-time is 
\newline \indent a Riemannian manifold.}} \newline \indent \indent{\it 
{Of course, this was Einstein's point of view, and his 
preemin-\newline \indent ent genius necessarily shapes our 
understanding of the theory he \newline \indent created.  However, 
I believe that the geometrical approach has driv-\newline \indent 
en a wedge between general relativity and the theory of elementary 
\newline \indent particles.}}\newline 

At the time this had a profound impact on my thinking about the
theory of general relativity. It suggested that there should exist 
a way to understand theories of gravitation that does {\underline
{not}} involve geometry.  I also believe that this statement stands
as a warning that should be recalled in this present epoch of 
``eka-general relativity'' (i.e. strings, superstrings, heterotic strings, 
D-branes, D-p-branes, M-theory, F-theory and whatever future
developments spring forth from the imagination of eka-relativists).

If we are to look elsewhere for the origins of theories of gravitation,
the first question becomes, ``What will replace Riemannian geometry?''  
The conclusion I came to is that within the confines of the 
relativistic field theory formulation of particle physics, the 
principle of gauge invariance is exquisitely suited to act as a 
logical foundation upon which to also anchor theories of gravitation.  
If this view is accepted, then we are logically freed from the confines 
of `geometry.'  

The implementation of the a gauge-theory based view of theories of 
gravitation hinges on being able to identify a Lie algebra which may 
be gauged.  This identification has been a question approached by 
authors too numerous to list long before I began my efforts in 
this direction. However, in the late seventies during my collaboration 
with Warren Siegel \cite{17}, I believe a solution to this question was 
unearthed that is the simplest of which I am aware and seems to be 
correct. The development of this particular approach can be found in a 
number of our papers from that period.  

It was the theory of supergravity and the struggle to reach its comprehensive 
and complete mathematically rigorous formulation in superspace that led 
to the formalism that I will discuss in this lecture.  Since the geometry 
and pregeometry of curved Wess-Zumino superspace was not so well studied 
in those early days of supergravity (and to a large extent even now), we were 
forced to introduce the usual supergeometrical structures only as devices 
of convenience. It was the gauge-theory structure that was initially the 
only clear signpost.  We thus unintentionally (and unconsciously) became 
Feynman's \cite{18} ``Venutian field theorists'' except that we were not 
located on the planet Venus but in a place called, ``Superspace.'' 

Finally, I believe that the hard won lessons of superspace supergravity
have been largely ignored in the effort of the eka-relativists.  The 
complete superfield structure of superspace supergravity has a layer 
of complication known as the ``pre-potential formulation'' (upon 
which I shall barely touch in these lectures) that is totally unexpected.
A more complicated theory that contains supergravity must be able 
to successfully reproduce this structure. This is an updated
version of the correspondence principle from quantum mechanics.
I have seen little sign of this restriction being fulfilled in many 
works beyond supergravity with the possible exception of Zwiebach's 
approach \cite{19} to bosonic string field theory.

$${~}$$

\noindent
{\it {Lecture}}

\subsection{Explicit Lie-Algebraic Basis of Gravitation}

~~~~We want to treat the symmetries of general relativity as {\it {similarly}} 
as possible as are the symmetries treated in Yang-Mills theory. This leads directly to the 
issue of what operators should play the role of $t_{\cal I}$?  Let us propose 
that the Lie algebra generators of the gauge group of all 4D gravitational
theories includes the follow set of operators
\be 
\Big( \, - i {\pa}_{\underline m} , \, {\cal M}_{\a}{}^{\g} , 
\, {\Bar {\cal M}}_{\Dot \a}{}^{\Dot \g} \, \Big) ~~~,
\label{eq:oex}\ee
acting on fields.

The first of these is recognizable as the momentum generator from elementary
first quantized theories, i.e. $ - i {\pa}_{\underline m} = P_{\underline m}$.
Accordingly we know that
\be
[\, P_{\underline m} \, , \, P_{\underline n} \, ] ~=~ 0 ~~~.
\ee
To understand the significance of ${\cal M}_{\a}{}^{\g}$ and ${\Bar {\cal M
}}_{\Dot \a}{}^{\Dot \g}$ we need to recall a result from our first
lecture. There we noted (\ref{eq:eig}) that there exist an infinite number
of sets of matrices whose commutator algebra is isomorphic to that of
the ``orbital angular momentum operator'' $L_{\underline m \, \underline n}$ 
of elementary first quantized theories.  The idea is to treat the sets of 
matrices ${\Tilde S}_{\underline m \, \underline n}$ in exactly the same 
manner as the sets ${\tilde t}_{\cal I}$ in a Yang-Mills theory.  This is 
done by recognizing that ${\cal M}_{\underline a \, \underline b}$ {\it {is}} 
the analog of $t_{\cal I}$ in a Yang-Mills theory.

Since ${\cal M}_{\underline a \, \underline b}$ has two ``vector'' indices,
we can use it to define two other abstract Lie algebra generators
${\cal M}_{\a \, \b}$ and ${\Bar {\cal M}}_{\Dot \a \, \Dot \b}$ via
the definition
\be
{\cal M}_{\underline a \, \underline b} ~=~ \Big[ \, C_{\Dot \a \, \Dot \b}
{\cal M}_{\a \, \b} ~+~ C_{\a \, \b} {\Bar {\cal M}}_{\Dot \a \, \Dot \b}
\, \Big] ~~~.
\label{eq:whg} \ee
The generator ${\cal M}_{\underline a \, \underline b}$ may be thought of
as the ``generator of spin angular momentum.'' The quantities ${\cal M}_{
\a \, \b}$ and ${\Bar {\cal M}}_{\Dot \a \, \Dot \b}$ may be roughly regarded
as the ``right-handed spin angular momentum generator'' and ``left-handed 
spin angular momentum generator respectively (c.f. equation (\ref{eq:fth})).  
We recall that any theory involving spinors possesses an intrinsic definition 
of handedness related to the chirality of the spinors. The spin angular 
momentum commutes with ordinary momentum thus
\be
[ \, {\cal M}_{\a \, \b} \, , \,  P_{\underline m} \, ] ~=~ 0 ~~ ,~~
[ \, {\Bar {\cal M}}_{\Dot \a \, \Dot \b} \, , \,  P_{\underline m} \, ] 
~=~ 0 ~~.
\ee
Since $ [ P_{\underline m} \, , \, x^{\underline n}  ] ~=~ - i 
\d_{\underline m} {}^{\underline n}$, it follows that 
\be
[ \, {\cal M}_{\a \, \b} \, , \,  x^{\underline m} \, ] ~=~ 0 ~~ ,~~
[ \, {\Bar {\cal M}}_{\Dot \a \, \Dot \b} \, , \,  x^{\underline m} \, ] 
~=~ 0 ~~.
\ee
But wait, isn't $x^{\underline m}$ a vector?  After all for a photon
$A_{\underline a}(x)$ we can use the definition of ${\cal M}_{\underline 
a \, \underline b}$ and the result of equation (\ref{eq:nin}) to show
\be
[ \, {\cal M}_{\a}{}^{\b} \, , \,  A_{\underline c} \, ] ~=~ \d_{\g}
{}^{\b} A_{\a \Dot \g} ~-~ \frac 12 \d_{\a} {}^{\b} A_{\underline c} ~~,~~
[ \, {\Bar {\cal M}}_{\Dot \a}{}^{\Dot \b} \, , \,  A_{\underline c} \, ] ~=~ 
\d_{\Dot \g} {}^{\Dot \b} A_{\g \Dot \a} ~-~ \frac 12 \d_{\Dot \a} {}^{
\Dot \b} A_{\underline c} ~~.~~
\ee
This seems to imply a contradiction!

The way out of this is to recognize that there are two distinct types
of ``vectors.'' We will call vectors like $P_{\underline m}$ ``curved
vectors'' and henceforth will denote such quantities by underlined Latin
letters ${\underline m}$, ${\underline n}$,... to the end of the
alphabet.  We will call vectors like $A_{\underline a}$ ``flat
vectors'' and henceforth will denote such quantities by underlined Latin
letters ${\underline a}$, ${\underline b}$,... , ${\underline \ell}$.
This is sometimes called the  ``early-late'' convention. In our 
approach {\it {all}} {\it {matter}} {\it {fields}} {\it {possess}}
{\it {only}} {\it {flat}} {\it {Lorentz}} {\it {indices}}. 
The realization of ${\cal M}_{\a \, \b}$ and ${\Bar {\cal M}}_{\Dot \a
\, \Dot \b}$ on $A_{\underline a}$ can also be used in two other ways.

We know that given two spinors $\psi_{\a}(x)$ and $\chi_{\Dot \g}(x)$,
their product $\psi_{\g} \chi_{\Dot \g}$ should transform like $
A_{\underline c}$,
\be
{\cal M}_{\a}{}^{\b} \Big( \psi_{\g} \chi_{\Dot \g} \Big) 
~=~ \d_{\g}
{}^{\b}  \Big( \psi_{\a} \chi_{\Dot \g} \Big)  ~-~ \frac 12 
\d_{\a} {}^{\b}  \Big( \psi_{\g} \chi_{\Dot \g} \Big) ~~~,
\ee
or
\be
 \Big( {\cal M}_{\a}{}^{\b} \psi_{\g}\Big) \, \chi_{\Dot \g} 
~+~  \psi_{\g} \, \Big(  {\cal M}_{\a}{}^{\b} \chi_{\Dot \g} \Big) 
~=~ \d_{\g}
{}^{\b}  \Big( \psi_{\a} \chi_{\Dot \g} \Big)  ~-~ \frac 12 
\d_{\a} {}^{\b}  \Big( \psi_{\g} \chi_{\Dot \g} \Big) ~~~,
\ee
where on the second line we have used the Leibniz property.
The solution we require to this is simply
\be
{\cal M}_{\a}{}^{\b} \psi_{\g} ~=~ \d_{\g} {}^{\b} \, \psi_{\a}
~-~ \frac 12 \d_{\a} {}^{\b}  \, \psi_{\g} ~~~,~~~
{\cal M}_{\a}{}^{\b} \chi_{\Dot \g} ~=~ 0 ~~~.
\ee
Upon complex conjugation these take the forms
\be
{\Bar {\cal M}}_{\Dot \a}{}^{\Dot \b} {\bar \psi}_{\Dot \g} ~=~ 
\d_{\Dot \g} {}^{\Dot \b} \, {\bar \psi}_{\Dot \a} ~-~ \frac 12 
\d_{\Dot \a} {}^{\Dot \b}  \, {\bar \psi}_{\Dot \g} ~~~,~~~
{\Bar {\cal M}}_{\Dot \a}{}^{\Dot \b} {\bar \chi}_{\g} ~=~ 0 ~~~.
\ee
With the results of (\ref{eq:wwu}) and (\ref{eq:wwv}) in hand, it 
is easy to derive the commutator algebra of ${\cal M}_{\a \, \b}$ 
and ${\Bar {\cal M}}_{\Dot \a \, \Dot \b}$. These generators are 
found to satisfy
\be
[ ~ {\cal M}_{\a}{}^{\b} \, , \, {\cal M}_{\g}{}^{\d} ~ ]
~=~ \d_{\g} {}^{\b} \, {\cal M}_{\a}{}^{\d} ~-~  
\d_{\a} {}^{\d} \, {\cal M}_{\g}{}^{\b} ~~~, \label{eq:wux}
\ee
\be 
[ ~  {\cal M}_{\a}{}^{\b}  \, , \, {\Bar {\cal M}}_{\Dot \g}{}^{\Dot \d} ~]
~=~ 0 ~~~, \label{eq:wuv}
\ee
\be
[ ~ {\Bar {\cal M}}_{\Dot \a}{}^{\Dot \b} \, , \,  {\Bar {\cal M}}_{\Dot \g}
{}^{\Dot \d} ~] ~=~ \d_{\Dot \g} {}^{\Dot \b} \, {\Bar {\cal M}}_{\Dot \a}
{}^{\Dot \d} ~-~  \d_{\Dot \a} {}^{\Dot \d} \, {\Bar {\cal M}}_{\Dot \g}
{}^{\Dot \b} ~~~. \label{eq:wug}
\ee
(Given these and the definition in (\ref{eq:whg}), we can always ``re-construct''
the commutator algebra of ${\cal M}_{\underline a \, \underline b}$.)

Using equations (\ref{eq:wux}), (\ref{eq:wuv}) and (\ref{eq:wug}) we 
have a complete definition of a Lie algebra for the set of operators 
in (\ref{eq:whg}). We now wish to ``gauge'' the algebra in (\ref{eq:oex}) 
as can be done for any Lie group for a Yang-Mills theory. For this purpose 
we introduce local parameters for our gauge group; $\xi^{\underline m}(x)$, 
$\l_{\a}{}^{\b}(x)$ and ${\bar \l}_{\Dot \a}{}^{\Dot \b}(x)$. For the 
infinitesimal parameter of the gauge transformation we define
\be
\eqalign{
K &\equiv~ i \xi^{\underline m} P_{\underline m} ~+~ \l_{\a}{}^{\b}(x) 
{\cal M}_{\b}{}^{\a} ~+~ {\bar \l}_{\Dot \a} {}^{\Dot \b}(x) {\Bar {\cal 
M}}_{\Dot \b}{}^{\Dot \a}  \cr
&=~ \xi^{\underline m} {\pa}_{\underline m} ~+~ \l_{\a}{}^{\b}(x) 
{\cal M}_{\b}{}^{\a} ~+~ {\bar \l}_{\Dot \a} {}^{\Dot \b}(x) {\Bar {\cal 
M}}_{\Dot \b}{}^{\Dot \a}  ~~~.
} \ee
The quantity $K$ is a local vector in the Lie algebra. The set of all
such possible local vectors is closed under commutation. If
\be
\eqalign{
K(\xi_1, \l_1, \, {\bar \l}_1)  &=~ \xi^{\underline m}_1 {\pa}_{\underline m} 
~+~ \l_{1 \, \a}{}^{\b}(x) {\cal M}_{\b}{}^{\a} ~+~ {\bar \l}_{1 \,\Dot \a} {}^{\Dot 
\b}(x) {\Bar {\cal M}}_{\Dot \b}{}^{\Dot \a}  ~=~ K_1 ~~~,
\cr
K(\xi_2, \l_2, \, {\bar \l}_2)  &=~ \xi^{\underline m}_2 {\pa}_{\underline m} 
~+~ \l_{2 \,\a}{}^{\b}(x) {\cal M}_{\b}{}^{\a} ~+~ {\bar \l}_{2 \, \Dot 
\a} {}^{\Dot \b}(x) {\Bar {\cal M}}_{\Dot \b}{}^{\Dot \a}  ~=~ K_2 ~~~,
}\ee
then there exists (with some restriction regarding the properties of
the derivatives of the local parameters) $\xi^{\underline m}_3 $,
$\l_{3 \, \a}{}^{\b}$ and ${\bar \l}_{3 \, \Dot \a} {}^{\Dot \b}$
such that
\be
[~ K(\xi_1, \l_1, \, {\bar \l}_1) \, , \, K(\xi_2, \l_2, \, {\bar \l}_2)
~ ] ~=~ K(\xi_3, \l_3, \, {\bar \l}_3) ~~~.
\ee
Furthermore, a Jacobi identity is satisfied
\be
[~ [ ~ K_1 \, , \, K_2 ~]\, , \, K_3 ~] ~+~ [~ [ ~ K_2 \, , \, K_3 ~]\, , 
\, K_1 ~] ~+~ [~ [ ~ K_3 \, , \, K_1 ~]\, , \, K_2 ~] ~=~ 0 ~~~.
\ee

\subsection{Gravitational Covariant Derivative and Gauge Transformations}

~~~~This local Lie algebra vector space is {\it {exactly}} like the corresponding
structure in a Yang-Mills theory. Therefore, we can define gauge transformation
acting on scalar, spinors or vector matter fields by
\be
\eqalign{
\Big[ \phi(x) \Big]' &\equiv~ e^K \phi(x) ~=~  e^K \phi(x) e^{-K} ~~~,\cr
\Big[ \psi_{\a}(x) \Big]' &\equiv~ e^K \psi_{\a}(x) ~~~, \cr
\Big[ \chi_{\Dot \a}(x) \Big]' &\equiv~ e^K \chi_{\Dot \a}(x) ~~~, \cr
\Big[ A_{\underline a}(x) \Big]' &\equiv~ e^K A_{\underline a}(x) ~~~.
} \label{eq:wvh}
\ee
Since we have chosen to represent the symmetries realized on any field
as in (\ref{eq:wvh}), this has an interesting consequence. In more
conventional treatments of general coordinate transformations, equations
such as ${\tilde x} = f(x)$ are used to describe such symmetries. Not
all such transformations can be represented using (\ref{eq:wvh}). Only
transformation that can be continously connected to the identity have
such representations. In the language of gauge theories, these are
``small gauge transformations.'' Transformation that do not possess
this property (e.g. ${\tilde x} = 1/x$) are in the language of gauge 
theories known as ``large gauge transformations.''

We require a covariant derivative with respect to this gauge group and this
can be introduced according to
\be
\nabla_{\underline a} ~ \equiv~ {\rm e}_{\underline a}{}^{\underline m} 
\pa_{\underline  m} ~+~ \fracm 12 \o_{\underline a \underline c }{}^{\underline  
d} \,{\cal M}_{\underline d }{}^{\underline c}
~~~, 
\ee
that includes the gauge fields ${\rm e}_{\underline a}{}^{\underline m}(x)$ and 
$\o_{\underline a \, \underline c \underline d}(x)$ with Lie-algebraic gauge 
transformations of the form
\be
\Big( \, \nabla_{\underline a} \Big)' ~=~ e^K \nabla_{\underline a} e^{- K} ~~~, 
~~~ K ~ \equiv ~ K^{\underline m} \pa_{\underline m} ~+~ \fracm 12 K_{\underline 
c }{}^{\underline d} \,{\cal M}_{\underline d }{}^{\underline c} ~~~, 
\label{eq:wvv} \ee
in terms of the local parameters $K^{\underline m}(x)$ and $K^{\underline 
c \underline d}(x)$.  (Since $\o_{\underline a \, \underline c \, \underline 
d} = - \o_{\underline a \, \underline d \, \underline c}$, we may also write
$\o_{\underline a \, \underline c \, \underline d} = C_{\Dot \g \, \Dot 
\d} \, \o_{\underline a \, \g \, \d} + C_{\g \d} \, \o_{\underline a \, 
\Dot \g \,\Dot \d}$).

There are several points to note regarding this definition. \newline

\indent (a.) Since $P_{\underline m}$ is one of the ``generators,'' this
covariant derivative does not \newline \indent ${~~~~~}$ possess a part 
that is independent of the Lie algebra generators. \newline

\indent (b.) The operator $\nabla_{\underline a}$ is a ``flat vector''. In
particular, \newline
${~~~~~~~~~~~~~~~~~~~~~~~~~~~}[ \, {\cal M}_{\underline a \, \underline b} 
\, , \,  \nabla_{\underline c} \, ] ~=~ ( \, \eta_{\underline a \underline 
c} \d_{\underline b}{}^{\underline d} ~-~ \eta_{\underline b \underline c} 
\d_{\underline a}{}^{\underline d} \,) \nabla_{\underline d} ~~,~~$
\newline \indent ${~~~~~}$ and this allows us to easily write actions for
spinors in the presence \newline \indent ${~~~~~}$ of the $K$-gauge group
(i.e. a minimal coupling procedure is of the  \newline \indent ${~~~~~}$ 
form ${\Bar \psi}{}^{\Dot \a} \pa_{\underline a} {\psi}{}^{\a} \to {\Bar 
\psi}{}^{\Dot \a} \nabla_{\underline a} {\psi}{}^{\a}$). \newline
${~}$ \newline
\noindent 
The gauge field for the translation generator $P_{\underline m}$ is ${\rm 
e}_{\underline a}{}^{\underline m}(x)$ and (as we shall see) this corresponds 
to the field that is usually called the ``inverse vierbein'' or ``inverse 
tetrad'' in geometrical approaches to general relativity.  The gauge field 
for the spin angular momentum generator ${\cal M}_{\underline a \, \underline 
b}$ is $\o_{\underline a \, \underline c \, \underline d}(x)$ and (once again 
we shall see) this corresponds to the field that is called ``the 
spin-connection\footnote{We must take caution to note that the spin-connection 
is {\it {not}} the usual Christoffel \newline ${~~~~~}$ connection.}.''
Since the $K$-gauge transformation law of $\nabla_{\underline a}$ is exact
as in a Yang-Mills theory, it follows that the transformation law of 
a covariant derivative of any matter field also takes the same form
as the transformation law of the field itself. For example for a scalar
field
\be
\Big[ \nabla_{\underline a} \phi(x) \Big]' ~=~ e^K \nabla_{\underline a}
\phi(x) ~~~.
\ee
If we start with a Lagrangian
\be
{\cal L} ~=~ {\cal L}( \phi, \, \pa_{\underline a} \phi ) ~~~,
\ee
which does not possess the local symmetries generated by the $K$ operator,
it follows that a new Lagrangian obtained by the minimal coupling 
replacement $\pa_{\underline a} \to \nabla_{\underline a}$ has the property
\be 
\Big[ {\cal L}( \phi, \, \nabla_{\underline a} \phi )  \Big]' ~=~ e^K {\cal L}
( \phi, \, \nabla_{\underline a} \phi ) ~=~ e^{\xi^{\underline m} \pa_{\underline 
m} } \, {\cal L}( \phi, \, \nabla_{\underline a} \phi )  ~~~.
\label{eq:wvg} \ee
The transformation law above is exactly the same as for any scalar field.

At this point the analogy with Yang-Mills theory seems to break down. If we 
look back at Yang-Mills theory then analogously,
$$
\Big[ \phi \Big]' ~=~ e^{i \l } \phi ~~,~~~ \Big[ \nabla_{\underline a}
 \Big]' ~=~ e^{i \l } \nabla_{\underline a} e^{- i \l }  ~~~, $$
\be 
\Big[ {\cal L}( \phi, \, \nabla_{\underline a} \phi )  \Big]' ~=~ 
{\cal L}( \phi, \, \nabla_{\underline a} \phi ) ~~~. \label{eq:wvn}
\ee
The difference can be seen in the operatorial pre-factor that appears
in (\ref{eq:wvg}) that does not appear in (\ref{eq:wvn}).  Thus, it at first 
seems to be impossible to write an action which possesses the local symmetries 
generated by $K$.  We will come back this shortly.

Let us turn our attention instead to the infinitesimal gauge field 
transformation laws implied by (\ref{eq:wvv}).
$$
\Big[ \nabla_{\underline a} \Big]' ~=~ \nabla_{\underline a}
~+~ [ \, K ~,~ \nabla_{\underline a} \,] ~~~ \to
$$
\be
\d_K \nabla_{\underline a} ~\equiv~ \Big[ \nabla_{\underline a} \Big]'
~-~ \nabla_{\underline a} ~=~ [ \, K ~,~ \nabla_{\underline a} \,] 
~~~. \ee
This final formula looks exactly like its Yang-Mills analog.  However, it
obscures the details of what is happening to the $K$-symmetry gauge 
field variables. After a straightforward calculation these are found to
be
$$
\d_K {\rm e}_{\underline a}{}^{\underline m} ~=~ \xi^{\underline n} 
\pa_{\underline n} {\rm e}_{\underline a}{}^{\underline m} ~-~ 
{\rm e}_{\underline a}{}^{\underline n} \pa_{\underline n}
\xi^{\underline m} ~-~ \l_{\underline a}{}^{\underline b}
{\rm e}_{\underline b}{}^{\underline m} ~~~, {~~~}
$$
\be
\eqalign{
\d_K \o_{\underline a \, \underline b}{}^{\underline c} &=~ \xi^{\underline 
n} \pa_{\underline n} \o_{\underline a \, \underline b}{}^{\underline d} 
~-~ {\rm e}_{\underline a}{}^{\underline n} \pa_{\underline n}  \l_{\underline 
b}{}^{\underline c} ~-~  \l_{\underline a}{}^{\underline d} \o_{\underline d 
\, \underline b}{}^{\underline c} {~~~~} \cr
&{~~~~~}-~ \l_{\underline b}{}^{\underline d} \o_{\underline a 
\, \underline d}{}^{\underline c} ~+~ \l_{\underline d}{}^{\underline c}
\o_{\underline a \, \underline b}{}^{\underline d}  ~~~.  
}  \label{eq:wxn} \ee
These transformation laws give us confidence that we have correctly
identified the Lie algebra that lies behind gravitational theories.  The 
gauge algebraic transformation law of (\ref{eq:wvv}) unambiguously leads 
to these results. If we set all the terms proportional to the spin angular
momentum generators to zero, then the result above has been derived
with no assumptions other than the fact that the exponential of the
the operator $i K^{\underline m}(x) P_{\underline m}$ must be well
defined.  If we restrict $ K^{\underline m}(x)$ to be a constant, then
the basic Lie algebraic structure of gravitational theories appears
via the usual translation generator of elementary quantum
theory.

The commutator of two of these gravitationally covariant derivatives
generate field strengths $t_{\underline a \, \underline b \, \underline c}$ and
$r_{\underline a \, \underline b \, \underline c \, \underline d}$
\be
\Big[ ~ \nabla_{\underline a} \, , \, \nabla_{\underline b} ~ \Big] ~=~ 
t_{\underline a \, \underline b}{}^{\underline c} \, \nabla_{\underline c}
~+~ \fracm 12 r_{\underline a \, \underline b \, \underline c }{}^{\underline d}\,
{\cal M}_{\underline d }{}^{\underline c} ~~~~.
\ee
Let us note that it would be just as natural to define the field strengths
via 
\be
\Big[ ~ \nabla_{\underline a} \, , \, \nabla_{\underline b} ~ \Big] ~=~ 
i \, {\cal F}_{\underline a \, \underline b}{}^{\underline m} \, 
P_{\underline m} ~+~ \fracm 12 {\cal F}_{\underline a \, \underline b 
\, \underline c }{}^{\underline d}\, {\cal M}_{\underline d }{}^{\underline 
c} ~~~~.
\ee
However, we have the following steps.
\be 
\nabla_{\underline a} ~=~ i {\rm e}_{\underline a} {}^{\underline m} P_{\underline 
m} ~+~ \fracm12  \o_{\underline a \, \underline d}{}^{\underline e} {\cal 
M}_{\underline e}{}^{\underline d} ~~\to ~~ {\rm e}_{\underline a} {}^{\underline m} 
P_{\underline m} ~=~ - i \, ( \, \nabla_{\underline a} ~-~  \frac 12 \o_{\underline 
a \, \underline d}{}^{\underline e} {\cal M}_{\underline e}{}^{\underline d}
\, )  ~~~.
\ee
The gauge field ${\rm e}_{\underline a} {}^{\underline m}(x) $ is a 4 $\times$ 4 matrix,
so we can define its inverse ${\rm e}_{\underline m} {}^{\underline a}(x) $ (note
that it is the location of the curved versus flat vector indices that distinguishes
one from the other) and use it via
\be
P_{\underline m} ~=~ - i \, {\rm e}_{\underline m} {}^{\underline a} \, (\,
\nabla_{\underline a} ~-~  \frac 12 \o_{\underline a \, \underline d}{}^{\underline 
e} {\cal M}_{\underline e}{}^{\underline d} \,)  ~~~.
\ee
This last result may be substituted into the expression containing ${\cal F
}_{\underline a \, \underline b}{}^{\underline m}$ and ${\cal F}_{\underline 
a \, \underline b \, \underline c}{}^{\underline d}$ to reach our initial
definition.

The field strengths are explicitly given by
\be
 \eqalign{
t_{\underline a \, \underline b \, \underline c} &=~ c_{\underline a \, 
\underline b \, \underline c} ~-~ \o_{\underline a \, \underline b \, \underline 
c} ~+~ \o_{\underline b \, \underline a \, \underline c} ~~~, \cr 
r_{\underline a \, \underline b \, \underline c \, \underline d} &=~ {\rm 
e}_{\underline a}{}^{\underline m} \pa_{\underline  m} \, \o_{\underline b 
\, \underline d \, \underline e} ~-~ {\rm e}_{\underline b}{}^{\underline 
m} \pa_{\underline  m} \, \o_{\underline a \, \underline d \, \underline 
e} ~-~ c_{\underline a \, \underline b}{}^{\underline c} \, \o_{\underline 
a \, \underline d \, \underline e} \cr
&{~~~}~+~ \o_{\underline a \, \underline c}{}^{\underline e} \, \o_{\underline 
b \, \underline e \, \underline d} ~-~ \o_{\underline b \, \underline c}{}^{
\underline e} \, \o_{\underline a \, \underline e \, \underline d} ~~~,
} \label{eq:txx}
\ee
with the anholonomity $(c_{\underline a \, \underline b}{}^{\underline c})$ defined 
by 
\be
c_{\underline a \, \underline b}{}^{\underline c} ~ \equiv~ \Big[ \, {\rm 
e}_{\underline a}{}^{\underline m} \pa_{\underline  m} \, {\rm e}_{\underline b} 
{}^{\underline n} ~-~ {\rm e}_{\underline b}{}^{\underline m} \pa_{\underline 
m} \, \, {\rm e}_{\underline a} {}^{\underline n} \, \Big] {\rm e}_{\underline 
n} {}^{\underline c} ~~~.
\ee
Due to its definition, the anholonomity satisfies the
identity
\be
{\rm e}_{[ \underline a} c_{\underline b \, \underline c ] }
{}^{\underline d} ~-~ c_{ [ \underline a \, \underline b | }
{}^{\underline k} c_{\underline k \, | \underline c ] }
{}^{\underline d} ~=~ 0 ~~~,  \label{eq:abi}
\ee
where ${\rm e}_{\underline a} \equiv {\rm e}_{\underline a}
{}^{\underline m} \pa_{\underline m}$.

The two field strengths $t_{\underline a \, \underline b \, \underline c}$ and
$r_{\underline a \, \underline b \, \underline c \, \underline d}$ are
well known from the study of differential geometry. The foremost
is called the torsion tensor and the latter is the Riemann curvature tensor.
Recall in the discussion of the supersymmetric Maxwell theory, the fact
that a gauge superfield appeared {\it {solely}} algebraically in the definition
of a field strength meant that it could be used to set the field strength 
to zero identically. It is natural to ask if this can be repeated?  

Before we answer this, let's find out how $t_{\underline a \, \underline b \, 
\underline c}$ and $r_{\underline a \, \underline b \, \underline c \, 
\underline d}$ behave under arbitrary variations of the gauge fields
${\rm e}_{\underline a}{}^{\underline m}$ and $\o_{\underline a \, 
\underline c \, \underline d}$. The results
\be 
\eqalign{ {~~~} 
\nabla_{\underline a} ~=~ {\rm e}_{\underline a} {}^{\underline m} \pa_{\underline 
m} &+~ \fracm12  \o_{\underline a \, \underline d}{}^{\underline e} {\cal 
M}_{\underline e}{}^{\underline d} ~~\to ~~ {~~~~~~~}\cr
\d \nabla_{\underline a} &=~ \d {\rm e}_{\underline a} {}^{\underline m} \pa_{\underline 
m} ~+~ \fracm12  \d \o_{\underline a \, \underline d}{}^{\underline e} {\cal 
M}_{\underline e}{}^{\underline d} \cr
&=~ \d {\rm e}_{\underline a} {}^{\underline m} {\rm e}_{\underline m} {}^{\underline b} 
{\rm e}_{\underline b} {}^{\underline n} 
\pa_{\underline n} ~+~ \fracm12  \d \o_{\underline a \, \underline d}{}^{\underline 
e} {\cal M}_{\underline e}{}^{\underline d} \cr
&\equiv~ h_{\underline a}{}^{\underline b} {\rm e}_{\underline b} {}^{\underline n} 
\pa_{\underline n} ~+~ \fracm12  \d \o_{\underline a \, \underline d}{}^{\underline 
e} {\cal M}_{\underline e}{}^{\underline d} \cr
&=~ h_{\underline a}{}^{\underline b} \,
( \nabla_{\underline b} ~-~ \frac 12 \o_{\underline b \, \underline d}{}^{\underline 
e} {\cal M}_{\underline e}{}^{\underline d} \,) 
 ~+~ \fracm12  \d \o_{\underline a \, \underline d}{}^{\underline 
e} {\cal M}_{\underline e}{}^{\underline d} \cr
&\equiv~ h_{\underline a}{}^{\underline b} \, \nabla_{\underline b} ~+~ 
\fracm12 h_{\underline a \, \underline d}{}^{\underline e} {\cal M}_{\underline 
e}{}^{\underline d} ~~~,
} \label{eq:wxh}  \ee
imply that the variation of the gravitationally covariant derivative is 
simply expressed in terms of the two quantities $ h_{\underline a}{}^{\underline 
b} = \d {\rm e}_{\underline a} {}^{\underline m} {\rm e}_{\underline m} {}^{\underline b}$
and $h_{\underline a \, \underline d \, \underline e} = \d \o_{\underline a 
\, \underline d \, \underline e} - h_{\underline a}{}^{\underline b}
\o_{\underline b \, \underline d \, \underline e}$.

By applying a variation operator starting at
\be \eqalign{
&\d \Big( \, \Big[ ~ \nabla_{\underline a} \, , \, \nabla_{\underline b} ~ \Big]
\, \Big) ~=~ \d t_{\underline a \, \underline b}{}^{\underline c} \, \nabla_{
\underline c} ~+~ t_{\underline a \, \underline b}{}^{\underline c} \, \d
\nabla_{\underline c} ~+~ \d \fracm 12 r_{\underline a \, \underline b \, \underline 
c }{}^{\underline d}\, {\cal M}_{\underline d }{}^{\underline c} ~\to \cr
&\Big[ ~ \d \nabla_{\underline a} \, , \, \nabla_{\underline b} ~ \Big]
~+~\Big[ ~ \nabla_{\underline a} \, , \, \d \nabla_{\underline b} ~ \Big]
~=~ \d t_{\underline a \, \underline b}{}^{\underline c} \, \nabla_{
\underline c} ~+~ t_{\underline a \, \underline b}{}^{\underline c} \, \d
\nabla_{\underline c} ~+~ \d \fracm 12 r_{\underline a \, \underline b \, \underline 
c }{}^{\underline d}\, {\cal M}_{\underline d }{}^{\underline c} ~~~,\cr
}\ee
we can rapidly derive the results
\be \eqalign{
\d t_{\underline a \, \underline b}{}^{\underline d} ~=~ 
&\nabla_{\underline a } h_{\underline b} {}^{ \, \underline c } ~-~ 
\nabla_{\underline b } h_{\underline a}{}^{ \, \underline c } 
~-~ h_{\underline a}{}^{\underline c} t_{\underline b \, \underline 
c}{}^{\underline d} ~+~ h_{\underline b}{}^{\underline c} t_{\underline 
a \, \underline c}{}^{\underline d} ~+~  t_{\underline a \, \underline 
b}{}^{\underline c} h_{\underline c}{}^{\underline d} \cr
&-~ h_{\underline a \, \underline b}{}^{ \, \underline d} ~+~ h_{
\underline b \, \underline a}{}^{ \, \underline d}  ~~~,\cr
\d r_{\underline a \, \underline b \, \underline c \, \underline 
d} ~=~ &\nabla_{\underline a } h_{\underline b \, \underline c
\, \underline d} ~-~ \nabla_{\underline b } h_{\underline a
\, \underline c \, \underline d} ~-~  t_{\underline a \, \underline 
b}{}^{\underline e} h_{\underline e \, \underline c
\, \underline d}  ~+~ h_{\underline c}{}^{\underline e} 
r_{\underline a \, \underline b \, \underline e \, \underline d}     \cr
&~-~ h_{\underline a}{}^{\underline e} r_{\underline b \, \underline e
\, \underline c \, \underline d} ~+~ h_{\underline b}{}^{\underline e} 
r_{\underline a \, \underline e \, \underline c \, \underline d}
~+~ h_{\underline d}{}^{\underline e} r_{\underline a \, \underline b \, 
\underline c \, \underline e}  ~~~. } \label{eq:wvz} \ee
We will make use of some of these later.

Let us now return to the issue of the spin-connection.  As an assumption, 
we can impose the condition that the $t_{\underline a \, \underline b}{}^{
\underline c}$ should vanish. The solution to this condition is given 
by choosing 
\be
\o_{\underline a \, \underline b\, \underline c} ~=~ \o_{\underline a \, 
\underline b\, \underline c} ({\rm e}) ~\equiv~  \fracm 12 \, [~ c_{\underline 
a \, \underline b\, \underline c} \, - \,  c_{\underline a \, \underline c \, 
\underline b} \,- \, c_{\underline b \, \underline c\, \underline a}~ ]  ~~~.
\label{eq:wve}\ee
So the gauge theory of gravity shares with supersymmetric Maxwell theory 
the feature that not all of the gauge fields introduced in the covariant 
derivative need be independent. With the choice $t_{\underline a \, 
\underline b}{}^{\underline c} = 0$ we find
\be
\Big[ ~ \nabla_{\underline a} \, , \, \nabla_{\underline b} ~ \Big] ~=~ 
\fracm 12 r_{\underline a \, \underline b \, \underline c }{}^{\underline d}\,
{\cal M}_{\underline d }{}^{\underline c} ~~~~,
\ee
and the resulting theory is ``torsionless.'' (In differential geometry
this is a ``Riemann geometry'' as compared to a ``Riemann-Cartan geometry''
where $t_{\underline a \, \underline b}{}^{\underline c} \ne 0$.) Since
in this expression $\o_{\underline a \, \underline b \, \underline c}$
is first order in derivatives of the remaining gauge field ${\rm e}_{\underline 
a}{}^{\underline m}$, the quantity $r_{\underline a \, \underline b 
\, \underline c \, \underline d}$ is second order in derivatives.

One of the curious differences between $r_{\underline a \, \underline b 
\, \underline c \, \underline d}$ and $f_{\underline a \, \underline b}{}^{ 
\cal I}$ in a Yang-Mills theory is that only in the former case is it
true that the ``group indices'' (i.e. the ${\underline c \, \underline
d}$ indices that were originally contracted with ${\cal M}_{\underline c \, 
\underline d}$ generator) are of the same type as the 4D space-time
flat vector indices.  This allows the following contraction
\be
r_{\underline a \, \underline c }~=~
r_{\underline a \, \underline d \, \underline c }{}^{\underline d}
~~~, \ee 
which has no analog in Yang-Mills theory. (This contracted quantity
is called the Ricci tensor.) Given this quantity it is possible
to form one more such contraction
\be
r ~=~ \eta^{\underline a \, \underline c} r_{\underline a \, \underline 
d \, \underline c }{}^{\underline d}  ~~~.
\ee
This object with no remaining free indices is the curvature scalar.

In Yang-Mills theory, there is a result that follows from the Jacobi
identity on the Yang-Mills covariant derivative $\nabla_{\underline 
a} = \pa_{\underline a} + i A_{\underline a}{}^{\cal I} t_{\cal I}$,
\be 
\eqalign{
0 ~=~ & [ \, [ \, \nabla_{\underline a} ~,~ \nabla_{\underline b} \, ] ~,~
\nabla_{\underline c} ~] ~+~ [ \, [ \, \nabla_{\underline b} ~,~ 
\nabla_{\underline c} \, ] ~,~ \nabla_{\underline a} ~] 
~+~ [ \, [ \, \nabla_{\underline c} ~,~ \nabla_{\underline a} \, ] ~,~
\nabla_{\underline b} ~] \cr
&{} \cr \to
0 ~=~ &  \nabla_{\underline a} F_{\underline b \, \underline c}{}^{ 
\cal I} ~+~ \nabla_{\underline b} F_{\underline c \, \underline a}{}^{ 
\cal I} ~+~ \nabla_{\underline c} F_{\underline a \, \underline b}{}^{ 
\cal I} ~~~.
}\ee
This equation is also called the Bianchi identity.  For the gravitationally
covariant derivative $ \nabla_{\underline a} ~=~ {\rm e}_{\underline a} 
{}^{\underline m} \pa_{\underline m} + \fracm12  \o_{\underline a \, \underline 
d}{}^{\underline e} {\cal M}_{\underline e}{}^{\underline d}$ a similar result
is obtained,
\be 
\eqalign{
0 ~=~ & [ \, [ \, \nabla_{\underline a} ~,~ \nabla_{\underline b} \, ] ~,~
\nabla_{\underline c} ~] ~+~ [ \, [ \, \nabla_{\underline b} ~,~ 
\nabla_{\underline c} \, ] ~,~ \nabla_{\underline a} ~] 
~+~ [ \, [ \, \nabla_{\underline c} ~,~ \nabla_{\underline a} \, ] ~,~
\nabla_{\underline b} ~] \cr
&{} \cr \to
0 ~=~ &  \nabla_{\underline a} t_{\underline b \, \underline c}{}^{ 
\underline d} ~+~ \nabla_{\underline b} t_{\underline c \, \underline a}{}^{ 
\underline d} ~+~ \nabla_{\underline c} t_{\underline a \, \underline b}{}^{ 
\underline d} \cr
&-~  t_{\underline a \, \underline b}{}^{\underline e} t_{\underline e \, 
\underline c}{}^{\underline d}~-~  t_{\underline b \, \underline c}{}^{
\underline e} t_{\underline e \, \underline a}{}^{\underline d}~-~  t_{
\underline c \, \underline a}{}^{\underline e} t_{\underline b \, 
\underline c}{}^{\underline d}  \cr
&+~ r_{\underline a \, \underline b \, \underline c }{}^{\underline d}
~+~ r_{\underline b \, \underline c \, \underline a }{}^{\underline d}
~+~ r_{\underline c \, \underline a \, \underline b }{}^{\underline d}
~~~, \cr
&{~} \cr 
0 ~=~ &  \nabla_{\underline a} r_{\underline b \, \underline c}
{}_{\underline k \, \underline \ell} ~+~ \nabla_{\underline b} r_{\underline c 
\, \underline a}{}_{\underline k \, \underline \ell} ~+~ \nabla_{\underline c} 
r_{\underline a \, \underline b}{}_{\underline k \, \underline \ell} \cr
&-~  t_{\underline a \, \underline b}{}^{\underline e} r_{\underline e \, 
\underline c}{}_{\underline k \, \underline \ell}~-~  t_{\underline b \, 
\underline c}{}^{\underline e} r_{\underline e \, \underline a}{}_{\underline
k \, \underline \ell}~-~ t_{\underline c \, \underline a}{}^{\underline e} 
r_{\underline b \, \underline c}{}_{\underline k \, \underline \ell}
 ~~~.
}\ee
Under the restriction that $t_{\underline a \, \underline b}{}^{\underline c}
= 0$, the first of these becomes a purely algebraic restriction on
$r_{\underline a \, \underline b}{}_{\underline c \, \underline d}$.
Under this same restriction the second set then takes a form that is 
identical to that of a Yang-Mills theory. So the similarities between 
the two types of covariant derivatives are very obvious.

\subsection{Field Equations and Actions}

~~~~However, we still have a few items to complete. We may start with the 
issue of the field equations. In Yang-Mills theory these are known to 
be
\be
\eta^{\underline a \, \underline b} \nabla_{\underline a} F_{\underline 
b \, \underline c}{}^{ \cal I} ~=~ 0 ~~~.
\ee
Like many equations of physical significance this is a second
order differential equation for the gauge field $A_{\underline a}
{}^{\cal I}$.  We have already seen that the condition that the
field strength $t_{\underline a \, \underline b}{}^{\underline c}
= 0$ implies that the remaining field strength $r_{\underline a \, 
\underline b \, \underline d \, \underline e}$ is second order in
derivatives of the gauge field ${\rm e}_{\underline a }{}^{\underline m}$.
So if this gauge field also obeys a second order differential
equation, this corresponds to something that must be linear in 
$r_{\underline a \, \underline b \, \underline d \, \underline e}$.
The Yang-Mills equations are derivable from an action principle.
It would be nice if this were also true for the field equation
for ${\rm e}_{\underline a }{}^{\underline m}$.  In order to find this,
we must go back and solve a problem which we left dangling, ``How
is $K$-gauge invariance to be realized at the level of actions?''

The infinitesimal version of (\ref{eq:wvg}) implies
\be 
\d_K {\cal L} (\phi, \, \nabla_{\underline a} \phi) ~=~
\xi^{\underline m}(x) \pa_{\underline m} {\cal L} (\phi, \, 
\nabla_{\underline a} \phi) ~~~,
\ee
and this is clearly not invariant. To fix this problem we go back 
and make an observation about the gauge field ${\rm e}_{\underline 
a }{}^{\underline m}$. Since it is a 4 $\times$ 4 matrix, its 
determinant may be calculated
\be
{\rm e} ~\equiv ~ det ({\rm e}_{\underline a }{}^{\underline m}) ~~~.
\ee
We also know the gauge variation of ${\rm e}_{\underline a }{}^{\underline m}$
given in (\ref{eq:wxn}) and this can be used to calculate the quantity
$h_{\underline a }{}^{\underline b}$ in (\ref{eq:wxh}). This takes the form
\be
h_{\underline a }{}^{\underline b} ~=~ \xi^{\underline n} (\pa_{\underline 
n} {\rm e}_{\underline a}{}^{\underline m}) {\rm e}_{\underline m}{}^{
\underline b} ~-~{\rm e}_{\underline a}{}^{\underline n} (\pa_{\underline n} 
\xi^{\underline m}) {\rm e}_{\underline m}{}^{\underline b} ~-~ 
\l_{\underline a}{}^{\underline b} ~~~. 
\ee
Next we use equation (\ref{eq:wxh}) applied to directly on ${\rm e}$ along 
with the identity $det(M) = exp[ Tr (ln M)]$ which is valid for any
matrix to derive,
\be 
\d_K \, {\rm e} ~=~ {\rm e} \, h_{\underline a }{}^{\underline a} 
~=~ {\rm e} \, \xi^{\underline m} \pa_{\underline m} \, {\rm e} ~-~ 
{\rm e} \, (\pa_{\underline m} \xi^{\underline m}) ~~~.
\ee
This then implies
\be
\d_K \, {\rm e}^{-1} ~=~ (\pa_{\underline m} \xi^{\underline m} {\rm e}^{-1} )
~~~.
\ee
This is a most fortunate result! Additionally we then see
\be 
\eqalign{
\d_K ({\rm e}^{-1} \, {\cal L} ) &=~ (\d_K {\rm e}^{-1} )\, {\cal L}
~+~ {\rm e}^{-1} \, (\d_K {\cal L}) \cr
&=~ [ (\pa_{\underline m} \xi^{\underline m} {\rm e}^{-1} ) ] {\cal L}
~+~ {\rm e}^{-1} \xi^{\underline m} \pa_{\underline m} {\cal L} \cr
&=~ \Big[ \pa_{\underline m}  (\xi^{\underline m} {\rm e}^{-1}  {\cal L}
) ~ \Big] ~~~.
}\ee
So if we integrate the quantity ${\rm e}^{-1} {\cal L}$, we find
\be 
\d_K \Big[ ~ \int d^4 x ~ {\rm e}^{-1} \, {\cal L} ~ \Big] ~=~
\int d^4 x \, \Big[ ~ \d_K ( {\rm e}^{-1} \, {\cal L}) ~ \Big]
 ~=~\int d^4 x ~ \Big[ \pa_{\underline m}  (\xi^{\underline 
m} {\rm e}^{-1}  {\cal L} ) ~ \Big] ~~~.
\ee 
In the middle result, we have used the fact that the operator
$\d_K$ is defined to act on fields and as such it may be
freely ``pushed'' past the integration measure $d^4 x$.
We thus reach the following understanding of gauge invariance.  
Although no Lagrangian is invariant under the action of $\d_K$, 
the integrated product ${\rm e}^{-1} {\cal L}$ changes only by surface 
terms under the action of the Lie-algebra based view of the
gravitational gauge group.  As long as such surface terms are
negligible, there is an effective gauge invariance of the
theory. (Note also that this is curiously similar to the case of
super translation invariance (\ref{eq:ehw},\ref{eq:ehv}).)

We may now complete the effort to write the Einstein field
equations derivable from an action. The required action
(the Einstein-Hilbert action) is simply as
\be
{\cal S}_{E\, H} ~=~ - { 3\over {\k^2}} \, \int d^4 x ~ {\rm e}^{-1}\,
r(\o(e),e) ~~~, \label{eq:whv}
\ee
written in the conventions of {\it {Superspace}}. We have written
it in this form to emphasize that the spin-connection $\o$ is
the one associated with the vanishing torsion. The variation of
this action (using the formula in (\ref{eq:wvz}) yields the Einstein
field equation
\be
{\d {\cal S}_{E\, H} \over h^{\underline a \, \underline b}} ~=~ 0 
~~\to ~~ {\cal E}_{\underline a \, \underline b} ~\equiv~
r_{\underline a \, \underline b} ~-~ \fracm 12 \, \eta_{
\underline a \, \underline b} r ~=~ 0 ~~~,
\ee
as the analog of the Yang-Mills free field equations.  (The 
tensor ${\cal E}_{\underline a \, \underline b}$ is often called 
the ``Einstein tensor.'') If we start with a more complicated 
action constructed from the sum of the Einstein-Hilbert action 
with other actions ${\cal S}_M$, then the field equations for 
${\rm e}_{\underline a }{}^{\underline m}$ take the forms
\be
r_{\underline a \, \underline b} ~-~ \fracm 12 \, \eta_{
\underline a \, \underline b} r ~=~  \frac {\k^2}3 \Big(
{\d {\cal S}_{M} \over h^{\underline a \, \underline b}} 
\Big) ~~~.
\ee
The quantity appearing on the rhs of this equation is the
energy-momentum tensor due to the fields in ${\cal S}_M$.
It is also possible to write one additional term that
depends solely on the gauge field ${\rm e}_{\underline a}
{}^{\underline m}$ and modifies its equation of motion by 
terms no more than quadratic in derivatives.  This is the
famous ``cosmological term''
\be
{\cal S}_{cos} ~=~ - { 3\over {\k^2}} \, \int d^4 x ~ {\rm 
e}^{-1}  ~~~.
\ee

\subsection{Principles for Matter Coupling}

~~~~The minimal coupling procedure works for matter fields such
as scalars or spinors. For example starting with our massless
spin-0 or Weyl spinor actions we are led to 
\be
{\cal S}_{scalar} ~=~ \int d^4 x ~ {\rm e}^{-1} \, \Big[ ~ -
\fracm 12 (\nabla^{\underline a} {\Bar A})\, (\nabla_{\underline a} 
A) ~ \Big] ~~~,
\ee
\be
{\cal S}_{spinor} ~=~ \int d^4 x ~ {\rm e}^{-1} \, \Big[ ~ - i \,
{\Bar \psi}{}^{\Dot \a} \, \nabla_{\underline a}\psi^{\a} ~ \Big]
~~~.
\ee

Note that we can even do some ``bizarre'' things once we have
the gravitationally covariant derivative ${\nabla_{\underline a}}$
and know how to use ${\rm e}^{-1}$ to construct invariant actions.
For example, a Schr\" odinger field subject to a force with 
potential $W(x)$, has the action obtained by the minimal coupling
prescription given by 
\be
{\cal S} ~=~ \int d^4 x ~ {\rm e}^{-1} \, \Big[ \,\, i \Big({\Bar 
\Psi} \nabla_0 \Psi ~-~ ( \nabla_0 {\Bar \Psi}) \Psi \Big) ~-~ 
\fracm1{2m} ( {\vec \nabla} {\Bar \Psi}) \, ( {\vec \nabla} { 
\Psi}) ~-~ W \, {\Bar \Psi} \, { \Psi} ~\Big] ~~~.~~ \ee
This action possesses local $P$-invariance! However, the local
SO(1,3)-invariance is broken down to local SO(3)-invariance
by the fact that the temporal component of the the gravitational 
covariant derivative ($\nabla_0$) is treated in a way that is not 
symmetrical with regards to the spatial components (${\vec 
\nabla}$). Note that the spin-connection is absent entirely from
this action if we choose ${\cal M}_{\a}{}^{\b} \Psi$ $=$
${\Bar {\cal M}}_{\Dot \a}{}^{\Dot \b} \, \Psi$ $=$ $0$.

The minimal coupling procedure, however, does not apply to matter
fields that possess their own gauge invariances not related to
the symmetries generated by the $K$-operator.  This is most
clearly seen by considering the case of Yang-Mills symmetries
simultaneously present with the $K$-symmetries.  This case is
simplest when treated by an appropriate modification of the
gravitationally covariant derivative to include Yang-Mills
fields
\be
\nabla_{\underline a} ~=~ {\rm e}_{\underline a} {}^{\underline m} 
\pa_{\underline m} ~+~ \fracm12  \o_{\underline a \, \underline 
d}{}^{\underline e} {\cal M}_{\underline e}{}^{\underline d}
~-~ i\,  A_{\underline a}{}^{\cal I} t_{\cal I} ~~~.
\ee
The commutator of this modified derivative defines the
Yang-Mills field strength via
\be
\Big[ ~ \nabla_{\underline a} \, , \, \nabla_{\underline b} ~ \Big] 
~=~ t_{\underline a \, \underline b}{}^{\underline c} \, \nabla_{
\underline c} ~+~ \fracm 12 r_{\underline a \, \underline b \, 
\underline c }{}^{\underline d}\, {\cal M}_{\underline d }{}^{
\underline c} ~-~ i \, f_{\underline a \, \underline b}{}^{\cal 
I} t_{\cal I} ~~~,
\ee
and leads to a modified definition of $f_{\underline a \, \underline 
b}{}^{\cal A}$
\be 
f_{\underline a \, \underline b}{}^{\cal I} ~\equiv~  {\rm e}_{\underline 
a}{}^{\underline m} \pa_{\underline m} \, A_{\underline b}{}^{\cal I}
~-~ {\rm e}_{\underline b}{}^{\underline m} \pa_{\underline  m} \, 
A_{\underline a}{}^{\cal I} ~-~ c_{\underline a \, \underline b}{}^{\underline 
c} \, A_{\underline c}{}^{\cal I} ~-~ i \,  f_{\cal A \, \cal B}{}^{\cal I}
A_{\underline a} {}^{\cal A} \, A_{\underline b} {}^{\cal B}
~~~. \ee
This result differs from that obtained by naive minimal coupling 
($\pa_{\underline a} \to {\rm e}_{\underline a}{}^{\underline m} 
\pa_{\underline m}$) by the term proportional to the anholonomity
$c_{\underline a \, \underline b}{}^{\underline c}$.  However,
once this definition is used the action for the Yang-Mills
field takes its familiar form,
\be
{\cal S}_{YM} ~=~ \int d^4 x ~ {\rm e}^{-1} \, \Big[ ~
- \fracm 18 \, f_{\underline a \, \underline b}{}^{\cal 
I} \, f^{\underline a \, \underline b}{}^{\cal 
I} ~\Big] ~~~.
\ee

\subsection{Implicit Lorentz-Invariant and Higher Derivative Gravitation}

~~~~From the view on the Lie algebraic origins of theories of 
gravitation, there are some issues that logically can be seen in 
their proper prospective.  One such point that can be seen  
debated (almost continuously) in the literature is the issue of 
torsion and whether it should be considered as fundamental.
The only role of the torsion is that it determines the
form of the spin-connection $\o_{\underline a \, \underline 
b \, \underline c}$. If we use the analogy with supersymmetric
field theories, the spin-connection is simply an auxiliary field. 
To see how irrelevant is this auxiliary field, consider the action 
given by
\be
{\cal S}_{C^2} ~=~ \int d^4 x ~ {\rm e}^{-1} \, \Big[ \, 
c_0 \, c^{\underline a \, \underline b \, \underline c}\,
c_{\underline a \, \underline b \, \underline c}\,
~+~ 
c_1 \, c^{\underline a \, \underline b \, \underline c}\,
c_{\underline c \, \underline a \, \underline b}\,
~+~ 
c_2 \, ( \, c^{\underline a}{}_{\underline b}{}^{\underline 
b}\,)^2 ~\Big] ~~~, \ee
where $c_0$, $c_1$ and $c_2$ are constants.  This action is 
also clearly independent of the spin-connection and is local a 
$P$-invariant. However, like our bizarre example of the 
Schr\" odinger field minimally coupled to gravity, it seems
as though this action is also not ${\cal M}$-gauge invariant.
In order to verify this, we may perform a local Lorentz
gauge transformation upon ${\rm e}_{\underline a}{}^{\underline
m}$ to find
\be \eqalign{ {~~~}
\d_{LL} {\rm e}_{\underline a}{}^{\underline m} &=~ - \, \l_{\underline a}
{}^{\underline b} \, {\rm e}_{\underline a}{}^{\underline m}
\to \cr
\d_{LL} {\cal S}_{C^2}
&=~  \int d^4 x ~ {\rm e}^{-1} \, \Big[ \, ~ ( \, 1 \, + \, 
4c_0 \,) \,   c^{\underline b \, \underline a \, \underline d} \, 
(\, {\rm e}_{\underline a} \l_{\underline b \, \underline d} 
\,) \cr
&{~~~~~~~~~~~~~~~~~~~~~~~~~} +~ ( \, 1 \, - \, 2 c_1 \,) \,
c^{\underline d \, [ \underline a \, \underline b ] } \, (\, 
{\rm e}_{\underline a} \l_{\underline b \, \underline d} \,) \cr
&{~~~~~~~~~~~~~~~~~~~~~~~~~}
 +~ 2 \, ( \, c_2 \, - \, 1\,) \, c^{\underline a}
{}_{\underline b}{}^{\underline b  } \, (\, {\rm e}_{\underline 
d} \l_{\underline b}{}^{ \,\underline d} \,) \cr
&{~~~~~~~~~~~~~~~~~~~~~~~~~} -~ 2 (\,( \,{\rm e}^{\underline a} ~-~ 
c^{\underline a }{}_{\underline b}{}^{\underline b  } ) \,
{\rm e}^{\underline c} \l_{\underline a \, \underline c} \,)
~\Big] ~~~. \label{eq:win}
} \ee
However, we have an identity of the form
\be
{\rm e}^{-1} (\,( \,{\rm e}^{\underline a} ~-~ 
c^{\underline a }{}_{\underline b}{}^{\underline b  } ) \,
X_{\underline a} \,) ~=~ \pa_{\underline m} ( \, {\rm e}^{-1} 
X_{\underline a} e^{\underline a \, \underline m} \,) ~~~,
\label{eq:wig}
\ee
substituting $X_{\underline a} = {\rm e}^{\underline c} \l_{\underline 
a \, \underline c}$ and it can be seen that for the special choice of 
coefficients given by $-4 c_0 = 2 c_1 = c_2 = 1$, the local Lorentz 
variation of the action is a total divergence.  In our Lie-algebraic 
motivated approach to understanding gravity we always neglect 
total divergence terms, so effectively for this choice ${\cal 
S}_{C^2}$ is both a $P$-gauge invariant and a ${\cal M}$-gauge 
invariant! 

A consequence of the result (\ref{eq:wig}) is that the rule for 
integration-by-parts takes the form
\be
\int d^4 x ~ {\rm e}^{-1} ~ U^{\underline a \, [X]}\,
{\rm e}_{\underline a} V_{ [X]} ~=~ - 
\int d^4 x ~ {\rm e}^{-1} ~ [ \, ( \, {\rm e}_{\underline a}
U^{\underline a \, [X]} \, )\,  V_{ [X]} ~-~
c_{\underline a \, \underline b}{}^{\underline b}
U^{\underline a \, [ X]} \, V_{ [X]} \, ] ~~~. \label{eq:ibp}
\ee
In this expression, we have used the notation $[X]$ to denote
any choice of Lorentz vector or spinor indices.

The equation of motion that follows from ${\cal S}_{C^2}$ is
clearly of the form of second order derivatives operating on
${\rm e}_{\underline a}{}^{\underline m}$ and is, in fact, the usual 
Einstein field equation for free space although it appears
in the form
\be
\eqalign{ {~~~}
&{\rm e}_{( \underline k} c_{\underline l ) \, \underline c  }
{}^{\underline c} ~-~ \fracm 12  c_{ (k \underline a | }{}^{
(\underline c \underline d )} c_{ | \underline l ) \, \underline 
c \,  \underline d}~-~ c_{ ( \underline k | \, \underline c }
{}^{\underline c} c_{| \underline k ) \,  \underline d }
{}^{\underline d} ~+~ \fracm 14 c_{  \underline c \, 
\underline d ( \underline k|} c^{\underline c \,  \underline d }
{}_{ | \underline l )} \cr
&-\, 2 {\rm e}^{-1} [ \, ({\rm e}_{\underline a} \, - \, c_{\underline a
\, \underline b}{}^{\underline b} )( \, \eta_{\underline k \underline l}
\, c^{\underline a}{}_{ \underline d}{}^{\underline d} \, - \, \fracm 12
c^{\underline a}{}_{( \underline k \, \underline l )} ) \, ] \cr
&- ~ \eta_{\underline k \, \underline l}
\Big[ \, - \fracm 14 \, c^{\underline a \, \underline b 
\, \underline c}\, c_{\underline a \, \underline b \, 
\underline c}\, ~+~ \fracm 12 \, c^{\underline a \, \underline b 
\, \underline c}\, c_{\underline c \, \underline a \, 
\underline b}\, ~+~ ( \, c^{\underline a}{}_{\underline 
b}{}^{\underline b}\,)^2 ~\Big]
= 0 ~~~.~~}
 \ee

The issue of how (or if) $\o_{\underline a \, \underline b \, 
\underline c}$ is specified becomes particularly important
when one considers theories (such as the effective gravitational 
interaction that emerges at the low-energy limit of various 
string theories\footnote{To my knowledge, {\underline 
{no}} {\underline {rigorous}} derivations of quantum 
nonlinear gravitational effective \newline ${~~~~~}$ actions have 
ever been found for any theory other than strings.}) which possess 
what are commonly referred to as higher curvature terms. A generic 
term of this form is given by 
\be
{\cal S}_{H.\, D.} ~=~ \int d^4 x ~ {\rm e}^{-1} \, \Big[ \,\, 
{f }( \nabla_{\underline a}, \, r_{\underline a \, \underline 
b \, \underline c \, \underline d} \, )  ~\Big] ~~~,~~~  
r_{\underline a \, \underline b \, \underline c \, \underline 
d} \, = \, r_{\underline a \, \underline b \, \underline 
c \, \underline d}(\o(e), \, e) ~~, \label{eq:wee}
\ee
where $f$ is a function that is quadratic or higher in the curvature 
tensor (and its possible contractions) and may depend on of the 
gravitationally covariant derivative.  Since $\o_{\underline a 
\, \underline b \, \underline c}(e)$ is solely a function of ${\rm e}_{
\underline a}{}^{\underline m}$, the only dynamics here are associated 
with the graviton. In fact, the action above must satisfy (up to 
a total derivative) 
\be
\int d^4 x ~ {\rm e}^{-1} \, \Big[ \,\, {f }( \nabla_{\underline 
a}, \, r_{\underline a \, \underline b \, \underline c  \, \underline 
d} \, )  ~\Big] ~=~ \int d^4 x ~ {\rm e}^{-1} \, \Big[ \,\, 
g( {\rm e}_{\underline a}, \, c_{\underline b \, \underline c \, \underline 
d } \, )  ~\Big]  ~~~~~,~~ 
\ee
for some function $g$.  The simplest example of this class of identities
is provided by
\be 
\eqalign{
f &=~ \eta^{\underline a \, \underline c}\,  \eta^{\underline b \, 
\underline d} ~ r_{\underline a \, \underline b \, \underline c  
\, \underline d} (\o({\rm e}),{\rm e}) ~~,~~\cr
g &=~ \Big[ \, - \fracm 14 \, c^{\underline 
a \, \underline b \, \underline c}\, c_{\underline a \, \underline b 
\, \underline c}\,~+~ \fracm 12 \, c^{\underline a \, \underline b \, 
\underline c}\, c_{\underline c \, \underline a \, \underline b}\, 
~+~ ( \, c^{\underline a}{}_{\underline b}{}^{\underline b}\,)^2 ~
\Big] ~~~.}
\ee

On the other hand, if we start with the action
\be
{{\cal S}'}_{H.\, D.} ~=~ \int d^4 x ~ {\rm e}^{-1} \, \Big[ \,\, 
{f }( {\widehat {\nabla}}_{\underline a}, \, {\widehat r}_{\underline 
a \, \underline b \, \underline c \, \underline d} \, )  ~\Big] 
~~~,~~~  {\widehat r}_{\underline a \, \underline b \, \underline c 
\, \underline d} \, = \, r_{\underline a \, \underline b \, 
\underline c \, \underline d}({\widehat \o}, \,e) ~~,  \label{eq:tzo}
\ee
where ${\widehat \o}_{\underline a \, \underline b \, \underline c}$ is 
left as an unspecified function, then in addition to the dynamics of 
${\rm e}_{\underline a}{}^{\underline m}$, one must also encounter dynamical 
equations for ${\widehat \o}_{\underline a \, \underline b \, \underline 
c}$! Thus, the physics described by ${\cal S}_{H.\, D.}$ is radically 
different from that described by ${{\cal S}'}_{H.\, D.}$ even though 
exactly the same function ${f}$ appears in the two actions.  

As mentioned above, the low-energy purely gravitational field approximation 
of string-type theories (i.e. where all fields other than ${\rm e}_{
\underline a}{}^{\underline m}$ are set to zero) takes the form
\footnote{The $\b$-function technique within the confines of bosonic and 
NSR $\s$-models is \newline ${~~~\,~}$ known to provide an explicit
way to derive this expansion.  The fact that ${\cal S}_{H.\, D.}$
\newline ${~~~\,~}$ and not ${{\cal S}'}_{H.\, D.}$ appears in string 
theories is obvious since a spacetime spin con-
\newline ${~~~\,~}$ nection is `removable' from the $\s$-model formalism.}
\be
{\cal S}_{stringy-grav.} ~\propto~ {\cal S}_{C^2} ~+~ {\cal S}_{H.\, D.} ~~~.
\ee
It is of interest to note that there exists a restricted form of
the higher derivative action that corresponds to a ``velocity
expansion.''  Note that the quantity $c_{\underline a \, \underline b \, 
\underline c }$ may be thought of as a velocity.  One may thus contemplate 
the special class of higher derivative terms of the form
\be
{\cal S}_{H.\, D.}({vel}) ~=~  \int d^4 x ~ {\rm e}^{-1} \,
\sum_{\ell = 3}^{\infty} \k^{ \ell - 4} \, L_{\ell} ~ {\cal P}^{(\ell) \, 
{\underline a}_1 ... {\underline a}_{\ell} \, {\underline b}_1 ... 
{\underline b}_{\ell} }_{~~~\, {\underline c}_1 ... {\underline c}_{\ell} } 
\, \Big( \prod_{j=1}^{\ell } c_{{\underline a}_j \, {\underline b}_j \, 
}{}^{{\underline c}_j } \Big)~~~.  \label{eq:gggg}
\ee
This expansion is roughly equivalent to the well-known chiral perturbation
theory (``soft pion'') representation of meson physics.  In this 
expression, $L_{\ell}$ are a set of dimensionless numbers and ${\cal 
P}^{(\ell) \, {\underline a}_1 ... {\underline a}_{\ell} \, {\underline 
b}_1 ... {\underline b}_{\ell} }_{~~~\, {\underline c}_1 ... {\underline 
c}_{\ell} } $ correspond to a set of constant tensors chosen in 
complete generality consistent with the $K$-gauge invariance of 
this action.

We close this section by noting that the quantity 
$c_{{\underline a} \, {\underline b} \, }{}^{{\underline c} }$
is the closest gravitational analog to the field strength 
$ f_{{\underline a} \, {\underline b} \, }{}^{\cal I}$ of
Yang-Mills theories. Accordingly, we might expect that many of the properties 
of Yang-Mills theories should have `echoes' for gravitational 
theories that may be recognized by using this analogy.

\newpage
\section{Fourth Lecture: Junior 4D, N = 1 Superfield Theory}
{\it {Introduction}}

~~~~~~In today's lecture, we will discuss the issue of quantization 
via functional methods.  The discussion will include the preliminary
development of superfunctional techniques, definitions of super
propagators and supergraphs through the examples of the chiral scalar
and ``gauge" multiplets.  Let me begin by saying something about 
what I am not going to do.

In going from being an official student to a practitioner (e.g. unofficial
student) of QFT, I learned or have learned a number of quantization
techniques:

${~~~~}$A. Canonical, \newline \indent
${~~~~}$B. Path Integral, \newline \indent
${~~~~~~~~}$(i) BRST, \newline \indent
${~~~~~~~~}$(ii) Batalin - type, \newline \indent
${~~~~~~~~~~~~}$(a.) BV, \newline \indent
${~~~~~~~~~~~~}$(b.) BFV. \newline \indent

Although the canonical method is considered the most ``rigorous," I know 
of no completely successful examples of its implementation within the 
realm of supersymmetrical theories.  I suspect that the reason for this 
is the fact that 4D, $N=1$ superspace makes fundamental use of 
relativistic spinors (i.e. the Grassmann coordinates belong to this 
representation).  The 4-momentum can be written as $P_{\underline m} 
= (E,\,{\vec p})$ and in the canonical approach $E$ or the Hamiltonian 
is treated differently from the other components.  Thus, there is a need 
to break 4D Lorentz invariance and hence break manifest supersymmetry.  
Instead, the discussion presented will very closely follow that of 
{\it {Superspace}}.

Accordingly, we will not discuss canonical approaches.  In this 
lecture, we will exclusively consider the path integral approach using 
at most BRST type techniques.  This material can be found in our
book {\it {Superspace}}. So I won't have anything to say about 
superfield Batalin-type (BV, BFV) quantization with its attendant 
use of ``the Master Equation" either.  I do point out, however, that 
M.~Grisaru et.~al.~have recently \cite{20} discussed such matters in 
connection with the quantization of the complex linear multiplet 
(which was introduced as the dual to the chiral scalar multiplet 
two lecture ago).  The results are most interesting and complicated.  
The class of Batalin-type quantization techniques are the most advanced 
known to date.  However, in the attempt to quantize the complex linear 
multiplet, the results derived do not seem easily applicable for the 
derivation of additional results in interacting theories.  This raises 
the question of the need for as yet unknown advances in the Batalin-type 
methods.
$${~}$$
\newpage
\noindent
{\it {Lecture}}
\subsection{A Little about Superfunctional Differentiation}

~~~~We can begin by noting that in ordinary four dimensional quantum field theory 
treated via functional methods, the Dirac delta function plays a fundamental 
role in the definition of functional differentiation.  Given some function 
$J(x)$ (dependent on the 4-vector  $x^{\underline m}$) we define
\be
\parvar{J(x)}{J(x')} ~\equiv~ \d^{(4)} (x ~-~ x') ~~~,
\label{eq:hzi} \ee
and since the operator $\parvar{~~~~}{J(x)}$ is essential, we need to
have a definition such that given a superfunction $J(z)$ we may define
\be
\parvar{J(z)}{J(z')} ~\equiv~ \d^{(4)} (x ~-~ x') \, \d^{(4)} (\q ~-~ {\q}') 
 ~~~.
\ee
But what is the meaning of $\d^{(4)} (\q ~-~ {\q}')$? For a single bosonic 
coordinate y, we know that
\be
\int_{-\infty}^{\infty} d y ~ f(y) \,  \d (y ~-~ y') ~=~   f(y') ~~~.
\ee
So we need something similar in a space with a single bosonic coordinate
y and a single fermionic one $\zeta$.  A real superfunction $f(y,\zeta)$ 
has the form
\be
f(y,\,\zeta) ~=~ a(y) ~+~ i \, \zeta \, \beta (y) ~~~. 
\ee
If there is a super Dirac delta function $\d ({\zeta} - {\zeta}')$ then
we expect
\be
\int d \zeta ~ f(y, \, \zeta) \,  \d ({\zeta} - {\zeta}') ~=~  f(y, \, 
{\zeta}') ~~~.
\ee
To find an explicit representation for $\d ({\zeta} - {\zeta}')$ we recall
\be
\eqalign{
\int  d \zeta ~ 1 &\equiv ~ \Big[\,  \partder{~\,}{\zeta} ~+~ i \, \fracm 12
\zeta \, \partder{~~}{y}  \, \Big] \, 1 {\Big |} ~=~ 0 ~~~, \cr
\int  d \zeta ~ \zeta &\equiv ~ \Big[\,   \partder{~\,}{\zeta} ~+~ i 
\, \fracm 12 \zeta \, \partder{~~}{y}   \, \Big] \, \zeta{\Big |} ~=~ 1 ~~~.
}\ee
We also note $ f(y, \, \zeta) \,({\zeta} - {\zeta}') \, = \, 
\zeta f(y, \, {\zeta}') \, - \, a(y) {\zeta}'$. Thus we have
\be
\int d \zeta ~ f(y, \, \zeta) \, ({\zeta} - {\zeta}') ~=~  f(y, \, 
{\zeta}') ~~~.
\ee
so that is it consistent to define $\d ({\zeta} - {\zeta}') = ({\zeta} - 
{\zeta}')  $.  In a superspace with more fermionic dimensions, we can 
proceed inductively to derive in our case
\be
\eqalign{
\d^{(4)} ( \q ~-~ {\q}' ) &\equiv~ \d^{(2)} ( \q ~-~ {\q}' ) \, 
\d^{(2)} ( {\bar \q} ~-~ {\bar \q}' ) ~~~, \cr
\d^{(2)} ( \q ~-~ {\q}' )   &\equiv~ \fracm 12 \,  ( \q ~-~ {\q}' )^{\a} \, 
( \q ~-~ {\q}' ){}_{\a} \,      ~~~,  \cr
\d^{(2)} ( {\bar \q} ~-~ {\bar \q}' )   &\equiv~ \fracm 12 \, 
( {\bar \q} ~-~ {\bar \q}' )^{\Dot \a}  \, ( {\bar \q} ~-~ {\bar \q}' 
){}_{\Dot \a}  ~~~.
}\ee
So given a general function $J(z)$, we define its functional derivative
as in (\ref{eq:hzi}).  Now the super operator $\parvar{~~~~}{J(x)}$ can be applied 
to superspace functionals. For example,
\be \eqalign{ {~~~}
{\cal S} &=~ \int d^4 x \, d^2 \q \, d^2 {\bar \q} ~ A(z) \, J(z) ~\equiv
\int d^8 z ~  A(z) \, J(z) ~~~ \cr
&\to~ \parvar{{\cal S}}{J(z')} ~=~ \int d^8 z ~  A(z) \, \d^{(8)} ( z ~-~ {z}' )
~=~ A(z') ~~~, \cr
&{~~~~~}{\rm{and~ if}} ~~ \d^{(8)} ( z ~-~ {z}' ) ~\equiv~ \d^{(4)} ( x ~-~ {x}' ) 
\, \d^{(4)} ( \q ~-~ {\q}' )  ~~~, \cr
&{~~~~~}{\rm{if}} ~~  \parvar{{J(z)}}{J(z')} ~\equiv~ \d^{(8)} ( z ~-~ {z}' ) 
~~~. }\ee

\subsection{On Superdeterminants} 

~~~~It sometimes occurs that we require the definition of a super
determinant (i.e. to calculate the Jacobian of a change of field
variables).  For this purpose, it is first useful to define a
supermatrix. For purposes of illustration, we return to the one
dimensional superspace described by $y$ and $\zeta$. We 
introduce a supervector and supermatrix
\be
z ~=~  \left(\begin{array}{c}
~y \\
~\zeta \\
\end{array}\right) ~~,~~~
M_s ~=~ \left(\begin{array}{cc}
~A & ~~B\\
{}~&~\\
~C & ~D \\
\end{array}\right) 
~~~, \ee
such that the supermatrix can act on the supervector $z$. We see
\be
z' ~=~ M_s \, z ~~\to ~~ \left(\begin{array}{c}
~y' \\ ~{\zeta}' \\
\end{array}\right) ~=~ \left(\begin{array}{cc}
~A & ~~B\\
{}~&~\\
~C & ~D \\
\end{array}\right) \, \left(\begin{array}{c}
~y \\ ~\zeta \\
\end{array}\right) ~~.
\ee
But $y^\prime$ will \underline{not} be bosonic if B is an ordinary 
(non-fermionic) number.  Similarly, $\zeta^\prime$ will \underline{not} 
be fermionic if C is an ordinary number.  Thus B and C must be Grassmann 
numbers.  How does one define a superdeterminant for $M_s$?  This problem 
was solved long ago by Arnowitt, Nath and Zumino \cite{200}. We will give 
below a definition that leads to their results.
\be
sdet(M_s) ~\equiv~ K_0 \, \int dy \, dy' \, d \zeta \, d {\zeta}' ~
exp [ \, - \, {\Tilde z}^t \, M_s \, z  \,] ~~~,~~~  {\Tilde z}^t
~\equiv~ ({\Tilde z}, \, {\Tilde \zeta}) ~~~.
\ee
where the constant $K_0$ is chosen so that $sdet({\rm I})=1$ By shifting
${\Tilde y}$ and $y$ according to ${\Tilde y} \, \to \, {\Tilde y} \, - \,
{\Tilde \zeta} C A^{-1}$ and $y \, \to \, y \, - \, A^{-1} B \zeta$ the 
exponential becomes
\be
exp [ \, - \, {\Tilde z}^t \, M_s \, z  \,] ~=~ exp [ \,
- {\Tilde y} \, A\,  y \,\, - ~ {\Tilde \zeta} \, (\, D ~-~ C \, 
A^{-1}\, B \, ) \zeta \, ] ~~~.
\ee
Instead we can choose to shift the Grassmann coordinates according to $
{\zeta}\, \to \,$ and $ {\Tilde \zeta}\, \to \, $ so that the exponential 
becomes
\be
exp [ \, - \, {\Tilde z}^t \, M_s \, z  \,] ~=~ exp [ \,
- {\Tilde y} \,  (\, A ~-~ B \, D^{-1}\, C \, )   y \,\, - ~ 
{\Tilde \zeta} \,D \, \zeta \, ] ~~~.
\ee
After performing either of these shifts, the integrals can be done 
and yield
\be
sdet(M_s) ~=~ {A \over  ~D ~-~ C \, A^{-1}\, B~  } ~=~
{ ~A ~-~ B \, D^{-1}\, C~   \over D} ~~~.
\ee
If we replace $A, \,B, \,C$, and $D$ by $2 \times 2$ matrices then $y$ and
$\zeta$ must become 2-component entities.  The same is true for ${\Tilde y}$
and ${\Tilde \zeta}$. The integrals then yield
\be
sdet(M_s) ~=~ {det(A) \over ~det[ D ~-~ C \, A^{-1}\, B ]~ } ~=~
{ ~det [ A ~-~ B \, D^{-1}\, C ]~  \over det(D) } ~~~.
\ee
This is the form of the general answer for a supermatrix. This formula
implies that either $A$ or $D$ should be nonsingular in order for the
superdeterminant to have a well defined meaning.

\subsection{Review of the Functional Quantization of the Scalar 
Field}

~~~~In order to orient ourselves, let's review ordinary 4D ``$\phi^4$" theory.
\be 
{\cal S} ~=~ \int d^4 x ~ \Big[ ~ \fracm 12 \, \phi (\, {}_{\bo} ~-~
m^2 \, ) \phi ~-~ \fracm{\lambda}{4 !} \, \phi^4 ~ \Big] ~~~.
\ee
We also add the source term
\be
{\cal S}_J ~=~  \int d^4 x ~ J \, \phi ~~~.
\ee
If we ignore momentarily the interaction term in the Lagrangian
\be
{\cal S}_{Int} ~=~ \int d^4 x ~ \Big[ ~-~ \fracm{\lambda}{4 !} \, \phi^4 
~ \Big] ~~~,
\ee
and derive the equation of motion for $\phi$ , we find
\be 
\eqalign{
 &(\, {}_{\bo} ~-~ m^2 \, ) \phi ~=~ - J ~~~, \cr
&{~~~~} \phi  ~=~ -\, (\, { J \over { {}_{\bo} ~-~ m^2 } } \, ) ~~~.
}\ee
and this may be substituted into $ {\cal S} - {\cal S}_{Int} \equiv W_0(J)$ 
and we find
\be
W_0(J) ~=~  \int d^4 x ~ \Big[ ~ - \, \fracm 12 \,  J \, 
 (\, { 1 \over { {}_{\bo} ~-~ m^2 } } \, )  \, J ~ \Big] ~~~.
\ee
The generator of 1PI Green's function is thus
\be
Z(J) ~=~ exp\Big\{ ~  \int d^4 x ~ \Big[ ~-~ \fracm{\lambda}{4 !} \, 
(\, \parvar{~~~~}{J(x)}\, )^4 ~ \Big\} \, exp\Big\{ \,  W_0(J) \,
 \Big\}  ~~~.
\ee

The propagator is just defined by
\be
\eqalign{
\Delta (\, x \, \,- \, \, x' \,) &=~  \parvar{{}^2 Z(J) }{J(x)\, \d J(x')} 
{\Big |}_{J = 0}
\cr
 &=~ { 1 \over { {}_{\bo} ~-~ m^2 } } \, \d^{(4)} ( \, x ~-~ x' \,) ~+~
 {\rm O} (\lambda ) ~~~.
}
\ee
If we ignore all higher order $\lambda$ terms this is the bare
propagator.  In a similar manner the 4-point function is just
\be
\eqalign{ {~~~}
G^{(4)} (\, x_1, \,  x_2, \, x_3, \, x_4 \,) ~=~  \parvar{{}^4 Z(J) }{J(x_1)\, 
\d J(x_2)\, \d J(x_3)\, \d J(x_4)} {\Big |}_{J = 0} ~~~.
}\ee

\subsection{Equation of Motion of the Chiral Scalar Superfield}

~~~~The simplest superfield theory is the free massive  chiral multiplet
described by the superfield action
\be
{\cal S}_c ~=~ \int d^8 z ~ {\Bar \Phi} \Phi ~-~ \Big[ ~ \int d^6 z ~ 
\fracm 12 \, m \Phi^2 ~+~ {\rm {h.\, c.}} ~ \Big] ~~~.
\ee
If we add chiral sources $J_c$ and ${\Bar J}_c$
\be
{\cal S}_J ~=~  \Big[ ~ \int d^6 z ~ J_c \,  \Phi ~+~ {\rm {h.\, c.}} ~ \Big] ~~~.
\ee
we would then like to treat this system just like the scalar ``$\phi^4
$-theory.''   First we would like to derive the superfield equation of motion.  
However, the superfield $\Phi$ satisfies the differential constraint ${\Bar 
D}{}_{\Dot \a} \Phi = 0 $ . This means that we \underline{cannot} freely vary 
the superfield $\Phi$.  As a simpler example of the problem we have encountered, 
let us consider an ordinary 4D theory that illustrates the nature of this 
problem and as well points the way to the solution.

The action for 4D Maxwell theory is
\be
{\cal S}_{Maxwell} ~=~ - \fracm 18 \, \int d^4 x ~ f^{\underline a \underline b}
\, f_{\underline a \underline b} ~~~,
\ee
and the field strength $f_{\underline a \underline b}$ satisfies the differential
constraint
\be
\pa_{\underline a} \, f_{\underline b \, \underline c} ~+~
\pa_{\underline b} \, f_{\underline c \, \underline a}  ~+~
\pa_{\underline c} \, f_{\underline a \, \underline b}  ~=~ 0 ~~~.
\ee
If we ignore the constraint on $f_{\underline a \underline b}$ and vary 
the action anyway we find
\be
\parvar{{\cal S}_{Maxwell}  }{f_{\underline b \, \underline c} } 
~=~ 0 ~~ \to ~~ f_{\underline b \, \underline c} ~=~ 0 ~~~.
\ee
Of course, the correct way to proceed is to first ``solve'' the constraint
\be
f_{\underline b \, \underline c} ~=~ \pa_{\underline b} \, A_{\underline c} ~-~
\pa_{\underline c} \, A_{\underline b} ~~~, 
\ee
and recognize that a variation of $f_{\underline b \underline c}$ takes the 
form $ \d f_{\underline b \, \underline c} ~=~ \pa_{\underline b} (\d A_{
\underline c} ) ~-~ \pa_{\underline c} ( \d A_{\underline b})$ and thus
\be
\parvar{{\cal S}_{Maxwell}  }{A_{\underline b} } 
~=~ 0 ~~ \to ~~ \pa^{\underline a} \, f_{\underline a \, \underline b} 
~=~ 0 ~~~.
\ee

These observations offer us a notion of how to correctly derive the
equation of motion for $\Phi$ . We first observe
\be
\eqalign{
D_{\a} D_{\b} &=~ - D_{\b} D_{\a} ~~\to ~~ D_{\a} D_{\b} ~=~ -\, C_{\a \, \b} \,
D^2 ~~~, \cr
\to~ D_{\a} D_{\b} D_{\g}  &=~ -\, C_{\b \, \g} \, D_{\a} \, D^2
~=~ -\, \fracm 12 \, C_{\b \, \g} \, D_{\a} \, D^{\d} D_{\d} \cr
&=~  \fracm 12 \, C_{\b \, \g} \, D_{\a} \, D_{\d} \, D^{\d}
~=~ -\, \fracm 12 \, C_{\b \, \g} \, D_{\d} \, D_{\a} \,D^{\d} \cr
&=~  \fracm 12 \, C_{\b \, \g} \, D_{\d} \, D^{\d} \, D_{\a}
~=~ \fracm 12 \, C_{\b \, \g} \, D_{\d} \, ( \d_{\a}{}^{\d} D^2 )
\cr
&=~ ~  \, \fracm 12 \, C_{\b \, \g} \, D_{\a} \, D^2 ~=~ - \, 
D_{\a} D_{\b} D_{\g}  ~~~. }\ee
\be
\eqalign{
&\to ~~ D_{\a} D_{\b} D_{\g} ~=~ - \, D_{\a} D_{\b} D_{\g} ~~ \to \cr
&{~~~} D_{\a} D^2 ~=~ 0 ~~ \to ~~  {\Bar D}_{\Dot \a} {\Bar D}{}^2 ~=~ 0 
~~~.}\ee

So that we see the equation ${\Bar D}{}_{\Dot \a} \Phi = 0 $ has a solution 
$\Phi = {\Bar D}{}^2 {\Bar U}  $ where $U$ is a complex unconstrained (i.e. it is 
free of all differential constraints) pre-potential superfield.  We use $U$ 
exactly like $A_{\underline a}$ in Maxwell theory.
\be
\eqalign{ {~~~}
{\cal S}_c &=~ \int d^8 z ~ ({\Bar D}{}^2 {\Bar U} ) \, ({D}{}^2 U ) ~
-~ \Big[ ~ \int d^6 z ~ \fracm 12 \, m \, ({\Bar D}{}^2 {\Bar U} )^2 
~+~ {\rm {h.\, c.}} ~ \Big] \cr
&{~~~~}+~  \Big[ ~ \int d^6 z ~ J_c \,  ({\Bar D}{}^2 {\Bar U} )  ~+~ {\rm {h. \, c.}} 
~ \Big] ~~~.
}\ee
If we vary with respect to $U$ we find
\be
\eqalign{ {~~~}
\d {\cal S}_c &=~ \int d^8 z ~  ({D}{}^2 \d U )\,  ({\Bar D}{}^2 {\Bar U} ) ~-~  
\int d^6 {\bar z} ~ m \, (D^2 {\d U} ) \,  (D^2 U )
\cr
&{~~~~}+~  \int d^6 {\bar z} ~ {\Bar J}_c \,  (D^2 {\d U} ) 
~~~.
}\ee
On the last two integrals, we can use the fact that $\int d^8 z \, = \, \int
d^6 {\bar z} D^2 $, thus
\be
\eqalign{ {~~~}
\d {\cal S}_c &=~ \int d^8 z ~ \Big[ ~ ({D}{}^2 \d U )\,  ({\Bar D}{}^2 {\Bar U} )~
-~  m \,  {\d U}  \,  (D^2 U ) ~+~   ~ {\Bar J}_c \, {\d U} ~ \Big] ~~~,
}\ee
\be
\eqalign{ {~~~} \to ~~
\d {\cal S}_c &=~ \int d^8 z ~ \Big[  (D^2 \d U )\,  \Phi 
~-~ m \, \, {\d U}  \,  {\Bar \Phi} ~+~  {\Bar J}_c \, {\d U} ~ \Big] 
~~~.
}\ee
To make further progress we have the identity
\be
\eqalign{
( D^{\a} D_{\a} A) \, B &=~ D^{\a} [ \, (D_{\a} A) \, B\, ] ~-~ 
 [ \, (D^{\a} A) \, (D_{\a} B) \, ]  \cr
&=~ A ( \, D^{\a} D_{\a}  B\, ) ~-~ D^{\a} [ \,  (D_{\a} A) \, B
~-~  A \,  (D_{\a} B) \, ] ~~~.}
\ee
Upon choosing $A = \d U$ and $B = \Phi$ this identity then implies
\be
\d {\cal S}_c ~=~ \int d^8 z ~ \d U \, \Big[  D^2  \Phi 
~-~ m \,   {\Bar \Phi} ~+~  {\Bar J}_c  ~ \Big] 
~~~.
\ee
Thus we derive the equation of motion from
\be
\parvar{{\cal S}_c  }{U(z')} ~=~ 0 ~~ \to ~~ D^2  \Phi 
~-~ m \,   {\Bar \Phi} ~=~ - \,  {\Bar J}_c  ~~~,
\ee
and acting on this with ${\Bar D}{}^2$  on both sides then leads to
\be
\eqalign{
&{\Bar D}{}^2 \, D^2  \Phi ~-~ m \,  {\Bar D}{}^2\,  {\Bar \Phi} ~=~ - \,  
{\Bar D}{}^2\, {\Bar J}_c  ~~~\cr
&\to~ {\Bar D}{}^2 \, D^2  \Phi ~-~ m \, ( \, m \, \Phi ~-~ J_c \, ) 
 ~=~ - \,  {\Bar D}{}^2\, {\Bar J}_c  ~~~\cr
&{\Bar D}{}^2 \, D^2  \Phi ~-~ m^2 \, {\Phi} ~=~ - \, (~ m \, J_c ~+~ 
{\Bar D}{}^2\, {\Bar J}_c ~)  ~~~.
}\ee

Now we need to do a little D-algebra
\be
\eqalign{
{\Bar D}{}^2 \, D^2 &=~ \fracm 14 \, {\Bar D}{}^{\Dot \a} {\Bar D}{}_{\Dot \a}
D{}^{\a} D{}_{\a} ~=~ - \, \fracm 14  \, {\Bar D}{}^{\Dot \a} {\Bar D}{}_{\Dot \a}
D{}_{\a} D{}^{\a}  \cr
&=~ \fracm 14 \, {\Bar D}{}^{\Dot \a} D{}_{\a}  {\Bar D}{}_{\Dot \a} D{}^{\a} 
~-~ i \, \fracm 14 \pa_{\underline a}  \, {\Bar D}{}^{\Dot \a} D{}^{\a}  \cr
&=~ - \, \fracm 14 \, {\Bar D}{}^{\Dot \a} D{}^{\a}  {\Bar D}{}_{\Dot \a} D{}_{\a} 
~+~ i \, \fracm 14 \pa_{\underline a}  \, D{}^{\a} {\Bar D}{}^{\Dot \a} 
~+~ \fracm 14 \, \pa^{\underline a} \pa_{\underline a} \cr
&=~  \fracm 14 \, {\Bar D}{}^{\Dot \a} D{}^{\a}  D{}_{\a}  {\Bar D}{}_{\Dot \a}
~-~ i \, \fracm 14 \pa_{\underline a}  \, {\Bar D}{}^{\Dot \a} D{}^{\a} 
~+~ i \, \fracm 14 \pa_{\underline a}  \,  D{}^{\a} {\Bar D}{}^{\Dot \a}
~+~ \fracm 14 \, \pa^{\underline a} \pa_{\underline a} \cr
&=~ \fracm 12 \, {\Bar D}{}^{\Dot \a} D{}^2  {\Bar D}{}_{\Dot \a}
~+~ i \, \fracm 12 \,  \pa_{\underline a}  \,  D{}^{\a} {\Bar D}{}^{\Dot \a}
~+~ {}_{\bo}  \cr
&{~~~}\to ~~  {\Bar D}{}^2 \, D^2 \,  \Phi ~=~  {}_{\bo} \, \Phi 
~~~.}\ee
So the equation of motion for $\Phi$ in the presence of sources $J$ and
${\Bar J}$ is
\be
( \, {}_{\bo} ~-~ m^ 2 \, ) \Phi ~=~ - \, ( \, m J_c ~+~ {\Bar D}{}^2 {\Bar 
J}_c \, ) ~~~.
\ee

\subsection{Functional Quantization of the Chiral Scalar Superfield}

~~~~At this point for $\phi^4$-theory, we simply ``inverted'' the operator on
the lhs by performing a formal division.  The justification for this is
that we could always choose to work in momentum space via the use of
Fourier transformations with respect to the $x$-space coordinates.
The most obvious generalization is to work in the Fourier transform
with respect to $z$-space.  However, this is not the most convenient
way to proceed. Instead the Fourier transform need only be taken with
respect to the $x$-subspace of superspace.  After going to momentum 
space, the equation of motion is of the form,
\be
( \, p^2 ~+~ m^ 2 \, ) \Phi ~=~  ( \, m J_c ~+~ {\Bar D}{}^2 {\Bar J}_c \, ) 
~~~,
\ee
where in this equation we regard $\Phi = \Phi(\q,\, {\bar \q}, \, p)$ and 
$J_c = J_c(\q,\, {\bar \q}, \, p)$.  As well the $D$-operator is written in 
$p$-space
\be
D_{\a} ~=~ \pa_{\a} ~+~ \fracm 12 \, {\bar \q}{}^{\Dot \a} p_{\underline a}
~~~.
\ee
With this understanding, the equation of motion for $\Phi$ can be inverted to
express $\Phi$ as a function of $J_c$ and ${\Bar J}_c$,
\be
\Phi ~=~ - \, [ \, {{m J_c ~+~ {\Bar D}{}^2 {\Bar J}_c} \over {\, {}_{\bo} 
~-~ m^ 2} } \, ] ~~~~~ {\rm {or}} ~~~~ {\Bar \Phi} ~=~ - \, [ \, { {m {\Bar 
J}_c ~+~ D^2 J_c } \over {\, {}_{\bo} ~-~ m^ 2} }
\, ] ~~~.
\ee

If we substitute the $x$-space version back into the original action we
obtain
\be
\eqalign{
W_0 (J_c, \, {\Bar J}_c) ~=~ - \, \int d^8 z ~\Big\{~ &{\Bar J}_c \, [ { {1 } 
\over {\, {}_{\bo} ~-~ m^ 2} } ] \, J_c  ~+~ \fracm 12 \,   J_c \, [ { {m \, 
D^2 } \over {\, {}_{\bo} \, ({}_{\bo} ~-~ m^ 2 \, )} } ] \, J_c \cr
&+~  \fracm 12 \,  {\Bar J}_c \, [ { {m \, {\Bar D}{}^2 } \over {\, {}_{\bo} 
\, ({}_{\bo} ~-~ m^ 2 \, )} }  ] \, {\Bar J}_c  ~ \Big\} ~~~.
}\ee
We can define a superfield 1PI generating functional via
\be
Z_0 ( J_c , \, {\Bar J}_c) ~=~ exp[ \, W_0 ( J_c , \, {\Bar J}_c) ~] ~~~.
\ee
This is valid, of course, only for the free massive theory.  In the case
of an interacting theory where we can write
\be
{\cal L}_{Tot} ~=~ {\cal L}_0 ~+~ {\cal L}_{Int} (\Phi, \, {\Bar \Phi} ) ~~~,
\ee
the generating functional becomes
\be
Z ( J_c , \, {\Bar J}_c) ~=~ exp[ ~ \int d^8 z ~ {\cal L}_{Int} (
\parvar{~~ }{J_c } ,\,\parvar{~~~~}{{\Bar J}}  ) ~] 
Z_0 ( J_c , \, {\Bar J}_c) ~~~.
\ee
In order for this definition to make sense, we require a definition of the
functional derivative taken with respect to chiral (as opposed to
general) superfields. For this purpose we note,
\be
\parvar{J_c (z) }{J_c (z')} ~=~ {\Bar D}{}^2 \d^{(8)} ( \, z \, - \, z' \, )
~~~. \label{eq:hvv}
\ee
In a theory only involving chiral superfields we can use this generating
function to derive superfield Feynman rules.

(a.) Propagators

\be
\begin{array}{ccc}
{\Bar \Phi} \, \Phi:  &{~~~~~~~~~~}{1 \over {~p^2 ~+~ m^2 }}\,  \d^{(4)} 
(\q \, - \, \q' ) &{~} ~~~, \\
\Phi \, \Phi:    &{~~~~~~~~~~}- \, { {m D^2 } \over {~p^2 ( \, p^2 ~+~ 
m^2 ) }}\,  \d^{(4)} (\q \, - \, \q' )    &{~}   ~~~,   \\
{\Bar \Phi} \, {\Bar \Phi}:  &{~~~~~~~~~~}- \, { {m {\Bar D}{}^2 } \over 
{~p^2 ( \, p^2 ~+~ m^2 ) }} \, \d^{(4)} (\q \, - \, \q' )  &{~} ~~~.
\end{array}
\ee

For simplicity's sake only, we can consider the massless limit where the 
latter two propagators vanish.  We may now look at what must be going 
on in terms of the component fields in the massless limit.  If we take 
$<0| {\Bar \Phi} \Phi  |0> |  $ this must correspond to the propagator 
$ <0| {\Bar A}(x) A(x') |0> $. If we act on this propagator with $D_{\a} 
\, {\Bar D}{}_{\Dot \a}$ we obtain 
\be
( {\Bar D}{}_{\Dot \a} {\Bar \Phi}) \, ( D_{\a} \Phi \,) : {~~~~~~~~~~~}
{{D_{\a} \, {\Bar D}{}_{\Dot \a}} \over {~p^2}}\,  \d^{(4)} 
(\q \, - \, \q' ) 
~~~,
\ee
and after applying the $|$-operation, this must correspond to the 
propagator $  <0| {\Bar \psi}{}_{\Dot \a}(x) \psi_{\a}(x') |0> $. Due 
to the interacting super Feynman rules, we can replace $D_{\a} \, 
{\Bar D}{}_{\Dot \a} \sim i p_{\underline a} $ and thus the propagator 
takes its usual form $i p_{\underline a} / p^2 $. Finally, there is a 
propagator for the $F$ field. We can find it by applying ${\Bar D}{}^2 
D^2 $ leading to
\be
( {\Bar D}{}^2 {\Bar \Phi}) \, ( D^2 \Phi \,) : {~~~~~~~~~~~}
{{\Bar D}{}^2 D^2  \over {~p^2}}\,  \d^{(4)} 
(\q \, - \, \q' ) 
~~~.
\ee
Now it might appear that the auxiliary field $F$ is propagating a massless
state (i.e. there seems to be a pole at $p^2=0$).  But once again the
way the super Feynman rules work we can replace ${\Bar D}{}^2 D^2 
\sim - p^2  $ so there is no pole at $p^2=0$.

\subsection{Feynman Rules for the Chiral Scalar Superfield}

~~~~To actually derive both the propagators as well as the rest of the super
Feynman rules we require a definition of $\d \Phi(z)/\d \Phi(z') $ where 
${\Bar D}_{\Dot \a} \Phi = 0 $.  For this purpose we simply note the
result in (\ref{eq:hvv}).  The following Feynman rules are obtained in
a straight forward way.\newline ${~}$ \newline

(a.) Vertices are read from the interaction Lagrangian in the usual
\newline ${~~~~~~~~~~~}$ way for ordinary four dimensional theory, 
with the additional \newline ${~~~~~~~~~~~}$ feature that for each 
chiral or antichiral line leaving a vertex \newline ${~~~~~~~~~~~}$ 
there is a factor of ${\Bar D}{}^2$ or $D^2$ acting on the corresponding 
\underline{purely} \newline ${~~~~~~~~~~~}$ chiral or antichiral 
vertices we omit a ${\Bar D}{}^2$ or $D^2$ factor from a- \newline 
${~~~~~~~~~~~}$ mong the ones acting on the propagators.

(b.) We integrate $d^2 \q \, d^2 {\bar \q}$ at each vertex and in momentum 
space \newline ${~~~~~~~~~~~}$ we have loop integrals $(2 \pi)^{-4} \int
d^4 p $ for each loop, and an overall \newline 
${~~~~~~~~~~~}$ factor $(2 \pi)^4$ $\d^{(4)}(\sum k_{ext})
$.

(c.) To obtain the effective action $\G$, we compute
one-particle-irreducible \newline ${~~~~~~~~~~~}$ graphs. For each external 
line with outgoing momentum $k_i$, we  \newline ${~~~~~~~~~~~}$ multiply by 
a factor $ \int d^4 k_i \,(2 \pi)^{-4} \Psi(k_i)$ where $\Psi$ stands for 
any  \newline ${~~~~~~~~~~~}$ of the fields in the effective action.  For 
each external chiral or  \newline ${~~~~~~~~~~~}$ antichiral line, we have 
a $\Phi$ or ${\Bar \Phi}$ factor, but no ${\Bar D}{}^2$ or $D^2$ factor

(d.) There may be symmetry factors associated with certain graphs.

\subsection{Review of the Functional Quantization of Yang-Mills 
Fields}

As usual with a gauge theory, the main problem is that gauge-invariance
implies that the gauge field propagator cannot be defined.  A quick
review of ordinary Yang-Mills theory will suffice to show the problem.
\be
\eqalign{
{\cal S}_{YM} &=~  - \fracm 18 \, \int d^4 x ~  tr [ \, f^{\underline a \, 
\underline b}  f_{\underline a \, \underline b}\, ]  ~~~, \cr
 f_{\underline a\, \underline b} &=~ \pa_{\underline a} A_{\underline b} 
~-~  \pa_{\underline b} A_{\underline a} ~-~i [ \, \, A_{\underline a}
\, , \,  A_{\underline b} ~] ~~~.
}\ee
The infinitesimal gauge transformation is
\be
\d_G  A_{\underline a} ~=~  \pa_{\underline a} \o ~-~ i \,
 [ \, A_{\underline a} \, , \, \o ~]  ~~~,
\ee
and the purely quadratic part of the action after an integration-by-parts
has the form $  A^{\underline a} [ \eta_{\underline a \underline b} 
p^2  \, - \, p_{\underline a} p_{\underline b} ] A^{\underline b}
\equiv A^{\underline a} \D_{\underline a \underline b} A^{\underline 
b}$ the object between these two fields is a $4 \times 4$ matrix
\be
\eqalign{
&\D_{\underline a \, \underline b} ~=~
\left(\begin{array}{cccc}
|p|^2 \, + \, (p_0)^2  & p_0 \, p_1 & p_0 \, p_2  & p_0 \, p_3 \\
p_1 \, p_0 & |p|^2 \, - \, (p_1)^2  & p_1 \, p_2  &  p_1 \, p_3 \\
p_2 \, p_0 & p_2 \, p_1 &  |p|^2 \, - \, (p_2)^2   &  p_2 \, p_3 \\
p_3 \, p_0 & p_3 \, p_1 & p_3 \, p_2  &  |p|^2 \, - \, (p_3)^2 \\
\end{array}\right) ~~~, \cr
&{~}|p|^2 ~=~ - \,  (p_0)^2 ~+~ (p_1)^2  ~+~ (p_2)^2 ~+~ (p_3)^2 
~~~.} \ee
After a direct calculation we find $ det (\D_{\underline a \underline b})
= 0$ so it cannot be inverted in order to define a propagator
for the gauge field. There is also an indirect way to see that this
operator cannot be inverted. It is generally true that if an operator
possesses a null space, then that operator has no inverse. 

This in turn implies the need for a gauge-fixing term which allows the
definition of a propagator for $A_{\underline a}$. However, this also 
permits the unphysical (gauge) component of $A_{\underline a}$ propagate.  
This is fixed by BRST ghost and antighost introduction. These fields
cancel out the effects of the unphysical components of $A_{\underline a}$.
So in the end the generator $Z(J)$ is defined by
\be
\eqalign{
Z(J) &=~ N' \, {\bf \int} \, [{\cal D} A_{\underline a}] \, [{\cal D} c]
\, [{\cal D} c'] ~ exp[~ {\cal S}_{eff} ~+~ {\cal S}_J ~] ~~~, \cr
{\cal S}_{eff} &=~ {\cal S}_{YM} \,-\, \int d^4 x \, \,tr [ ~{1 \over 
\a}  \, [ F(A_{\underline a})]^2 \,+\, i  \int d^4 x' \, c'(x') (\parvar{ 
F(\d_G A)}{\o(x')} ) c(x') \, 
  \, ] ~~~, \cr
{\cal S}_J &=~ \int d^4 x ~ tr[ \, A_{\underline a} J^{\underline a} \,] 
~~~,
}\ee
where $F(A_{\underline a})$ is a function that is not gauge invariant.

\subsection{Functional Quantization of Yang-Mills Superfields}

~~~~We now wish to repeat this for the 4D, N = 1 supersymmetric Yang-Mills
theory. The gauge invariant action for the 4D, N = 1 supersymmetric 
Yang-Mills theory may be written as
\be
\eqalign{
{\cal S}_{YM} &=~ \fracm 12 \Big\{ ~ tr[ \, \int d^6 z ~ W^2 
 \, ]  ~+~ {\rm {h.\,c.}}~ \Big\}  \cr
 &=~ - \, \fracm 14 \Big\{ ~ tr[ \, \int d^8 z ~(\, e^{-V} D^{\a} e^V ) 
\,( {\Bar D}{}^2 ( e^{-V} D^{\a} e^V )) \, \,]  ~+~ {\rm {h.\,c.}}~~ 
\Big\}  \cr
&\approx~  \fracm 12  \, tr[ \, \int d^8 z ~(\, V  D^{\a}\,
{\Bar D}{}^2  D_{\a} V ) ~] ~+~ {\rm 0} (V^3) ~~~.
}\ee
To reach the second line we have used the non-abelian definition of 
$W_{\a}$ (i.e. $W_{\a} =  [{\Bar D}{}^2 ( e^{-V} D^{\a} e^V )] $ along 
with the definition $\int d^8 z = \int d^6 z \, {\Bar D}{}^2$.
So the operator between the quadratic $V$-terms is given 0 = $
D^{\a} {\Bar D}{}^2  D_{\a}$ .  Due to the identities,
\be
D^{\a}\,{\Bar D}{}^2  D_{\a} D^2 ~=~ 0 ~~~,~~~ D^{\a}\,{\Bar D}{}^2  
D_{\a} {\Bar D}{}^2 ~~~,
\ee
this operator is seen to possess null-spaces and is not invertible.  For 
the first of these, we already saw a derivation.  For the second one we 
have
\be
\eqalign{
D^{\a}\,{\Bar D}{}^2 D_{\a} {\Bar D}{}^2 &=~ \fracm 12 D^{\a}\,{\Bar D}{}^2  
D_{\a} {\Bar D}{}^{\Dot \b} {\Bar D}{}_{\Dot \b} ~=~ - \fracm 12 
D^{\a}\,{\Bar D}{}^2  D_{\a} {\Bar D}{}_{\Dot \b} {\Bar D}{}^{\Dot \b} \cr
&=~ \fracm 12 D^{\a}\,{\Bar D}{}^2 {\Bar D}{}_{\Dot \b}  D_{\a} 
{\Bar D}{}^{\Dot \b} ~-~ i \, \fracm 12 \pa_{\a \Dot \b}  
D^{\a}\,{\Bar D}{}^2 {\Bar D}{}^{\Dot \b} ~=~ 0 ~~~,
}\ee
since $  {\Bar D}{}^2{\Bar D}{}^{\Dot \b} = 0$.

The gauge variation of the superspace Yang-Mill gauge field
is
\be
\eqalign{
\d_G \, V  &=~ - i ( \,  \L \, - \, {\Bar \L } \,) ~-~ i \fracm 12 L_V 
( \,  \L \, + \, {\Bar \L } \,)  ~+~ i f(\fracm 12 L_V) 
( \,  \L \, - \,  {\Bar \L } \,) ~~~,
}\ee
where the transcendental function $f(x)$ is $f(x) = 1 - x \, coth(x)$.
This transformation law may be compared with that of ordinary
Yang-Mills theory (\ref{eq:ett}) which when written in term of forms becomes
\be
\d_G  A ~=~  d \o ~-~ i \,[ \,  A \, , \, \o ~]  
~=~  d \o ~-~ i \,L_A \,  \o   ~~~.
\ee
Written in this form the similarities are obvious. The only major 
difference between the gauge variations of ordinary Yang-Mills theory 
and 4D, N = 1 superspace  Yang-Mills theory is that the latter involves
a transcendental {\it {not}} linear function of the $L$-operator.

So like ordinary Yang-Mills theory the supersymmetric theory requires
a gauge fixing term be introduced. The simplest choice is
\be
{\cal S}_{GF} ~=~ - { 1 \over \a} tr [ ~ \int d^8 z ~ (D^2 V ) \, ( 
{\Bar D}{}^2 V) ~] ~~~,
\ee
and this is the direct analog of $ - { 1 \over \a} tr [ ~ \int d^4
x  ~ (\pa^{\underline a} A_{\underline a})^2 ]$. Above we have found the
explicit form of $\d_G V $, and this can be used to calculate $\d_G 
{\cal S}_{GF}$. After replacing the gauge parameter by chiral ghosts 
$c$ and introducing the anti-ghost chiral superfields $c'$ one finds
\be
{\cal S}_{FP} ~=~ - { 1 \over {2\a}} tr [ ~ \int d^8 z ~
(c' \, + \, {\bar c}' )  L_V \, [~ ( \,  c \, + 
\,  {\Bar c }\, ) \, + \, coth(\fracm 12 L_V) \,( \,  c \, - \,  
{\Bar c} \,)  ~] ~~~.
\ee
where the antighosts $c'$ and not related to the ghosts $c$.  This term 
defines the propagator and interactions for the ghost and antighost 
superfields.  The purely quadratic $V$-terms are now 
\be
- \, \fracm 12 V \, [~ D^{\a} {\Bar D}{}^2 D_{\a} ~-~ {1 \over \a} ( \,
 D^2 {\Bar D}{}^2 \, + \,{\Bar D}{}^2  D^2 ) \,\, ] V ~~~.
\ee
If we pick $\a = 1$ (Fermi-Feynman gauge) and after some more $D$-algebra
this becomes
\be
- \, \fracm 12 ~V \, {}_\bo \, V ~~~,
\ee
and if we look at the part of the action that is pure quadratic in
ghost-antighost we find
\be
{\cal S}_{FP}^{(2)} ~=~ \int d^8 z tr [ ~ {\bar c}' \, c ~-~  c' \, 
{\bar c} ~] ~~~.
\ee
So we conclude that the $V$ propagator is
\be
V \, V: ~~~~~~~- \, {1 \over p^2} \d^{(8)} (\, z \, - \, z' \,) 
~~~. \ee
and since $c$ and $c'$ are chiral superfields
\be
\eqalign{
&{\bar c}' \, c: ~~~~~~~- \, {1 \over p^2}  \d^{(4)} (\, \q \, - \, \q' \,)  \cr
&{\bar c}\, c': ~~~~~~~- \, {1 \over p^2}   \d^{(4)} (\, \q \, - \, \q' \,)  
}\ee

The real magic of using supergraphs only begins at this point. The formal
mathematical expressions that we have seen in this lecture can be associated
with diagrams precisely like ordinary quantum field theory. In particular
the graphs associated with the structures that we have seen are called
``Feynman supergraphs'' and a single Feynman supergraph is equivalent to 
the Feynman graphs of all the component fields contained in the superfield.
This is a topic treated in numbers of text books on the subject.

\newpage
\section{Fifth Lecture: Senior 4D, N = 1 Superfield Theory}
{\it {Introduction}}

~~~~~~In this final lecture we discuss the theory of 4D, N = 1 supergravity.
The presentation includes both aspects of superfield and component
descriptions. In terms of superfields, it is shown that 4D, N = 1 
supergravity is a natural synthesis of what we have seen as the
structure of supersymmetric Yang-Mills theory and ordinary gravity.
So a complete structure of constraints as well as their solution,
very similar to that of the supersymmetric Yang-Mills case, appears. 
$${~}$$

\noindent
{\it {Lecture}}
\subsection{Supergravitation Orientation}

~~~~In lecture two, the superspace structure of 4D, N = 1 supersymmetric
Yang-Mills theory was presented.  A superspace connection $\G_{\underline A}
{}^{\cal I} $ was defined and used to introduce a supergeometrical 
Yang-Mills covariant derivative,
\be
\eqalign{
\nabla_{\underline A} &\equiv~ D_{\underline A} ~-~ i \G_{\underline A}
{}^{\cal I} t_{\cal I}  ~~~, \cr
D_{\underline A} &\equiv~ \Big( ~D_{\a},\, {\Bar D}{}_{\Dot 
\a} ,\, \pa_{\underline a} ~ \Big) ~~~, \cr
{\G}_{\underline A}{}^{\cal I} &\equiv~ \Big( ~{\G}_{\a}{}^{\cal I} ,\, 
{\G}{}_{\Dot \a} {}^{\cal I} ,\, \G_{\underline a}{}^{\cal I} ~ \Big)
~~~, } \label{eq:hsv} \ee
where $t_{\cal I}$ denote Lie algebraic generators. This can be used to define 
field strength superfields $F_{\underline A \, \underline B}
{}^{\cal I}$.
\be
\Big[ ~ \nabla_{\underline A} \, , \, \nabla_{\underline B} 
~ \Big\} ~=~  C_{\underline A \, \underline B}{}^{ \underline C}
\, \nabla_{\underline C} ~-~ i \,  F_{ {\underline A} \, 
{\underline B}}{}^{\cal I} t_{\cal I}~~~.
\ee

In lecture three, we saw that the theory of general relativity, although
formulated by Einstein in purely geometric terms, has a natural interpretation 
in terms of concepts that arise from Yang-Mills gauge theory. The most
important point of our discussion was that we were able to identify the
Lie algebra generators that underlie general relativity
\be
t_{\cal I} ~~\to~~ \Big( ~ P_{\underline m} , \, {\cal M}_{\underline
a \, \underline b} ~ \Big) ~~~, 
\ee
where $P_{\underline m}$ is the usual momentum operator from quantum mechanics 
and (or alternately ${\cal M}_{\a \, \b}$ and ${\Bar {\cal M}}{}_{\Dot \a \, 
\Dot\b}$) are the abstract Lie algebra generators of spin-angular momentum 
operators also from quantum mechanics.  This particular suggestion for the 
Lie-algebraic basis for general relativity was first made in 1978 
\cite{6}.  Once this Lie-algebraic basis for general relativity is 
accepted, the structure of a covariant derivative follows
\be
\nabla_{\underline a} ~ \equiv~ {\rm e}_{\underline a}{}^{\underline m} 
\pa_{\underline  m} ~+~ \fracm 12 \o_{\underline a \underline c }{}^{\underline  
d} \,{\cal M}_{\underline d }{}^{\underline c}
~~~, 
\ee
and its commutator is used to define field strengths
\be
\Big[ ~ \nabla_{\underline a} \, , \, \nabla_{\underline b} ~ \Big] ~=~ 
t_{\underline a \, \underline b}{}^{\underline c} \, \nabla_{\underline c}
~+~ \fracm 12 r_{\underline a \, \underline b \, \underline c }{}^{\underline 
d}\, {\cal M}_{\underline d }{}^{\underline c} ~~~,
\ee
where the first is ``the torsion tensor" and the second is ``the 
Riemann curvature tensor."

In the 4D, N = 1 superspace Yang-Mills theory, the field strength 
$F_{\underline A \, \underline B}{}^{\cal I}$ can be separated into 
various irreducible Lorentz representations
\be
F_{\underline A \, \underline B}{}^{\cal I} ~=~ 
\left(\begin{array}{ccc}
~F_{\a \, \b}{}^{\cal I} & ~~F_{\a \, \Dot\b}{}^{\cal I} &  ~~F_{
\a \, \underline b}{}^{\cal I}  \\
~F_{\Dot \a \, \b}{}^{\cal I} & ~~F_{\Dot \a \, \Dot \b}{}^{\cal I} &  
~~F_{\Dot \a \, \underline b}{}^{\cal I}  \\
~F_{\underline a \, \b}{}^{\cal I} & ~~F_{\underline a \, \Dot \b}{}^{\cal 
I} &  ~~F_{\underline a \, \underline b}{}^{\cal I}  \\
\end{array}\right) ~~~,
\ee
and a minimal theory exists by imposing algebraic conditions that some of 
these irreducible Lorentz representations are constrained to vanish: 
$F_{\a \, \b}{}^{\cal I} = F_{\Dot \a \, \Dot \b}{}^{\cal I} = F_{\Dot 
\a \, \b}{}^{\cal I} = F_{\a \, \Dot \b}{}^{\cal I} = 0$.  The solution 
to these constraints implies that all of the connection superfields can 
be defined in terms of a more fundamental object, the pseudo-scalar
pre-potential superfield $V^{\cal I}$.

In general relativity, also one of the field strengths is constrained to
vanish: $ t_{\underline a \, \underline b \, \underline c} = 0 $.  The 
solution to this condition implies that the spin-connection can be 
defined in terms of the anholonomity. So superspace Yang-Mills theory and
general relativity are both gauge theories where some of the a priori
non-vanishing field strengths are constrained to vanish.

\subsection{Superspace Supergravity Covariant Derivatives}

~~~~From all of this it should be clear that in order to describe a 4D, N = 1 
theory that is analogous to general relativity, we need to combine the 
features of the two distinct theories discussed above.  In other words, 
the Lie algebraic generators of the 4D, N = 1 supersymmetric covariant
Yang-Mills derivative need to be replaced by the analog of $(
P_{\underline m} , \, {\cal M}_{\underline a \, \underline b})$. 
From lecture one we know these turn out to be $( Q_{\mu}, \, 
{\Bar Q}{}_{\Dot \mu}, \, P_{\underline m} , \, {\cal M}_{\underline
a \, \underline b})$. However, for convenience sake and without loss of
generality, we make the equivalent replacement of $( D_{\mu}, \, 
{\Bar D}{}_{\Dot \mu}, \, P_{\underline m} , \, {\cal M}_{\underline
a \, \underline b})$ in the superspace supergravity covariant derivative
\be
\nabla_{\underline A} ~ \equiv~ {\rm E}_{\underline A}{}^{\underline M} 
D_{\underline  M} ~+~ \fracm 12 \o_{\underline A \, \underline c }{}^{
\underline  d} \,{\cal M}_{\underline d }{}^{\underline c}
~~~, 
\ee
where instead of $( Q_{\mu}, \, {\Bar Q}{}_{\Dot \mu}, \, P_{\underline m}) $
we use $  ( D_{\mu}, \, {\Bar D}{}_{\Dot \mu}, \, \pa_{\underline m}) $. In 
this definition, the quantity ${\rm E}_{\underline A}{}^{\underline M}
(\q, \,{\bar \q},\,x)$ is known as the ``super vielbein" and $\o_{\underline 
A \, \underline c \, \underline  d}(\q, \,{\bar \q},\,x)$ is known as the 
``super spin-connection.''

The graded commutator of this operator is used to define the field
strengths
\be
\Big[ ~ \nabla_{\underline A} \, , \, \nabla_{\underline B} ~ \Big\} 
~=~  T_{\underline A \, \underline B}{}^{ \underline C} \, \nabla_{
\underline C} ~+~ \fracm 12 R_{\underline a \, \underline b \, \underline 
c }{}^{\underline d}\, {\cal M}_{\underline d }{}^{\underline c} ~~~~,
\ee
where $T_{\underline A \, \underline B}{}^{ \underline C}$ is called ``the 
torsion super-tensor" and $R_{\underline a \, \underline b \, \underline 
c }{}^{\underline d}$ is called `` the Riemann curvature supertensor." These 
may be expressed in terms of ${\rm E}_{\underline A}{}^{\underline M}$,
$ \o_{\underline A \, \underline c \,\underline  d}$ and
their derivatives.  For the torsion
\be
\eqalign{
T_{\underline A \, \underline B}{}^{ \underline C} &=~
C_{\underline A \, \underline B}{}^{ \underline C} ~-~ 
\o_{\underline A \, \underline B }{}^{\underline  C} ~+~
(-1)^{\underline A \, \underline B} \,
\o_{\underline B \, \underline A }{}^{\underline  C} ~~~, \cr
{\rm E}_{\underline A} &\equiv~ {\rm E}_{\underline A}
{}^{\underline M} D_{\underline M} ~~~,~~~ 
\Big[ ~ {\rm E}_{\underline A} \, , \, {\rm E}_{\underline B} ~ \Big\} 
~=~ C_{\underline A \, \underline B}{}^{ \underline C} \, {\rm E}_{
\underline C} ~~~,\cr
C_{\underline A \, \underline B}{}^{ \underline C}&=~
[ \, {\rm E}_{\underline A}{}^{\underline M}\,( D_{\underline M}
{\rm E}_{\underline B}{}^{\underline N} ) \, - \, (-1)^{AB}
{\rm E}_{\underline B}{}^{\underline M}\,( D_{\underline M}
{\rm E}_{\underline A}{}^{\underline N} )\,] \,
{\rm E}_{\underline N}{}^{\underline C}  ~~~,
}\ee
where $C_{\underline A \, \underline B}{}^{ \underline C}$ is the 
super-anholonomity.  In writing the torsion supertensor it appears that
$\o_{\underline A \, \underline B }{}^{\underline  C}$ is a general super 
matrix in its ${\underline B}$ and ${\underline C}$ indices. This is 
not the case. Instead we have
\be
\o_{\underline A \, \underline B}{}^{\underline C} ~=~ 
\left(\begin{array}{ccc}
~\o_{\underline A \,\b}{}^{ \, \g} & ~~0 &  ~~0 \\
~~0 & ~~\o_{\underline A \,\Dot \b}{}^{ \, \Dot \g} &  ~~0  \\
~~0 & ~~0 &  ~~\o_{\underline A \, \underline b}{}^{\underline c} \\
\end{array}\right) ~~~,
\ee
and this is a consequence since we only included ${\cal M}_{\underline a \, 
\underline b}$ (or equivalently ${\cal M}_{\a \, \b}$ and ${\Bar {\cal 
M}}{}_{\Dot \a \, \Dot \b}$) in the superspace supergravity covariant 
derivative. (Here we may also write $\o_{\underline A \, \underline b
\, \underline c} = C_{\Dot \b \, \Dot \g} \o_{\underline A \, \b
\, \g} + C_{\b \, \g} \o_{\underline A \, \Dot \b \, \Dot \g}$ as 
is our usual custom.) This further implies that for some 
values of the indices ${\underline A}$, ${\underline B}$ and 
${\underline C}$, the torsion is identical to the anholonomity. 
For example the spin connection superfield does not appear in
\be
\eqalign{
T_{\a \, \b}{}^{ \Dot \g} &=~ C_{\a \, \b}{}^{ \Dot \g} ~~~,~~~
T_{\a \, \b}{}^{ \underline c} ~=~ C_{\a \, \b}{}^{ \underline c}~~~, \cr
T_{\a \, \Dot \b}{}^{ \underline c} &=~ C_{\a \, \Dot \b}{}^{ \underline c}
~~~,~~~ T_{\a \, \underline b}{}^{ \Dot \g} ~=~ C_{\a \, \underline b}{}^{
\Dot \g }~~~, \cr
T_{\underline a \, \underline b}{}^{ \g} &=~ C_{\underline a \, 
\underline b}{}^{ \g} ~~~ ,
}\ee
and as well for the complex conjugates of these irreducible Lorentz
components of the torsion supertensor.

However, for other components of the torsion supertensor, the 
spin-connection superfield appears as in
\be
\eqalign{
T_{\a \, \b}{}^{\g} &=~ C_{\a \, \b}{}^{\g}
~+~ \o_{\a \, \b}{}^{\g} ~+~  \o_{\b \, \a}{}^{\g}
 ~~~,~~~ \cr
T_{\a \, \Dot \b}{}^{ \Dot \g} &=~ C_{\a \, \Dot \b}{}^{ \Dot \g}
~+~ \o_{\a \, \Dot \b}{}^{ \Dot \g} ~~~, \cr
T_{\a \, \underline b}{}^{ \g} &=~ C_{\a \, \underline b}{}^{ \g}
~+~ \o_{\underline b\, \a}{}^{ \g} ~~~, \cr
T_{\a \, \underline b}{}^{ \underline c} &=~ C_{\a \, \underline b}{}^{
\underline c} ~-~ \o_{\a \, \underline b}{}^{ \underline c}
~~~,  \cr
T_{\underline a \, \underline b}{}^{ \underline c} &=~ 
C_{\underline a \, \underline b}{}^{ \underline c}
~-~ \o_{\underline a \, \underline b}{}^{ \underline c}
~+~ \o_{\underline b \, \underline a}{}^{ \underline c}
~~~, }\ee
and in all the complex conjugates of these equations.

\subsection{Superspace Supergravity Conventional Constraints}

~~~~In general relativity, the constraint $t_{\underline a \, \underline b
\, \underline c} = 0 $ allowed the spin connection to be expressed in terms 
of the anholonomity.  A similar stratagem works here where
\be
\eqalign{
T_{\a \, \b}{}^{\g} &=~0 ~~, ~~\to ~~ \o_{\a \, \b \, \g} ~=~ -
\fracm 12 ~[ \, C_{\a \, \b \, \g} ~-~ C_{\b \, \g \, \a} ~+~ 
C_{\g \, \a \, \b} \, ] ~~~,~~~ \cr
T_{{\Dot \a} \, {\Dot \b}}{}^{{\Dot \g}} &=~0 ~~, ~~\to ~~ \o_{{\Dot \a} 
\, {\Dot \b} \, {\Dot \g}} ~=~ - \fracm 12 ~[ \, C_{{\Dot \a} \, {\Dot \b} 
\, {\Dot \g}} ~-~ C_{{\Dot \b} \, {\Dot \g} \, {\Dot \a}} ~+~ C_{{\Dot \g} 
\, {\Dot \a} \, {\Dot \b}} \, ] ~~~,~~~ \cr
T_{\a \, ( \Dot \b \, \Dot \g )} &=~0 ~~, ~~\to ~~ \o_{\a \, \Dot \b
\, \Dot \g} ~=~ - \, \fracm 12 C_{\a \, (\Dot \b \, \Dot \g )} ~~~,~~~ \cr
T_{\underline a \, ( \b \, \g ) } &=~0~~, ~~\to ~~ \o_{\underline b
\, \a}{}^{ \g} ~=~  \,  C_{\underline a \, ( \b \, \g )}
 ~~~, \cr
T_{{\underline a} \, {\underline b}}{}^{{\underline c}} &=~0 ~~, 
~~\to ~~ \o_{{\underline a} \, {\underline b} \, {\underline c}} ~=~ 
\fracm 12 ~[ \, C_{{\underline a} \, {\underline b} \, {\underline c}} 
~-~ C_{{\underline b} \, {\underline c} \, {\underline a}} ~+~ 
C_{{\underline c} \, {\underline a} \, {\underline b}} \, ] ~~~.
} \label{eq:hgv} \ee
Thus, $\o_{\underline A \, \underline b \, \underline c}$ is completely 
determined just as in general relativity.  

At this stage, the only independent degrees of freedom in $\nabla_{
\underline A} $ that remain are those in ${\rm E}_{\underline A} 
{}^{\underline M}$.  But should all of these be independent?  Recall the 
following analogy
\be
\nabla_{\underline A}^{(YM)} ~=~ D_{\underline A} ~-~ i 
\G_{\underline A}{}^{\cal I} t_{\cal I} ~~\sim~~ {\rm E}_{\underline A}
~=~ {\rm E}_{\underline A}{}^{\underline M} D_{\underline M}
~~~.
\ee
In the Yang-Mills theory, the vectorial connection $\G_{\underline a}
{}^{\cal I}$ was defined in terms of $\G_{\a}{}^{\cal I}$, $\G_{\Dot \a}
{}^{\cal I}$ and their derivatives.  This is motivation to attempt to
do the same  for supergravity. It turns out that this is possible
but in a slightly subtle way. We note
\be
\eqalign{
[ \, {\rm E}_{\a} ~,~ {\rm E}_{\Dot \a} \, \} &=~ C_{\a \, \Dot \a}
{}^{\underline c} \, {\rm E}_{\underline c} ~+~  C_{\a \, \Dot \a}
{}^{\g} \,{\rm E}_{\g} ~+~  C_{\a \, \Dot \a}{}^{\Dot \g} \,{\rm E}_{\Dot 
\g} \cr 
~~\to C_{\a \, \Dot \a} {}^{\underline c} \,{\rm E}_{\underline c} &=~ 
[ \, {\rm E}_{\a} ~,~ {\rm E}_{\Dot \a} \, \} ~-~  C_{\a \, \Dot \a}
{}^{\g}\, {\rm E}_{\g} ~-~  C_{\a \, \Dot \a}{}^{\Dot \g} \,{\rm E}_{\Dot 
\g} \cr 
~~\to C_{\a \, \Dot \a} {}^{\underline c} \,{\rm E}_{\underline c} &=~ 
[ \, {\rm E}_{\underline A}{}^{\underline M}\,( D_{\underline M}
{\rm E}_{\underline B}{}^{\underline N} ) \, - \, (-1)^{AB}
{\rm E}_{\underline B}{}^{\underline M}\,( D_{\underline M}
{\rm E}_{\underline A}{}^{\underline N} )\,] \, D_{\underline N} \cr
&{~~~}~-~  C_{\a \, \Dot \a}
{}^{\g} \,{\rm E}_{\g} ~-~  C_{\a \, \Dot \a}{}^{\Dot \g} \,{\rm E}_{\Dot 
\g} ~~~.\cr 
}\ee
If $C_{\a \, \Dot \a} {}^{\underline c}$ is a constant, then this 
equation defines ${\rm E}_{\underline c}$ as a function of ${\rm E}_{\g}$, 
${\rm E}_{\Dot \g}$ and their derivatives. The value of this constant can 
be deduced by considering the ``flat limit" ${\rm E}_{\underline c}
\, \to \, \pa_{\underline c}$, ${\rm E}_{\g} \, \to \, D_{\g}$ and ${\rm 
E}_{\Dot \g} \, \to \, {\Bar D}{}_{\Dot \g}$ which possesses no supergravity 
fields. This tells us
\be
C_{\a \, \Dot \a} {}^{\underline c} ~=~ i \d_{\a}{}^{\g} \, \d_{\Dot \a}
{}^{\Dot \g} ~~~. \label{eq:hnz}
\ee
The set of constraints in (\ref{eq:hgv}) and (\ref{eq:hnz}) have been named 
``conventional constraints'' \cite{5}. This name is suggested for a couple 
of reasons. First, these are conventional in the sense that they define 
some {\it a} {\it{priori}} independent gauge fields in terms of others and 
are thus a `convention' for the definitions of some gauge fields. These 
constraints are also conventional in the sense that they are the usual type 
of constraint that was seen in global supersymmetrical Yang-Mills
theory and general relativity.

\subsection{Superspace Supergravity Representation-Preserving Constraints}
 
~~~~Continuing we have the ``integrability conditions" \cite{6} coming from 
chiral matter superfields. In the case of the superspace Yang-Mills theory 
these conditions arose due the demand that ``covariant chiral'' superfields
should exist even in the presence of gauge symmetries. We may invoke the
same argument in the presence of the symmetries of supergravity.
In the absence of supergravity, chiral superfields are defined by $
{\Bar D}{}_{\Dot \a} \Phi = 0$.  However, in the presence of supergravity 
we know that $\Phi$ can undergo K-gauge transformations
\be
\Big( \Phi \Big)^{\prime} ~=~ e^K \ \Phi ~~~,~~~
K~\equiv~ K^{\underline M} D_{\underline M} ~=~ 
K^{\m} D_{\m} ~+~  K^{\Dot \m} {\Bar D}{}_{\Dot \m}
~+~  K^{\underline m} \pa_{\underline m} ~~~.
\ee
These transformations are precisely the generalization of those that
we saw in our discussion of the Lie algebraic formulation of theories
of ordinary gravitation.

It is clearly indicated that in order to possess this as a symmetry, the
definition of a chiral superfield in the presence of a supergravity
should be $\nabla_{\Dot \a} \Phi = 0$. (Since $\Phi$ is a Lorentz scalar,
the connection superfields do not appear in this condition.) We can
thus repeat the derivation of the `integrability constraints'' but
now applied to the superspace supergravity covariant derivative.
\be
\eqalign{ {~~~}
&\nabla_{\Dot \a} \Phi ~=~ 0 ~~\to ~~ \nabla_{\Dot \a} \nabla_{\Dot 
\b} \Phi ~=~ 0 ~~\to ~~ [ \, \nabla_{\Dot \a} \, , \,  \nabla_{\Dot 
\b} \,\} \Phi ~=~ 0 ~~\to \cr
&T_{\a \, \Dot \a} {}^{\underline c} \,  ({\nabla}_{\underline c}  \Phi  \,) 
~+~  T_{\a \, \Dot \a}{}^{\g} \, ( {\nabla}_{\g}  \Phi  \,)  ~+~  T_{\a \, 
\Dot \a}{}^{\Dot \g} \, ( {\nabla}_{\Dot \g}  \Phi  \,) ~=~ 0 ~~\to \cr
&T_{\a \, \Dot \a} {}^{\underline c} \, ({\nabla}_{\underline c}  \Phi  \,) 
~+~  T_{\a \, \Dot \a}{}^{\g} \,( {\nabla}_{\g}  \Phi  \,) ~=~ 0 ~~\to \cr
&C_{\a \, \Dot \a} {}^{\underline c} \, ({\nabla}_{\underline c}  \Phi  \,) 
~+~  C_{\a \, \Dot \a}{}^{\g} \,( {\nabla}_{\g}  \Phi  \,) ~=~ 0 ~~~.}\ee
This is just like the Yang-Mills case and has a similar solution 
\be
C_{\a \, \Dot \a} {}^{\underline c} = 0 ~~,~~ C_{\a \, \Dot \a}{}^{\, \g} 
= 0 ~~~. \label{eq:hnh}
\ee
In turn these constraints imply that
\be
[ \, {\rm E}_{\a} ~,~ {\rm E}_{\b} \, \} ~=~  C_{\a \, \b}
{}^{\g} \,{\rm E}_{\g} ~~~.
\ee
This equation means that the differential operators ${\rm E}_{\a}$
form a closed set under graded commutation. There is a result from
differential geometry due to Frobenius which implies that the
solution to these constraints is also like supersymmetric
Yang-Mills theory
\be
\eqalign{ {~~~} 
{\rm E}_{\a} &=~ \Psi \, N_{\a} {}^{\m} \, e^{\fracm 12 U} D_{\m} 
e^{- \fracm 12 U} ~~,~~  U ~\equiv ~ U^{\m} D_{\m} ~+~ {\Bar 
U}{}^{\Dot \m} {\Bar D}{}_{\Dot \m} ~+~ U^{\underline m} \pa_{\underline m} ~~~.
}\label{eq:eiv} \ee
In writing this solution, we have directly gone to a gauge that 
is analogous to that in (\ref{eq:nhx}).

The superfield $U^{\underline M}$, first suggested by Ogievetsky and 
Sokatchev \cite{21} who called it the ``axial vector superfield,'' 
is to supergravity as $V^{\cal I}$ is to supersymmetric Yang-Mills 
theory.  In the expression for ${\rm E}_{\a}$, $\Psi$ is an arbitrary 
complex scalar superfield (called the``superfield scale compensator'' 
or ``the density-type conformal compensator") and $N_{\a}{}^{\m}$ 
(called ``the superfield Lorentz compensator") is an arbitrary 
complex matrix superfield restricted to satisfy the constraint 
\be 
det (\, N_{\a} {}^{\m} \,) ~=~ 1 ~~~.
\ee
(Note that $N_{\a} {}^{\m}$ is not a supermatrix.) These two 
superfields have no analogs within the context of supersymmetric 
Yang-Mills theory.  Although a complete proof is beyond the level 
of exposition in these lectures, in addition it turns out that 
$U^{\m}$, ${\Bar U}{}^{\Dot \m}$ and $N_{\a}{}^{\m}$ may all be 
``gauged away'' in an appropriate W-Z gauge. Thus these are 
extraneous. The constraints in (\ref{eq:hnh}) are also called 
``representation-preserving constraints'' since they are required 
in order for the chiral superfield representation of rigid superspace
to exist in the presence of superspace supergravity.

The result in (\ref{eq:eiv}) is very revelatory. The curved 
Salam-Strathdee superspace (also called a Wess-Zumino superspace) 
definitely possess a ``supergeometry'' described by the superspace 
supergravity covariant derivative $\nabla_{\underline A}$ which is 
expressed in terms of the vielbein ${\rm E}_{\underline A}{}^{
\underline M}$ and superfield spin-connection $\o_{\underline 
A ~ \underline d \, \underline e}$. All of this is the simplest
sort of extrapolation of Riemannian geometry to superspace that 
should be expected. After imposing conventional constraints, all
geometrical quantities are expressed solely in terms of ${\rm 
E}_{\a}{}^{\underline M}$. After imposing (\ref{eq:hnh}) we enter
the domain of ``pre-geometry.'' This is a realm that, although
well understood for 4D, N = 1 superspace, remains essentially not
understood for most superspaces and therefore by implication for 
all theories that contain supergravity. Gaining an understanding
of pre-geometry is the most important challenge to superstrings, 
heterotic strings and all other such constructions. For it is only
with the understanding of pre-geometry that the superspace analog
of the equivalence principle reveals itself completely.

\subsection{Superspace Supergravity Conformal-Breaking Constraints}

~~~~From our discussion in the first lecture we know that a general superfield, 
such as $\Psi$ is not the smallest representation of supersymmetry.  It
would be nice to try to make $\Psi$ into some type of chiral superfield
\cite{22}.  This can actually be done if the constraints below are imposed.
\be
T_{\a \, \underline b} {}^{\underline c} ~=~ T_{\Dot \a \, \underline b}
{}^{\underline c} ~=~ 0 ~~~.
\ee
The solution to this final constraint is of the form
\be
\Psi ~=~ \Psi (U^{\underline m}, \, \varphi) ~~~,
\ee
where $\varphi$ is a chiral superfield in the sense that $ {\Bar D}{}_{\Dot 
\m} \varphi = 0 $. The explicit form of the function $\Psi (U^{\underline m}, 
\, \varphi)$ can be found in the last two books of the bibliography. 

\subsection{The WZ Gauge and Supergravity Component Fields}
  
~~~~All of the physics of (this version of) supergravity comes solely from
$U^{\underline m}$ and $\varphi$.  As in the Yang-Mills case, there 
exist a WZ gauge wherein
\be
\eqalign{ {~~~}
U^{\underline m}(z) &=~ \q^{\a} \bar\q^{\Dot \a} ~{\rm e}_{\underline 
a}{}^{\underline m}(x)  ~-~ ( \, i {\bar \q}{}^2  \q^{\a} \psi^{\underline 
b}{}_{\a}(x) \, + \, {\rm {h. \, c.}} \,) \, {\rm e}{}_{\underline
b}{}^{\underline m}(x) ~+~ \cr
&{~~~~~} \q^2  {\bar \q}{}^2 ( \,A^{\underline a} (x) \, - \, \fracm 13 \,
\e^{\underline a \, \underline b \, \underline c \, \underline d}
c_{\underline b \, \underline c \, \underline d} ~) \, {\rm e}{}_{\underline
a}{}^{\underline m}(x)
~~~. \label{eq:hnn}
}\ee
and in this same gauge \cite{17}
\be
\varphi ~=~ {\rm e}^{-\fracm 13} \, \{ ~ 1 ~-~ \q^2 \, [ \, S(x) ~+~ i \, 
P(x) \, ]  ~ \} 
~~~.  \label{eq:fzz}
\ee
Thus, the component fields of (minimal) 4D, N = 1 supergravity are

${~~~~}$ ${\rm e}_{\underline a}{}^{\underline m}(x)  $ graviton spin - 2

${~~~~}$ $\psi_{\underline a}{}^{\m}(x)$ gravitino spin - 3/2

${~~~~}$ $A_{\underline a} (x) $ axial-vector auxiliary field

${~~~~}$ $S(x)$    scalar auxiliary field spin-0

${~~~~}$ $P(x)$   pseudoscalar auxiliary field spin-0

These are the fields that remain in the WZ gauge and are therefore the
only component fields that occur in the component-level formulations.
A complete set of constraint as imposed in our discussion was first
suggested by J. Wess and B. Zumino \cite{23}.  In recognition of this 
they are often called the ``Wess-Zumino supergravity constraints.''

\subsection{Graded Commutator Algebra of the Supergravity Supercovariant  
Derivative and Supergravity Action} 

~~~~Independently, Warren Siegel and I developed the complete description 
\cite{17} of $\nabla_{\underline A}$ in terms of $U{}^{\underline m}$ and 
$\varphi$ in the period of 1977-1979.  We also used a slightly different 
set of constraints from Wess and Zumino,
\be
T_{\underline a \, \underline b}{}^{\underline c} ~=~ 0 ~~\to~~ 
R_{\a \, \Dot \a \, \underline c \, \underline d} ~=~ 0 ~~~.
\ee
This choice only determines $\o_{\underline a \, \underline c \, \underline 
d}  $ in a slightly different way.  In our system, the algebra of the 
supergravity covariant derivative takes the form
\be
\eqalign{ 
[ \nabla_{\a }, \nabla_{\b} \}  &=~ - 2 \Bar R \,{ \cal M}_{\a \b}
~~~~~, ~~~~~ 
[ \nabla_{\a }, { \Bar {\nabla}}_{\dot \a} \}  ~=~ i {\nabla}_{ \underline a}
~~~~~, \cr
[ \nabla_{\a }, \nabla_{\underline b} \}  &=~ - i \, C_{ \a \b} [~ \Bar R \, 
{\Bar {\nabla}}_{\dot \b} ~-~ G^{\g} {}_{\dot \b } \nabla_{\g} ~ ] 
~-~ i \, ( {\Bar \nabla}_{\dot \b} \Bar R )  { \cal M}_{\a \b} ~~ \cr
&~~~~~+~ i \, C_{\a \b} [~ {\Bar W}_{\dot \b \dot \g} {}^{\dot \d}
 { \Bar {\cal M}}_{\dot
\d} {}^{\dot \g} ~-~ ( \nabla^{\d} G_{\g \dot \b}) {\cal M}_{
\d} {}^{ \g}  ~ ]  ~~~, \cr
[ {\nabla}_{\underline a }, \nabla_{\underline b} \}  &=~ \{ ~ [~ C_{\dot \a 
\dot \b} W_{ \a \b} {}^{ \g} ~+~ \fracm 12  C_{ \a \b} ({\Bar \nabla}_{( 
\dot \a} G^{\g}{}_{ \dot \b )}) ~-~ \fracm 12 C_{\dot \a \dot \b} \,( 
\nabla_{ ( \a} R  \,)  \,~ ] \nabla_{\g} \cr
&~~~~~+~  i \fracm12  C_{\a \b } G^{\g} {}_{ ( \dot \a} \nabla_{ \g 
\dot \b)} ~-~ \fracm 12 C_{\a \b} \, ( \, {\Bar \nabla}_{( \dot \a} 
\nabla_{\g}  G_{\d}{}_{ \dot \b )}\, ) \, ~] {\cal M}^{\g \d}   \cr
&~~~~~-~ [~ C_{\dot \a \dot \b} W_{ \a \b \g \d} ~-~ \fracm 12 C_{\a 
\b} \, ({\Bar \nabla}_{( \dot \a} \nabla_{\g} G_{\d}{}_{ \dot \b )})~] 
{\cal M}^{\g \d}   \cr
&~~~~~- i \fracm 12 C_{\dot \a \dot \b} C_{\g ( \a}\,  [~ \fracm 14 (\, 
\nabla_{\b ) } {}^{\dot \e} G_{\d \dot \e} ~+~  \nabla_{\d}{}^{\dot \e} 
G_{\b ) \dot \e}) ~\, ] {\cal M}^{\g \d} \cr
&~~~~~- \fracm 12 C_{ \dot \a \dot \b } C_{\g ( \a } C_{\b ) \d } \, 
[ \, {\Bar \nabla}^2 \Bar R ~+~ 2 R \Bar R ~\, ] {\cal M}^{\g \d} ~+~ 
{\rm {h.\, c.}}  ~ \}  ~~~,} \label{eq:uzw}
\ee
the superfields $W_{\a \b \g}$, $G_{\underline a}$ and $R$ are to supergravity 
as $W_{\a}{}^{\cal I}$ is to supersymmetric Yang-Mills theory.  The superspace 
Bianchi identities exist in direct analogy with ordinary gravity and
supersymmetrical Yang-Mills theory
\be 
\eqalign{
0 ~=~ & (-1)^{A C} \, [ \, [ \, \nabla_{\underline A} ~,~ \nabla_{\underline 
B} \, ] ~,~\nabla_{\underline C} ~] ~+~ (-1)^{B A} \, [ \, [ \, \nabla_{
\underline B} ~,~ \nabla_{\underline C} \, ] ~,~ \nabla_{\underline A} ~] 
 \cr &+~(-1)^{C B} \,[ \, [ \, \nabla_{\underline C} ~,~ \nabla_{\underline 
A} \, ] ~,~\nabla_{\underline B} ~] ~~~,}
\ee
and they imply the following equation for the superfields $W_{\a \b \g}$, 
$G_{\underline a}$ and $R$
$$
G_{\a \dot \b } ~=~ {\Bar G}_{\a \dot \b } ~~,~~ 
{\Bar {\nabla}}_{\dot \b} R ~=~ {\Bar {\nabla}}_{\dot \b} 
W_{\a \b \g} ~=~ 0~~, 
$$
$$
\nabla^{\a} R ~+~ {\Bar {\nabla}}_{\dot \a} G^{\a \dot \a}
 ~=~ 0  ~~,~~ 
\nabla^{\b} W_{\b \g \d } ~+~ i \fracm 12 {\nabla}_{( \g \dot \d } 
G_{ \d )}{}^{\dot \d} ~=~ 0 ~~~, $$
\be 
 \fracm1{4!}{\nabla}_{( \a} W_{\b \g \d) } ~= ~ W_{\a \b \g \d} ~~.
\ee

In ordinary gravity, the Einstein-Hilbert action takes the form in 
(\ref{eq:whv}) and in 4D, N =1 supersymmetric Yang-Mills theory the action 
takes the form in (\ref{eq:nfn}).  So it is obviously of interest to know 
what is the form of the 4D, N = 1 supergravity action.  As first suggested 
by Siegel \cite{22}, it is
\be
{\cal S}_{SG} ~=~ - \fracm3{\k^2} \, \int d^4 x \, d^2 \q \, d^2 {\bar \q}
~ {\rm E}^{-1} (U^{\underline m}, \, \varphi) ~~~,~~~
{\rm E} ~\equiv~ sdet({\rm E}_{\underline A}{}^{\underline M} )
~~~. \label{eq:uzv}
\ee
As indicated in the equation above ${\rm E}(U^{\underline m}, \, 
\varphi)$ is a function of $U^{\underline m}$ and $\varphi$ alone. Variation 
of this action produces two equations
\be
\eqalign{
{\rm e}_{\underline a}{}^{\underline m} \parvar{{\cal S}_{SG}}
{U^{\underline m}} &= ~ 0 ~~~\to~~~ G_{\underline a} ~=~ 0 ~~~, \cr
\parvar{{\cal S}_{SG}}{\varphi} &= ~ 0 ~~~\to~~~ {\Bar R} ~=~ 0 
~~~~~.
}\ee
Of course written in terms of superfields, the supergravity action looks
(but hides lots) very simple.  To get more insight into what's going on,
a peek at components is warranted.

\subsection{Component Supergravity Action}

~~~~After using some techniques that will be explained later, we find
(\ref{eq:uzv}) leads to
\be
\eqalign{ {~~~}
{\cal S}_{SG} &=~ \fracm1{\k^2} \, \int d^4 x ~ {\rm e}^{-1} \, 
\Big[ \, r(\o({\rm e},\psi,A),{\rm e}) \, - \, i \, \e^{\underline 
a \, \underline b \, \underline c \, \underline d} \, {\Bar \psi}{}_{
\underline a \, \Dot \b} \, {\psi}_{\underline c \, \underline d \, 
\b} {~~~~~~}\cr
&{~~~~~~~~~~~~~~~~~~~~~~~~~~}~-~ 3 \, ( \, S^2 ~+~ P^2 \,) ~ \Big] ~~~.
} \label{eq:fzv} \ee
This result (in a slightly different form) was first derived without 
its auxiliary field structure by Freedman, Ferrara and van Nieuwenhuizen 
\cite{24} and later with the auxiliary field by Ferrara and van Nieuwenhuizen 
\cite{25}.  Above the spin-connection is given by
\be
\eqalign{ {~~~}
\o({\rm e},\psi,A) &=~ \o_{\underline a \, \underline b \, \underline 
c}({\rm e}) ~+~ i \, \fracm 12 (~  {\psi}{}_{[ \underline b \, \a} {\Bar 
\psi}{}_{\underline c ]  \, \Dot \a} ~+~ {\psi}{}_{[ \underline a \, \b} 
{\Bar \psi}{}_{\underline c ]  \, \Dot \b} ~-~ {\psi}{}_{[ \underline a 
\, \g} {\Bar \psi}{}_{\underline b ]  \, \Dot \g} ~) \cr
&{~~~~}-~ \fracm 12 \, \e_{\underline a \, \underline b \, \underline 
c \, \underline d} \, A^{\underline d}
~~~, 
} \label{eq:fzg} \ee
where $\o_{\underline a \, \underline b \, \underline c}(
{\rm e})$ is the spin-connection (\ref{eq:wve}) in ordinary general relativity
and the ``curl'' of the gravitino is explicitly of the form
\be
\eqalign{ {~~~}
{\psi}_{\underline a \, \underline b}{}^{\g} &\equiv~ {\rm e}_{[
\underline a}{}^{\underline m} \,\Big( \pa_{\underline m} 
{\psi}{}_{\underline b ]}{}^{\g} ~-~ \o_{\underline m \, \b}{}^{\g}
({\rm e},\psi,A) \, {\psi}{}_{\underline b ]}{}^{\b} \Big) ~-~ 
c_{\underline a \, \underline b}{}^{\underline d} \, {\psi}{}_{
\underline d} {}^{\g}
\cr
&\equiv~ {\rm e}_{[ \underline a}{}^{\underline m} \,\Big( {\cal D}_{
\underline m}({\rm e},\psi,A) {\psi}{}_{\underline b ]}{}^{\g} \Big) 
~-~c_{\underline a \, \underline b}{}^{\underline d} \, {\psi}{}_{\underline 
d} {}^{\g} \cr
&\equiv~ \Big( {\cal D}_{[ \underline a}({\rm e},\psi,A) {\psi}{}_{\underline 
b ]}{}^{\g} \Big) ~-~ c_{\underline a \, \underline b}{}^{\underline d} \, 
{\psi}{}_{\underline d} {}^{\g} ~~~~.}
\ee
We can also define
\be
\o_{\underline a \, \underline b \, \underline c}({\rm e},\psi,A)
~=~ \o_{\underline a \, \underline b \, \underline c}({\rm e},\psi) 
~-~ \fracm 12 \, \e_{\underline a \, \underline b \, \underline c 
\, \underline d} \, A^{\underline d} ~~~,
\ee
and use this to re-express the component level action in the form
\be
\eqalign{
{\cal S}_{SG} &=~ \fracm1{\k^2} \, \int d^4 x ~ {\rm e}^{-1} \, \Big[ \, 
r(\o({\rm e},\psi),{\rm e}) \, - \,i \, \e^{\underline a \, \underline b \, 
\underline c \, \underline d} \, {\Bar \psi}{}_{\underline a \, \Dot \b} 
\, {\psi}_{\underline c \, \underline d \, \b} {~~~~~~~~~~}\cr
&{~~~~~~~~~~~~~~~~~~~~~~~~~~}~-~ 3 \, ( \, A^{\underline a}
\, A_{\underline a} ~+~ S^2 ~+~ P^2 \,) ~ \Big] ~~~.
}\ee
This is the almost universal form in which the supergravity action appears
in the literature. Let us end this section by noting that this action 
has one final form in which it may be written,
\be
\eqalign{
{\cal S}_{SG} &=~ \fracm1{\k^2} \, \int d^4 x ~ {\rm e}^{-1} \, \Big[ \, 
- \fracm 14 \, c^{\underline a \, \underline b 
\, \underline c}\, c_{\underline a \, \underline b \, 
\underline c}\, ~+~ \fracm 12 \, c^{\underline a \, \underline b 
\, \underline c}\, c_{\underline c \, \underline a \, 
\underline b}\, ~+~ ( \, c^{\underline a}{}_{\underline 
b}{}^{\underline b}\,)^2 \cr
&{~~~~~~~~~~~~~~~~~~~~~~~~~~} \, - \,i \, \e^{\underline a \, 
\underline b \, \underline c \, \underline d} \, {\Bar \psi}{}_{\underline 
a \, \Dot \b} \, {\Hat {\psi}}_{\underline c \, \underline d \, \b} 
~-~ 3 \, ( \, A^{\underline a} \, A_{\underline a} ~+~ S^2 ~+~ P^2 \,) 
\cr
&{~~~~~~~~~~~~~~~~~~~~~~~~~~}~+~ \psi^4 -{\rm {terms}} ~~ \Big] ~~~.
} \label{eq:ftv} \ee
The gravitino field strength that appears here is defined by
\be
{\psi}_{\underline c \, \underline d \, \b} ~=~ {\Hat {\psi}}_{
\underline c \, \underline d \, \b} ~+~ \psi^3 -{\rm {terms}} ~~~.
\ee
The $\psi^4 -{\rm {terms}}$ in the action arise from this
substitution as well as writing
\be
r(\o({\rm e},\psi),{\rm e}) ~=~ r(\o({\rm e}),{\rm e}) ~+~
\psi^4 -{\rm {terms}} ~~~.
\ee
This form of the action is noteworthy for a couple of reasons. First
this is the form of the supergravity action as first noted by Ferrara,
Freedman and van Nieuwenhuizen. Secondly, if we recall the component
fields in (\ref{eq:hnn}) and (\ref{eq:fzz}), it is apparent that the 
variable $\o_{\underline a \, \underline b \, \underline c}$ does not 
occur at all among the component fields! In other words the actual
evaluation of (\ref{eq:uzv}) really leads to the action in (\ref{eq:ftv}). 
The action in (\ref{eq:fzv}) is just a very convenient way to re-write 
the final result. 

The most important point about this observation is that any
manifestly supersymmetric quantum theory that begins with the
action of (\ref{eq:uzv}) {\it {can}} {\it {never}} lead to a
quantum effective action of the form in (\ref{eq:tzo}). Instead
the pure graviton part of the supersymmetric quantum gravity
effective action must take the form of (\ref{eq:wee}).

\subsection{Local Symmetries of the Component Supergravity Action}

The supergravity fields transform under a number of local symmetries.
All possess infinitesimal local Lorentz transformation of the form
\be
\d_{L.L.}( \l) {\cal F} ~=~ [ ~ \l_{\a}{}^{\b}(x) {\cal M}_{\b}{}^{\a}
~+~ {\bar \l}{}_{\Dot \a}{}^{\Dot \b}(x) {\Bar {\cal M}}{}_{\Dot\b}
{}^{\Dot \a} \, , \, {\cal F} ~ ] ~~~,
\ee
where ${\cal F} \equiv \{ {\rm e}_{\underline a}{}^{\underline m},
\, \psi_{\underline a}{}^{\b} , \, A_{\underline a} , \, S , \,
P\}  $. Similarly they almost all transform as
\be
\d_{G.C.}( \xi) {\cal F} ~=~ \xi^{\underline m} \pa_{\underline m}
{\cal F} ~~~,
\ee
under an infinitesimal coordinate transformation. The exception is 
${\rm e}_{\underline a}{}^{\underline m}$ for which
\be
\d_{G.C.}( \xi) {\rm e}_{\underline a}{}^{\underline n} ~=~ 
\xi^{\underline m} \pa_{\underline m} {\rm e}_{\underline a}{}^{\underline 
n} ~-~ {\rm e}_{\underline a}{}^{\underline m} \pa_{\underline m} 
\xi^{\underline n} ~~~.
\ee
But the most interesting variations are those under local supersymmetry
\be
\eqalign{
&\d_Q(\e) \,{\rm e}_{\underline a}{}^{\underline m} ~=~ i \,
( \, \e^{\b} \, {\Bar \psi}{}_{\underline a}{}^{\Dot \b}
~+~  {\bar \e}^{\Dot \b} \, {\psi}{}_{\underline a}{}^{\b} \,)\, 
{\rm e}_{\underline b}{}^{\underline m} ~~~,  \cr
&\d_Q(\e) \,{\psi}_{\underline a}{}^{\b} ~=~ {\cal D}_{\underline
a}({\rm e}, \psi , A) \e^{\a} ~+~ i \, ( \, \e^{\b} \, {\Bar 
\psi}{}_{\underline a}{}^{\Dot \b} ~+~  {\bar \e}^{\Dot \b} 
\, {\psi}{}_{\underline a}{}^{\b} \,) \, \psi_{\underline b}{}^{\a}  
\cr
&{~~~~~~~~~~~~~~~~~}+~ i \, {\bar \e}{}_{\Dot \a} \, \d_{\a} {}^{\b} 
\, ( \, S ~+~ i \, P \,) ~-~ i \e_{\a} A^{\b}{}_{\Dot \a} ~~~,\cr
&\d_Q(\e)\, A_{\underline a} ~=~  - \fracm 12 \, \Big[
\, \e^{\g} \, {\Bar f}{}_{\g \Dot \b \, \a}{}^{\Dot \b}{}_{\Dot \a} ~+~
\fracm 13 \e_{\a} \, {\Bar f}{}_{\b \Dot \a \,}{}^{ \, \b}{}_{\Dot \d}
{}^{\, \Dot \d} \,\,  \Big] ~+~ { \rm { h. \, c.}} ~~~,\cr
&\d_Q(\e) \, ( \, S ~+~ i \, P \,) ~=~   - \fracm 13 \, \e^{\a}
\, f_{\a \Dot \b \,\, \g}{}^{\Dot \b}{}_{\, \g} ~~~.
}\ee
In all of these expressions $\e^{\a}$ and ${\bar \e}{}^{\Dot \a}$ are 
functions of $x$ (i.e. local parameters of supersymmetry).

In the last two lines above, the quantity $f_{\underline a
\, \underline b}{}^{\g}$ is the ``supercovariantized curl''
of the gravitino (also known as the supercovariantized gravitino
field strength) which is explicitly given by
\be
\eqalign{
f_{\underline a \, \underline b}{}^{\g} ~=~ &\psi_{\underline a
\, \underline b}{}^{\g} ~+~ i \, ( \, \psi_{\underline a \, \b} \,
A^{\g}{}_{\Dot \b} ~-~ \psi_{\underline b \, \a} \,
A^{\g}{}_{\Dot \a} \, )  \cr
&~-~ i \, ( \, S ~+~ i \, P \,) \,\, ( \, {\Bar \psi}{}_{\underline a \, 
\Dot \b} \, \d_{\b}{}^{\g} ~-~ {\Bar \psi}{}_{\underline b \, \Dot 
\a} \, \d_{\a}{}^{\g} \, ) ~~~. }
\ee
The most interesting property of this quantity is that under a
local supersymmetry variation, it transforms covariantly (i.e.
without terms proportional to spacetime derivatives of $\e$
or ${\Bar \e}$). In a similar manner, the Riemann curvature
tensor also possesses a ``supercovariantized'' version given
by
\be
\eqalign{ {~~~}
g_{\underline a \, \underline b \, \underline c \, \underline d}
&=~ C_{\Dot \g \, \Dot \d} \,\, g_{\underline a \, \underline b \, 
\g \, \d} ~+~ C_{\g \,\d}\,\,  g_{\underline a \, \underline b \, 
\Dot \g \, \Dot \d} ~~~, \cr
g_{\underline a \, \underline b \, \g \, \d} &=~ r_{\underline 
a \, \underline b \, \g \, \d} ~+~ (\, S \,- \, i P \,) \,
{\psi}_{\underline a \, ( \g |} {\psi}_{\underline b \, | \d) }\cr
&{~~~~~~}+~ i \Big\{ \, \Big[ \,  \fracm 1{12} {\Bar \psi}
{}_{\underline a \, \Dot \g } \, f_{(b | \Dot \k \, |\g| }{}^{\Dot 
\k}{}_{ |\d)}  ~+~ \fracm 14 {\psi}_{\underline a \, \b}  \, {\Bar 
f}{}_{( \g | \Dot \e \,\,|\d)}{}^{\Dot \e}{}_{\, \Dot \b} \cr
&{~~~~~~~~~~~~~~~~}+~ 
\fracm 16 {\psi}_{\underline a \, ( \g } C_{\d ) \, \b} \, 
{\Bar f}{}_{\k  \Dot \b\,}{}^{\, \k}{}_{\Dot \k}{}^{\, \Dot \k}
\, \Big] ~-~ ( \underline a ~ \iff ~ \underline b) ~ \Big\} 
~~~~.}\ee
In this expression above, the quantity $ r_{\underline 
a \, \underline b \, \g \, \d}$ is defined by
\be
r_{\underline a \, \underline b \, \underline c \, \underline d}
~=~ C_{\Dot \g \, \Dot \d} \,\, r_{\underline a \, \underline b \, 
\g \, \d} ~+~ C_{\g \,\d}\,\,  r_{\underline a \, \underline b \, 
\Dot \g \, \Dot \d} ~~~.
\ee
For this supergravity theory $r_{\underline a \, \underline b \, 
\underline c \, \underline d}$ is defined as in (\ref{eq:txx})
but where the spin-connection is defined by (\ref{eq:fzg}).

If one computes the commutator of two local supersymmetry variations on
any field, the answer is
\be
\eqalign{
[ ~ \d_Q (\e_1 ) \, , \, \d_Q (\e_2 ) ~ ] &=~  \d_{G.C.} (\xi_{12} ) 
~+~  \d_{L.L.} (\l_{12} ) ~+~  \d_{Q} (\e_{12} ) ~~~,{~~~~~~} \cr
\xi_{12} {}^{\underline m} & = ~ - \, i \, ( \, \e^{\a}_1 \, {\bar 
\e}{}^{\Dot \a}_2 ~+~  {\bar \e}^{\Dot \a}_1 \, {\e}^{\a}_2 \,)\, {\rm 
e}_{\underline a}{}^{\underline m} ~~~,\cr
\l_{12} {}^{\underline b \, \underline c} & = ~ 
- \, i \, ( \, \e^{\a}_1 \, {\bar \e}{}^{\Dot \a}_2 ~+~  {\bar 
\e}^{\Dot \a}_1 \, {\e}^{\a}_2 \,)\,\, \o_{\underline a}{}^{\underline b
\, \underline c}({\rm e}, \, \psi , \, A) \cr
&{~~~~~}+ \, 2 [ ~ {\e}^{( \b}_1 \, {\e}^{\g )}_2 \, C^{\Dot \b \, \Dot \g}
 \, \,(\, S \,+ \, i P \,) ~+~ { \rm {h. \, c.}} ~ ] ~~~,\cr
\e_{12} {}^{\b} & = ~ 
- \, i \, ( \, \e^{\a}_1 \, {\bar 
\e}{}^{\Dot \a}_2 ~+~  {\bar \e}^{\Dot \a}_1 \, {\e}^{\a}_2 \,)\, 
{\psi}_{\underline a}{}^{\b}
 ~~~.
 }\ee
As can be seen the composition laws for $\xi_{12}$, $\l_{12}$
and $\e_{12}$ depend on the fields of the supergravity multiplet.
For this reason the commutator algebra for the symmetries of 
supergravity theories are often described as being ``field-dependent.''

\subsection{Supersymmetric Matter Coupling}

~~~~Coupling matter superfields to the supergravity multiplet works
essentially like that described for the component fields coupling
of gravity. The key point is to note that local superspace
possesses a measure analogous to $\int d^4 x {\rm e}^{-1}$.
Arnowitt, Nath and Zumino \cite{200} first suggested such 
integration measures should be written as
\be
\int~d\mu ~\equiv~ \int d^{N_B} x \, d^{N_F} \q ~ {\rm E}^{-1} ~=~
\int d^{N_B} x \, d^{N_F} \q ~ [ sdet \, ({\rm E}_{\underline A} 
{}^{\underline M} 
 (\q , x)\,) \,]^{-1} ~~~,
\ee
for a superspace of $N_{B}$ bosonic coordinate and $N_F$ fermionic
coordinates. For the case of our interest $N_{B} = N_F = 4$.
A minimal coupling philosophy suggests that if we start with
a supersymmetric action of the form
\be
{\cal S}_{Matter} ~=~ \int d^4 x \, d^2 \q \, d^2 {\bar \q} ~
{\cal L} ( \Phi, \, {\Bar \Phi}, \, D_{\underline A}) ~~~, 
\ee
its coupling to supergravity may be simply obtained by the
replacement
\be
{\cal S}_{Matter} ~=~ \int d^4 x \, d^2 \q \, d^2 {\bar \q} ~
{\rm E}^{-1} \, 
{\cal L} ( \Phi, \, {\Bar \Phi}, \, \nabla_{\underline A}) ~~~. 
\ee
It is also possible to consider non-minimal coupling by
considering the class of actions described by
\be
{\cal S}_{Non-min.-Matter} ~=~ \int d^4 x \, d^2 \q \, d^2 {\bar \q} ~
{\rm E}^{-1} \, 
{\cal L} ( \Phi, \, {\Bar \Phi}, \, \nabla_{\underline A},
\,
W_{\a \b \g} , \, G_{\underline a}, \, R) ~~~. \label{eq:hhh}
\ee

However, one somewhat surprising feature of superspace supergravity
is that it possesses more than one local measure over which
integrations may be performed to construct invariant objects. 
The existence of the ``chiral measure'' is a local generalization 
of the ``F-terms'' invariants that we met in the first lecture.  
In addition to ${\rm E}^{-1}$, there exists \cite{17} an object 
${\cal E}^{-1}$ that permits the definition
of a measure
\be
\int~d \mu_c ~\equiv~ \int d^4 x \, d^2 \q ~ {\cal E}^{-1} 
~~~. \ee
This measure will lead to an invariant if it is applied to a
Lagrangian superfield ${\cal L}_c$ which satisfies the
superspace supergravity chirality condition ${\Bar \nabla}
{}_{\Dot \a} {\cal L}_c = 0$. The existence of such a measure
is critical so that the superpotential term possesses an
extension in the presence of supergravity,
\be
{\cal S}_{Pot} ~\equiv~ \int d^4 x \, d^2 \q ~ {\cal E}^{-1} \,
W(\Phi) ~~~. \ee
Similarly, the chiral measure is important to be able to
write the action of supersymmetric Yang-Mills theory in the
presence of supergravity,
\be
{\cal S}_{YM} ~\equiv~ \frac 14
Tr \Big[ ~ \int d^4 x \, d^2 \q ~ {\cal E}^{-1} \,
W^{\a} \, W_{\a} ~+~ {\rm {h. \, c.}} ~ \Big]  ~~~. \ee
In writing this last equation, we should recognize however,
that the superfield $W_{\a}$ is modified from is rigid definition
because in the presence of supergravity a combined 
supergravity-Yang-Mills supercovariant derivative
\be
\nabla_{\underline A} ~=~ {\rm E}_{\underline A}
~+~ \fracm 12 \, \o_{\underline A \, \underline c}{}^{\underline d} {\cal
M}_{\underline d}{}^{\underline c} ~-~ i\, \G_{\underline A}{}^{\cal I}
\, t_{\cal I} ~~~,
\ee
must be used to define it. 

\subsection{Theory of Local Superspace Integrations and 
Component Results }

~~~~In the case of rigid supersymmetry, we were efficiently able to
derive component results from superspace ones via an equation of
the form
\be
\int d^{4} x \, d^2 \q \, d^2 {\bar \q} ~ {\cal L} ~\equiv~ \fracm 12 \,
\Big\{ \int d^{4} x ~ [\, D^2 \, {\Bar D}{}^2  \,  {\cal L} \, | ~] 
~+~{\rm {h. \, c.}} ~ \Big\} ~~~,
\ee
where
\be
D_{\a}~\equiv~ \partial_{\alpha} ~ +~ i \frac{1}{2} \bar{\q}^{
\Dot{\alpha}} \partial_{\underline{a}}  ~~~,~~~{ \Bar D}_{\Dot{\alpha}} 
~\equiv~ \bar{\partial}_{\Dot{\alpha}}~+~i \frac{1}{2} \q^{\alpha} 
\partial_{\underline{a}}  ~~~.
\ee
In the presence of supergravity, it is important to maintain this
efficiency and level of completeness in our understanding of the
theory. This suggests that there should exist for a general
superspace an operator ${\cal D}^{N_F}$ 
such that
\be
\int d^{N_B + N_F} z ~ {\rm E}^{-1} {\cal L} ~=~ \int d^{N_B} z ~
{\rm e}^{-1} \, [~ {\cal D}^{N_F} {\cal L} {\bf |} ~] ~~~,
\ee
independent of the superfield $\cal{L}$ that appears in this equation
and where
\be
{\rm e}^{-1} ~ \equiv ~ [det\, ({\rm e}_{\underline a} {}^{\underline m} 
( x)\,)] ^{-1}  ~~~,~~~ {\cal D}^{N_F} {\cal L} | ~ \equiv ~ 
\lim_{\q \to 0}^{} ~ (\, {\cal D}^{N_F} {\cal L} 
) ~~~. \ee
Also this operator should be such that it can be written as
\be
\int d^{N_B} z ~{\rm e}^{-1} \,  \Big[ ~ {\cal D}^{N_F} {\cal 
L} \, | ~ \Big] ~=~ \int d^{N_B} z ~{\rm e}^{-1} \, \Big[ ~ 
\sum_{i = 0}^{N_F} \, c_{(N_F - i)} \, (\nabla \, \cdot\,  \, 
\cdot\, \, \cdot\, \nabla)^{N_F - i}  \, {\cal L} | ~ \Big]
~~~,
\ee
in terms of some field-dependent coefficients $c_{(N_{F}-i)}$
and powers of the spinorial superspace supergravity covariant 
derivative $\nabla_{\a}$. It is a remarkable fact that although
the operator ${\cal D}^{N_F}$ has been known on a case-by-case
basis for many examples, until quite recently \cite{26} there 
existed no systematic way to construct these field dependent 
coefficients. 

It is now known that there are two such methods to find these
coefficients by the use of; (a.) the super differential calculus
of super p-forms and (b.) a normal-coordinate expansion of
the $\q$ coordinate of superspace. Both of these methods lead
to the same result which we now describe. The critical point
is the form of an operator ${\cal D}^2$ that takes the form
\be
{\cal D}^2 ~\equiv ~ \nabla^2 ~+~ i  {\Bar \psi}{}^{\underline 
a}{}_{\dot \a} \nabla_{\a} ~+~ 3 {\Bar R} ~+~  \fracm 12 C^{\a 
\b} {\Bar \psi}_{\underline a}{}^{( \dot \a} \, {\Bar \psi
}_{\underline b  }{}^{\dot \b )} 
~~~.
\ee
Using this we define
\be
\eqalign{ {~~~~}
\int d \m ~{\cal L} &\equiv~ \int d^4 x \, {\rm e}^{-1} 
~ \Big[ \, {\cal D}^4 ~ {\cal L} \, | ~ \Big] \cr
&\equiv~ \fracm 12\Big\{ ~ \int d^4 x \, {\rm 
e}^{-1} ~ \Big[ \, {\cal D}^2 \, ~({\Bar \nabla}^{2} ~+~ R) \, {\cal 
L} \, | ~ \Big] ~+~ {\rm {h.c.}} ~\Big\} ~~~,} \label{eq:uhx}
\ee
where the operator ${\cal D}^4$ is defined by
\be
{\cal D}^4 ~=~ \fracm 12 \, \Big[ ~ {\cal D}^2 \, ~({\Bar \nabla}^{2} 
~+~ R) ~+~ {\Bar {\cal D}}^2 \, ~({ \nabla}^{2} ~+~ {\Bar R}) ~\Big] 
~~~. \ee
Similarly, for the case of the chiral measure we define
\be
\int d \m_c ~ {\cal L}_c ~=~ \int d^4 x \, {\rm e}^{-1} ~ \Big[ \,
{\cal D}^2 {\cal L}_c \, | ~ \Big] ~~~.
\ee
By using the formulae above (that we have named ``density projectors'') 
the problem of finding the explicit component forms of any local 
superspace action can be reduced to the problems of a series of 
differentiations, etc.

In the case of supersymmetric Yang-Mills theory, we saw how the
component fields also emerged from the field strength supertensors
associated with the theory. A similar occurrence takes place
for supergravity. In particular looking back at the form of
the superspace supergravity commutator algebra, three supertensors
are apparent; $W_{\a \b \g}$, $G_{\underline a}$ and $R$. It is
of interest to know the correspondence of these to the component
fields of supergravity. After some long calculations starting
from the solution to constraints, one can verify
\be
G_{\underline a} { |} ~=~ A_{\underline a} ~~~,~~~ R { |}
~=~ (\, S \, + \, i P \, ) ~~~.
\ee
In a similar manner it can be shown that
\be
T_{\underline a \, \underline b}{}^{\g} | ~=~ f_{\underline a 
\, \underline b}{}^{\g} ~~~,~~~ R_{\underline a \, 
\underline b \, \g}{}^{\d} | ~=~ g_{\underline a \, 
\underline b \, \g}{}^{\d}~~~. \ee
The lowest component of the superfield $W_{\a \b \g}$ just 
corresponds to 
\be
W_{\a \, \b \, \g} | ~=~ \fracm 1{12} C^{\Dot \a \, \Dot \b}
\, f_{( \a | \Dot \a \, | \b |\Dot \b \, | \g)} 
~~~.
\ee

Finally, to be able to evaluate most actions completely
in terms of component fields, it is useful to know the 
identities,
\be
\eqalign{
\nabla_{\a} R &=~ - \, \fracm 13 \, T_{\a \Dot \a \, \, \g }
{}^{\Dot \a \, \g}   ~~, \cr
\nabla_{\a} G_{\underline b} &=~ - \, \fracm 12 \, \Big[ ~ \, 
T_{\a \Dot \a \, \, \, \b } {}^{\Dot \a \,}{}_{\Dot \g}  
 - \, \fracm 13 \, C_{\a \, \b} \, T_{\g \Dot \b}{}^{ \, \,  
\g } {}^{\Dot \d \,}{}_{\, \Dot \d} ~ \Big] 
~~~. }\ee
${}$\newline
Now we can explain the derivation of (\ref{eq:fzv}) from first principles.
Since we know that the action for old minimal supergravity is
just given by the integral of the volume element of the full
superspace, it follows from (\ref{eq:uhx}) that
\be
\int d \m ~ 1 ~=~  \fracm 12\Big\{ ~ \int d^4 x \, {\rm 
e}^{-1} ~ \Big[ \, {\cal D}^2 \,  R \,  \, | ~ \Big] ~+~ 
{\rm {h.c.}} ~\Big\} ~~~.
\ee
It remains to simply evaluate the differential operator ${\cal 
D}{}^2$ acting on $R$. The first derivative expression is given 
just above. For, the second derivative, once again using the 
technique known as ``solving the supergravity Bianchi identities'' 
the appropriate expression can be derived.  After completing 
this, we arrive at the action in (\ref{eq:uhx}).

We would be remiss if we did not note that the superspace geometry
and its corresponding supergravity theory discussed here is but one of
a number of such theories. It is called ``old minimal supergravity''
to distinguish it from other possibilities. This is an off-shell,
irreducible local supersymmetry representation. All other 
off-shell, irreducible versions of supergravity only differ
from this one by the set of constraints on their supergeometries
and the auxiliary component fields that appear.

\subsection{4D, N = 1 Supergeometry from Heterotic and Superstrings}

~~~~Presently many physicists regard the most fundamental of theories 
as constructs that go well beyond the theory of supergravity. However,
a feature most common to many extensions beyond supergravity is that 
it is recovered in a certain limit\footnote{This is another modern 
version of the correspondence principle.}. Although it is popular now to 
discuss a most fundamental theory in terms of ``M-theory,'' I will be 
a bit more conservative and limit my considerations to superstrings 
and heterotic strings.

These are mathematical constructions that are vastly more complicated
than is the construction of supergravity.  However, it is my personal
view that for even these theories, their ultimate formulations are
presently a mystery.  The solution of this mystery is the central
problem of ``covariant superstring (and heterotic string) field
theory.'' For although there has been progress in understanding 
superstring and heterotic string theory, the simple notion of a 
`stringy' field strength is at present unknown. My experience with 
the ultimate formulation of supergravity suggests that there still 
remains to be given ``geometrical'' formulations of superstring and 
heterotic string theory. I expect this belief to ultimately be realized 
at some point in the future. Then it is natural to expect that the 
supergeometry of 4D, N = 1 supergravity will be contained in the 
eka-geometry of some more fundamental theory just the non-Riemannian 
geometry of gravity is contained in the superspace geometry of 
Wess-Zumino superspace.

Rather than wait until these putative geometrical formulations
of superstring and heterotic strings are found, it is possible
to attempt to find the superspace supergravity limit of these by 
looking for ``special versions'' of 4D, N = 1 supergravity
directly. Although there is a large literature on this topic,
in 1988 \cite{27} a special version we now called $\b FFC$ supergravity 
was proposed to be this limit. Although additional time will
likely pass before a rigorous derivation proves that this is
the unique 4D, N = 1 supergravity theory associated with
eka-supergravity, strong evidence for this has started to
appear recently \cite{28}. So we will end our discussion of supergravity
with an eye toward the future.

A most distinctive feature of this theory is that it contains an
additional local symmetry (a U(1) gauge symmetry) for which there
does {\it {not}} appear a fundamental gauge field. In our book
``Superspace'' we call such a phenomenon a ``fake'' gauge symmetry.
Thus, the starting point of the $\b FFC$ theory is a superspace
supercovariant derivative of the form
\be
\nabla_{\underline A} ~=~ {\rm E}_{\underline A}
~+~ \o_{\underline A \, \underline c}{}^{\underline d} {\cal
M}_{\underline d}{}^{\underline c} ~-~ i\, \G_{\underline A}
\,{\cal Y} ~~~.
\ee
This looks very similar to the form of the supergravity-Yang-Mills
covariant derivative in (\ref{eq:hsv}). It possesses a crucial difference.
For an internal symmetry of the Yang-Mills theory we have
\be
[ \, t_{\cal I} ~, ~ \nabla_{\underline A} \, \} ~=~ 0 ~~~,
\ee
whereas for the generator ${\cal Y}$ instead we define
\be
\eqalign{
[ \, {\cal Y} ~, ~ \nabla_{\a} \, \} &=~ - \, \fracm 12 
\, \nabla_{\a} ~~~, \cr
[ \, {\cal Y} ~, ~ {\Bar \nabla}{}_{\a} \, \} &=~  \, \fracm 12 
\, {\Bar \nabla}{}_{\Dot \a} ~~~, \cr
[ \, {\cal Y} ~, ~ \nabla_{\underline a} \, \} &=~ 0 ~~~.
}\ee
So in a sense this generator has more in common with ${\cal M}
{}_{\underline a \, \underline b}$ for which it is also true 
that $[ {\cal M}{}_{\underline a \, \underline b}  \, ,  \, 
\nabla_{\underline A}  \} ~\ne ~ 0$. The complete set of generators
that do {\it {not}} commute with $\nabla_{\underline A}$ is called
the ``holonomy group.'' Thus the holonomy group of $\b FFC$ supergravity
is SL(2,C) $\otimes$ U(1). The appearance of this U(1) factor is
highly significant. Every known 4D, N = 1 truncation of heterotic 
or superstring theory possesses such a U(1) symmetry on its
world sheet.

The commutator algebra of $\b FFC$ supergravity is considerably
different from that of the minimal theory and is given by
\be
\eqalign{
[ \nabla_{\a }, \nabla_{\b} \}  &= 0   ~~, ~~ \cr
[ \nabla_{\a }, { \Bar {\nabla}}_{\dot \a} \}  &=~  
i {\nabla}_{ \underline  a} ~+~ H_{\b \dot \a} {\cal M}_{\a } {}^{\b} 
~-~ H_{\a \dot \b} {\Bar {\cal M}}_{\dot \a } {}^{\dot \b}~+~ 
H_{\underline  a} {\cal Y}  ~~~,~~ \cr
[ \nabla_{\a }, \nabla_{\underline  b} \}  &=~  i (\nabla_{\b} H_{\g 
\dot \b}) \, [~{\cal M}_{\a} {}^{ \g} ~+~ \d_{\a} {}^{\g} \, {\cal Y} ~]  
~~ \cr
&~~~~~+~ i  [~ C_{\a \b} \, {\bar W}_{\dot \b \dot \g} {}^{\dot \d}
~-~ \fracm 13 \d_{\dot \b} {}^{\dot \d} (\, 2 \nabla_{\a} H_{\b \dot 
\g} ~+~ \nabla_{\b} H_{\a \dot \g} \, ) ~ ] {\bar {\cal M}}_{\dot \d} 
{}^{\dot \g}  ~~~, \cr
[ \nabla_{\underline  a }, \nabla_{\underline  b} \}  &=~ \big \{ ~ 
\fracm 12  C_{\a  \b } [~ i \, H^{\g} {}_{ ( \dot \a} \nabla_{ \g \dot 
\b)} ~-~ ( \nabla^{\g} {\Bar \nabla}_{ ( \dot \a} H_{\g | \dot \b )} ) 
\, {\cal Y} 
~] \cr
&{~~~~}+~  [~ C_{\dot \a \dot \b} ( W_{ \a \b} {}^{ \g} ~-~ 
\fracm 16 ( {\Bar \nabla}^{\dot \g} H_{ ( \a \dot \g}) \d_{ \b )} 
{}^{\g}) ~-~ \fracm 12 C_{\a  \b} ( {\Bar \nabla}_{ ( \dot \a } 
H^{\g} {}_{\dot \b ) })~] \nabla_{\g} \cr
&{~~~~}-~  C_{\dot \a \dot \b} \, [~ W_{ \a \b \g \d} ~+~ i 
\fracm 14 C_{\g (\a | } ({\nabla}_{| \b ) }{}^{ \dot \e}  
H_{ \d \dot \e})  \cr 
&{~~~~}+ \fracm 1{12} C_{\g (\a }  C_{\b ) \d } ({\nabla}^{\e} 
{\Bar \nabla}^{\dot \e } H_{ \e \dot \e})~] {\cal M}^{\g \d} \cr 
&{~~~~}+ \fracm 12 C_{\a \b } [~{\nabla}_{ \g} {\Bar \nabla}_{ ( 
\dot \a } H_{\d} {}_{\dot \b ) } ~] {\cal M}^{\g \d} ~+~ {\rm 
{h.\, c.}} ~~~ \big \} ~~, 
 } \label{eq:uuv}
\ee
As we have described the theory above, we see that only the components
of the torsion supertensor ($T_{\a  \Dot \b \underline c}$ and 
$T_{\underline a \underline b \underline c}$) are non-zero.

The irreducible Bianchi identities associated with this
theory are given by,
$$
 \nabla^{\underline  a} H_{\underline  a} ~=~0 ~~,~~ {\Bar 
\nabla}_{\dot \b} W_{\a \b \g} ~=~ 0 ~~,~~  \nabla^{\b}\nabla_{\b} 
H_{\underline  a} ~=~ 0~~~, 
$$
\be
\nabla^{\a} W_{\a \b \g} ~=~ - \fracm 16 \nabla_{( 
\b} { \Bar \nabla}^{\dot \g} H_{\g ) \dot \g} ~-~  
\fracm 12 { \Bar \nabla}^{\dot \g} \nabla_{( \b} H_{\g ) \dot 
\g}~~~~.~~ 
\ee
As in our previous discussion, the effects of the constraints above 
are to determine now both $\o_{\underline A \, \underline a \underline 
b}$ and $\G_{\underline A}$ in terms of other quantities.  

In addition to the supergravity covariant derivative, there is
also the super 2-form $B_{\underline  A \, \underline  B}$ whose 
super exterior derivative $H_{\underline  A \, \underline  B \, 
\underline  C}$ satisfies the conditions,
$$ 
H_{\a \b \g} ~=~ H_{\a \b \dot \g} ~=~ H_{\a \b \g} ~=~ H_{\a \b 
\underline  c} ~=~ H_{\a \dot \b \underline  c} ~-~ i \fracm 12
C_{\a \g} C_{\dot \b \dot \g} ~=~ 0  ~~~,  $$
\be
{H}_{\a \underline  b \underline  c} ~=~ 0 ~~,~~ {H}_{\underline  a 
\underline  b \underline  c} ~=~ i \fracm 14  [~ C_{\b \g} C_{ \dot 
\a ( \dot \b } {H}_{\a \dot \g )} ~-~ C_{ \dot \b \dot \g } C_{ \a ( 
\b } {H}_{ \g ) \dot \a} ~ ]  ~~. \label{eq:uuw}
\ee
There is an interesting similarity of the components of the super 
3-form field strength and the components of the torsion supertensor. 
Namely, the only nonvanishing components of the latter are $H_{\a
\Dot \b \underline c}$ and $H_{\underline a \underline b \underline 
c}$. 

There is one more supertensor required to give a complete description
of 4D, N = 1 $\b$FFC supergravity.  We denote this additional 
tensor by $L$.  The spinorial derivatives of this superfield are 
particularly important since
\be
\eqalign{
\nabla_\a L &=~ -\fracm 43 \, \chi_\a  ~ ~~~~, ~~~~ 
\nabla_\a\chi_\b  ~=~ C_{\a\b}{\Bar R} ~~,~~\cr
 \nabla_\a{\bar \chi}_{\dot \a}  ~&=~ \fracm14 H_{\a \dot \a} 
~+~ \fracm 12 G_{\a \dot \a} ~-~ i \fracm{3}8 ~\nabla_{\underline  a} L
~ ~~.~~} 
\ee
Thus, the three basic field strength superfields of 4D, N = 1 $\b$FFC
supergravity are $W_{\a \b \g}$, $H_{\underline  a}$ and $L$.
However, note the total absence of $L$ in (\ref{eq:uuv}) and
(\ref{eq:uuw})

It is appropriate here to make some comments about how the component
fields arise in $\b$FFC supergravity in contrast to their appearance
in old minimal supergravity.  At lowest order in $\q$, the superfields
$W_{\a \b \g}$ and $H_{\underline  a}$ correspond to the 
supercovariantized ``curls'' of the gravitino and axion component 
fields.  The lowest component of $W_{\a \b \g \d}$ corresponds to 
the supercovariantized Weyl tensor (as usual the remaining components 
of the Riemann curvature are at higher orders in the superfields). 
The lowest order components of $L$ and $\chi_{\a}$ are the dilaton 
and dilatino, respectively. Finally the lowest order components 
of $G_{\underline a}$ and $R$ correspond to the usual auxiliary 
fields of old minimal supergravity.

Given the structure of $\b$FFC supergeometry, its action
is given by
\be
\eqalign{
{\cal S}_{\b FFC \, SG} &=~ - {~3~~ \over {\k^2}} \int \, d^4 x \, d^2 
\q \, \, d^2 {\bar \q} ~ { E}^{-1} \,  e^{f(L)} ~~~, } 
\ee
where $f$ is any function of $L$.  Even though the component 
fields of $\b FFC$ supergravity correspond to 
those of minimal supergravity ($({\rm e}_{\underline 
a}{}^{\underline m}(x), \, \psi_{\underline a}{}^{\b}(x) ,$ 
$ \, A_{\underline a}(x), S(x) , \, P(x))$  plus a super 
2-form multiplet $(\varphi(x), b_{\underline a \,\underline 
b}(x), \, \chi_{\b}(x)$) it is a remarkable fact that for 
$\b FFC$ supergravity,
\be
 - {~3~~ \over {\k^2}} \int \, d^4 x \, d^2 
\q \, d^2 {\bar \q} ~ {E}^{-1} ~=~ 0 ~~~.
\ee
We emphasize this result holds for the $\b FFC$ supervielbein 
that satisfies (\ref{eq:uuv}) not the one that satisfies 
(\ref{eq:uzw}). We believe the fact that the $\b FFC$ supergravity
action cannot be written without the use of $L$ (the field 
strength of the tensor multiplet) is a good indication that
the two multiplets are intimately related. That such a relation
should be present is well representative of heterotic and
superstring theory where it is known that both multiplets
describe the ``stringy'' supergravitational sector of the theories.

Within the confines of $\b FFC$ supergravity coupling to matter
fields is distinctly different than in the old minimal theory.
For example, a $\b FFC$ SG-chiral multiplet coupled system
has the form
\be
\eqalign{
{\cal S} &=~ - {~3~~ \over {\k^2}} \int \, d^4 x \, d^2 
\q \, \, d^2 {\bar \q} ~ { E}^{-1} \, \Big[ \,  e^{f(L)} \,
 \, + \, K(\Phi, \,{\Bar \Phi} ) \, \Big]  \cr
&{~~~~}+~ \Big\{~ \int \, d^4 x \, d^2 \q \,  ~ {\cal E}^{-1} \, W(\Phi)
~+~ {\rm {h.\, c.}} ~\Big\} ~~~, } 
\ee
and when the coupling of the spin-0 fields is evaluated, it can be 
shown to possess no ``improvement terms'' of the form $A {\Bar A} 
\eta^{\underline a \, \underline c} r_{\underline a \, \underline 
b \, \underline c}{}^{  \underline b}$. Evaluation of this superfield 
action using the old minimal supergeometry does to ``improvement 
terms.'' Finally, the chiral volume of $\b FFC$ supergravity is 
proportional to the mass-like term for the gravitino,
\be
\int \, d^4 x \, d^2 \q \,  ~ {\cal E}^{-1} ~=~  \fracm 12 \, 
\int \, d^4 x \, C^{\a \b} {\Bar \psi}_{\underline a}{}^{( 
\dot \a} \, {\Bar \psi}_{\underline b  }{}^{\dot \b )} 
~~~.
\ee

\newpage
\begin{center}
{\it {Epilogue}}
\end{center}
$${~}$$
~~~~We have come to the end of this series of lectures. I have
a great desire to say, ``Free at last, free at last,...,'' but
these have only begun to brush the surface of the current
developments in modern relativistic theories of fundamental
objects. I have enjoyed lecturing to an enthusiastic set of aspiring
physicists. Beyond the simply mastery of numbers of topics that 
were touched upon in the lectures there looms superstring and 
heterotic string field theory and even more fearsome `beasts.' 

The role of superstrings and heterotic strings is to `control'
many aspects of the theories we have discussed. For example,
super and heterotic strings are very restrictive and arbitrary 
choices for gauge groups (i.e. $V^{\cal I}$), numbers and 
representations of matter (i.e. $\Phi^{\rm I}$ and perhaps 
even $\S^{\rm I}$) and `stringy' supergravity (i.e. $\b FFC$ SG) 
are not in general consistent. Thus, 4D, N = 1 super and heterotic 
string theory only permit special choices. In the realm of quantum 
theories, 4D, N = 1 superstrings and heterotic strings are the only
known models where nonlinear expressions like (\ref{eq:wze}),
(\ref{eq:fff}), (\ref{eq:ggg}), (\ref{eq:gggg}), (\ref{eq:wee}) 
and (\ref{eq:hhh}) may derived to arbitrary order in any of the
fields. So the long dreamt about reconciliation between quantum 
theory and gravitation has apparently been accomplished. From
this point of view, it remains to simply discover seemingly
the answer to a few things.

Even beyond strings, more recently, there has been suggested
the existence of `M-theory' and relatives. However, I believe
that in our rush to move even further from the known physics
of fundamental particles there have been left unsolved major 
problems even in the realm of string theory. The principal
among these is that we do not possess a fundamental understanding
of `pre-geometry' beyond the realm of supergravity. As we have
seen in these lectures, Salam-Strathdee superspace provides
a comprehensive concept upon which to investigate supersymmetrical
theories. It is a fact that for even the simplest
superstring and heterotic string theories, there does not
exist the analog of Salam-Strathdee superspace.  While many
physicists have regarded this as of not much import and 
have continued to increase our understanding of ever more 
imaginative mathematical constructions, it is my opinion
that until we possess a truly geometrical understanding
of strings we will not have achieved the same comprehensive
level of understanding of these theories as that which is 
Einstein's legacy embodied by the ``equivalence principle.''
This lack in our understanding will continue to offer a deep 
challenge to the theoretical understanding of fundamental 
physics.

\pagebreak

\section*{Acknowledgments}
~~~~~~I wish to acknowledge the support for my research program
under grants \# PHY-96-43219 from the U.S. National Science 
Foundation and \# CRG-93-0789 from NATO.

\section*{Bibliography}

~~~~~~For the student wishing to continue to study supersymmetry 
there are by now many textbooks available. At the introductory 
level, I highly recommend \newline

${~~}$[1.] J. Wess and J. Bagger, ``Supersymmetry and 
Supergravity," Prince-  \newline \indent ${~~~~~~}$ ton 
Univ. Press, Princeton, NJ (1983). \newline

${~~}$[2.] P. West, ``Introduction to Supersymmetry and 
Supergravity," World  \newline \indent ${~~~~~~}$
Scientific, Singapore (1986). \newline
${~~}$ \newline

At the comprehensive level, there are only two books that treat 
the subject in real depth.\newline

${~~}$[3.] S.J.Gates, Jr., M.T.Grisaru, M. Ro\v cek and W.~Siegel, 
``Superspace \newline \indent ${~~~~~~}$
or One Thousand and One Lessons in Supersymmetry," Benjamin 
\newline \indent ${~~~~~~}$ Cummings 
(Addison-Wesley), Reading, MA (1983). \newline

${~~}$[4.] I. Buchbinder and S. Kuzenko, ``Ideas and Methods 
of Supersym-  \newline \indent ${~~~~~~}$ metry and Supergravity 
or A Walk Through Superspace,"  IOP  \newline \indent ${~~~~~~}$
Publishing, Bristol and Philadelphia, (1995). \newline

The first of these is out of print, so you may have to look in your
library or write to the publisher.  These latter two books nicely
compliment each other covering slightly different matters.  If you
master these things, you will be an expert.

\end{document}